\documentclass[usenatbib, fleqn]{aa}
\usepackage{graphicx}
\usepackage{amsmath}
\usepackage{multicol}
\usepackage{color}
\usepackage{scrextend}
\usepackage[pdftex, linktocpage=true]{hyperref}
\hypersetup{
	colorlinks= true, 
	urlcolor= blue, 
	linkcolor=blue, 
	citecolor=blue 
}
\usepackage[normalem]{ulem}
\usepackage{fancyhdr}
\usepackage{caption}
\usepackage{subcaption}
\usepackage{float}
\usepackage{wrapfig}
\usepackage{pdflscape}
\usepackage{makecell}
\usepackage{amssymb}
\usepackage{booktabs}
\usepackage{cuted}
\usepackage[normalem]{ulem}

\usepackage[flushleft]{threeparttable}
\newcommand\Tstrut{\rule{0pt}{2.6ex}}         
\newcommand\Bstrut{\rule[-0.9ex]{0pt}{0pt}}
\usepackage{txfonts}
\usepackage{multirow}

\usepackage{setspace}
\defcitealias{Mehdipour_Prep}{Paper I}
\defcitealias{Wang_Prep}{Paper II}
\defcitealias{Mao_Prep}{Paper III}

\begin{document}
	
	\title{Transient obscuration event captured in NGC 3227 \\
		IV. Origin of the obscuring cloud variability}
	
	\author{S. Grafton-Waters \inst{\ref{MSSL}} \and
		J. Mao \inst{\ref{Hiroshima}, \ref{Strathclyde}, \ref{SRON}} \and
		M. Mehdipour \inst{\ref{STSI},\ref{SRON}} \and
		G. Branduardi-Raymont \inst{\ref{MSSL}} \and
		M. Page \inst{\ref{MSSL}} \and
		J. Kaastra \inst{\ref{SRON}, \ref{Leiden}} \and
		Y. Wang \inst{\ref{SRON}, \ref{Leiden}, \ref{Hefei}, \ref{Hefei_School}} \and
		C. Pinto \inst{\ref{Palermo}} \and
		G. A. Kriss \inst{\ref{STSI}} \and
		D. J. Walton \inst{\ref{Cambridge}} \and
		P.-O. Petrucci \inst{\ref{Grenoble}} \and
		G. Ponti \inst{\ref{INAF_Brera}, \ref{MaxPlanck}} \and
		B. De Marco \inst{\ref{Barcelona}} \and
		S. Bianchi \inst{\ref{Roma}} \and
		E. Behar \inst{\ref{Israel}}\and
		J. Ebrero \inst{\ref{ESAC}}
	}
	
	\institute{
		Mullard Space Science Laboratory, University College London, Holmbury St. Mary, Dorking, Surrey, RH5 6NT, UK \\ \label{MSSL} \email{sam.waters.17@ucl.ac.uk} \and
		Department of Physics, Hiroshima University, 1-3-1 Kagamiyama, Higashi-Hiroshima, Hiroshima 739-8526, Japan \label{Hiroshima} \and
		Department of Physics, University of Strathclyde, Glasgow, G4 0NG, UK \label{Strathclyde} \and
		SRON Netherlands Institute for Space Research, Niels Bohrweg 4, 2333 CA Leiden \label{SRON} \and
		Space Telescope Science Institute, 3700 San Martin Drive, Baltimore, MD 21218, USA \label{STSI} \and
		Leiden Observatory, Leiden University, PO Box 9513, 2300 RA Leiden, the Netherlands \label{Leiden} \and
		CAS Key Laboratory for Research in Galaxies and Cosmology, Department of Astronomy, University of Science and Technology of China, Hefei 230026, China \label{Hefei} \and
		School of Astronomy and Space Science, University of Science and Technology of China, Hefei 230026, China \label{Hefei_School} \and
		INAF-IASF Palermo, Via U. La Malfa 153, I-90146 Palermo, Italy \label{Palermo} \and
		Institute of Astronomy, University of Cambridge, Madingley Road, Cambridge CB3 0HA, UK \label{Cambridge} \and
		Univ. Grenoble Alpes, CNRS, IPAG, 38000 Grenoble, France \label{Grenoble} \and
		INAF-Osservatorio Astronomico di Brera, Via E. Bianchi 46, 23807 Merate (LC), Italy \label{INAF_Brera} \and
		Max Planck Institute fur Extraterrestriche Physik, 85748 Garching, Germany \label{MaxPlanck} \and
		Departament de Física, EEBE, Universitat Politècnica de Catalunya, Av. Eduard Maristany, 16, Barcelona 08019, Spain \label{Barcelona} \and
		Dipartimento di Matematica e Fisica, Università degli Studi Roma Tre, via della Vasca Navale 84, 00146 Roma, Italy\label{Roma} \and
		Department of Physics, Technion-Israel Institute of Technology, 32000 Haifa, Israel \label{Israel} \and
		Telespazio UK for the European Space Agency (ESA), European Space Astronomy Centre (ESAC), Camino Bajo del Castillo, s/n, 28692 Villanueva de la Cañada, Madrid, Spain \label{ESAC}
	}

	\date{Received date / Accepted date}
	
	\abstract
	{ Obscuration events in type I active galactic nuclei (AGN) have been detected more frequently in recent years. The strong flux decrease in the soft X-ray band between observations has been caused by clouds with large column densities transiting our line of sight (LOS) and covering the central AGN. Another event has been captured in NGC 3227 at the end of 2019, which was observed with \textit{XMM-Newton}, \textit{NuSTAR}, and the \textit{Hubble} Space Telescope.}
	{We aim to determine the nature and origin of the observed spectral variability in the 2019 obscuration event. }
	{We study the evolution of the obscurer by splitting the two XMM-Newton observations from 2019 into timing bins of length $\sim 10$ ks. We used the \texttt{SPEX} code to analyse the 0.35 - 10 keV EPIC-PN spectra of each timing bin.}
	{In the first observation (Obs 1), there is a strong anti-correlation between the column density ($N_{\rm H}$) of the obscurer and the continuum normalisations of the X-ray power law and soft Comptonisation components ($N_{\rm pow}$ and $N_{\rm comt}$, respectively). The power-law continuum models the hard X-rays produced by the corona, and the Comptonisation component models the soft X-ray excess and emission from the accretion disk. 
		Through further testing, we conclude that the continuum is likely to drive the observed variability, but we cannot rule out a possible contribution from $N_{\rm H}$ of the obscurer if it fully transverses across the ionising source within our LOS during the observation. The ionisation parameter ($\xi$) of the obscurer is not easily constrained, and therefore it is not clear whether it varies in response to changes in the ionising continuum. The second observation (Obs 2) displays a significantly lower count rate due to the combination of a high $N_{\rm H}$ and covering fraction of the obscurer, and a lower continuum flux.}
	{The observed variability seen during the obscuration event of NGC 3227 in 2019 is likely driven by the continuum, but the obscurer varies at the same time, making it difficult to distinguish between the two possibilities with full certainty.}
	
	
	\titlerunning{Origin of the obscuring cloud variability in NGC 3227}
	\authorrunning{Grafton-Waters et al.}
	\maketitle

	\section{Introduction}
	
	Obscuration events in active galactic nuclei (AGN) have started being detected more readily in recent years. The first major event was seen in NGC 5548 \citep{Kaastra2014} where the \textit{XMM-Newton} soft X-ray flux in 2013 was 25 times weaker compared to the 2002 Chandra data. Other AGN that have undergone obscuring events include NGC 3783 \citep{Mehdipour2017, Kriss2019}, Mrk 335 \citep{Longinotti2013, Longinotti2019, Parker2019}, NGC 985 \citep{Ebrero2016}, NGC 3227 \citep{Turner2018}, and ESO 033-G002 \citep{Walton2021}. In addition, Rossi X-ray Timing Explorer (RXTE) monitoring by \cite{Markowitz2014} detected 12 new events in eight objects, significantly increasing the number observed in AGN. It is possible that many AGN have undergone obscuration at some point in their lifetime and these events are more common than first thought \citep{Kaastra2018}. For example, multiple obscuration events were seen in NGC 3783 between 1993 and 2016 \citep{Kaastra2018}, including the December 2016 event lasting 32 days \citep{Mehdipour2017}, while NGC 3227 has undergone many occultations lasting days, weeks and months \citep{Lamer2003, Markowitz2014, Beuchert2015, Turner2018}. Comparatively, some AGN, such as Seyfert 2 or extreme Compton-thick AGN, are persistently obscured due to their geometry or a viewing angle that prevents our line of sight (LOS) from reaching the innermost regions near the supermassive black hole (SMBH). Obscuring clouds have different properties compared to the warm absorbers \citep[WAs; e.g.][]{Kaastra2002, Grafton-Waters2020}, ultra-fast outflows \citep[UFOs; e.g.][]{Risaliti2005, Tombesi2013, Nardini2015}, and the emission line regions \citep[e.g.][]{Kinkhabwala2002, Grafton-Waters2021}, such as column densities ($N_{\rm H, WA} = 10^{24} - 10^{26}$ m\textsuperscript{-2}) and outflow velocities ($v_{\rm out, WA} =$ few $100 - 1000$ km s\textsuperscript{-1}; $v_{\rm out, UFO} = 0.05 - 0.4 c$).
	
	Since the obscuring event in NGC 5548 \citep{Kaastra2014}, we started a \textit{Swift} \citep{Gehrels2004} monitoring programme to trigger joint Target of Opportunity (ToO) observations of Seyfert 1 AGN with \textit{XMM-Newton} \citep{Jansen2001}, the \textit{Hubble} Space Telescope \citep[HST;][]{Green2012}, and \textit{NuSTAR} \citep{Harrison2013}. NGC 3783 was triggered in December 2016 as the \textit{Swift} hardness ratio was large enough to signify an obscuration event, caused by a high-velocity wind crossing our LOS \citep{Mehdipour2017}. The column densities of the two obscuring components in NGC 5548 and NGC 3783 were $N_H = 10^{26} - 10^{27}$ m\textsuperscript{-2} \citep{Kaastra2014, Mehdipour2017}. These obscuring clouds were likely to be located within the broad line region (BLR), originating from the accretion disk \citep{Kaastra2014, Mehdipour2017, Kriss2019}.
	
	The type 1.5 AGN, NGC 3227 \citep[z = 0.003859;][]{deVaucouleurs1991}, has shown strong variability in both X-ray and UV bands, particularly during periods of high flux \citep{Markowitz2009, Arevalo2014}, with a `steeper-when-brighter' trend consistent with other AGN \citep{Lobban2020}. The intrinsic X-ray continuum in NGC 3227 has been found to vary across many timescales, from days to months \citep[e.g.][]{George1998, Gondoin2003, Uttley2005, Lobban2020}. However, obscuration events can cause additional contributions to the observed variability in sources such as NGC 3227 \citep[e.g.][]{Beuchert2015}. This is similar to the explanations for the variability in NGC 4151, which are both driven by the intrinsic continuum and absorbing material within the LOS \citep[e.g.][]{Keck2015, Beuchert2017}. 
	
	
	
	Two \citep{Markowitz2009, Markowitz2014} and three \citep{Crenshaw2001, Beuchert2015} ionised absorbers have been found in NGC 3227 during different epochs. The variable absorption event in NGC 3227 during 2008, analysed by \cite{Beuchert2015}, was caused by a mildly ionised component ($\log \xi \sim 1.2 - 1.4$ erg cm s\textsuperscript{-1}) with increasing covering fraction ($C_f$; from 0.7 to 0.9) and column density ($N_{\rm H}$; from 5 to $18 \times 10^{26}$ m\textsuperscript{-2}), possibly located within the BLR or inner torus \citep{Beuchert2015}. Furthermore, an \textit{XMM-Newton} and \textit{NuSTAR} campaign in 2016 found a rapid variability event lasting a day \citep{Turner2018}. To explain the observed spectral hardening and a depth change of the unresolved transition array (UTA; at $\sim 16 - 17$ \AA) in the RGS data, a component covering 60\% of the X-ray continuum was required in the modelling (with $N_H = 5 \times 10^{26}$ m\textsuperscript{-2}; $\log \xi = 2$; $v_{out} \sim -800$ km s\textsuperscript{-1}). This obscurer was also associated with clouds within the BLR.
	
	During the Swift Cycle 16 ToO monitoring programme, an obscuration event in NGC 3227 was captured at the end of 2019, triggering observations with \textit{XMM-Newton}, \textit{NuSTAR} and HST. Strong spectral hardening and intrinsic X-ray variability was found between the two 2019 observations taken three weeks apart, along with strong soft X-ray absorption compared to the 2016 observations \citep[see Figs. 1 and 2 in][]{Mehdipour_Prep}. In the first paper of this series \citep[][hereafter Paper I]{Mehdipour_Prep}, we determined the broadband continuum, giving rise to the spectral energy distribution (SED) of NGC 3227. Analysis of the archival 2006 and 2016 \textit{XMM-Newton} observations were carried out by \cite{Wang_Prep} (hereafter Paper II), where we used the broadband SED from \citetalias{Mehdipour_Prep} to model the WA wind. Four WA components were found with ionisation parameters\footnote{The ionisation parameter is defined as $\xi = \frac{L_{ion}}{n_H r^2}$, where $L_{ion}$ is the ionising luminosity (measured between 13.6 eV and 13.6 keV), $n_{\rm H}$ is the hydrogen number density, and $r$ is the distance of the plasma from the black hole.} ranging from $\log \xi = -1$ to $3$, and outflow velocities in the range $v_{\rm out} = -100$ to $-1300$ km s\textsuperscript{-1}, located between the BLR and narrow line region (NLR). A missed obscuring event in 2006 was found for the first time in \citetalias{Wang_Prep}, while the event in 2016 \citep{Turner2018} was re-evaluated. The 2006 obscurer was explained with one component, but the 2016 obscurer required two: $\log \xi = 1 - 1.9$ and $N_H = 10^{26}$ m\textsuperscript{-2} for both epochs, and $\log \xi = 2.8$ and $N_H = 10^{27}$ m\textsuperscript{-2} was needed for 2016 only. The obscurers were located within the BLR. In \cite{Mao_Prep} (Paper III, hereafter), we studied the 2019 obscuring event in detail, focusing on the time-averaged spectra through modelling the RGS, EPIC-PN, and \textit{NuSTAR} data. Different models were applied to each 2019 observation, with one or two photoionisation components required to account for the obscurer. The estimated location for the obscurer was comparable to the BLR and inner torus. In this paper, we study the variability of the obscurer and continuum in NGC 3227, focusing on the EPIC-PN data only.
	
	The outline of this paper is as follows. We describe how we reduced the data for analysis in Sect. \ref{Sec:Data_Reduction}, while in Sect. \ref{Sec:Observation_analysis} we investigate the light curves of each observation and explain how we split each observation into equal length timing bins. The spectral fitting and modelling process is explained in detail in Sect. \ref{Sec:Spectral_Fitting}, before we present the results and discuss our findings in Sect. \ref{Sec:Results_Discussion}. Finally, we give our concluding remarks in Sect. \ref{Sec:Conclusions}. 
	
	\begin{figure*}
		\centering
		\begin{subfigure}{0.5\linewidth}
			\includegraphics[width=1\linewidth]{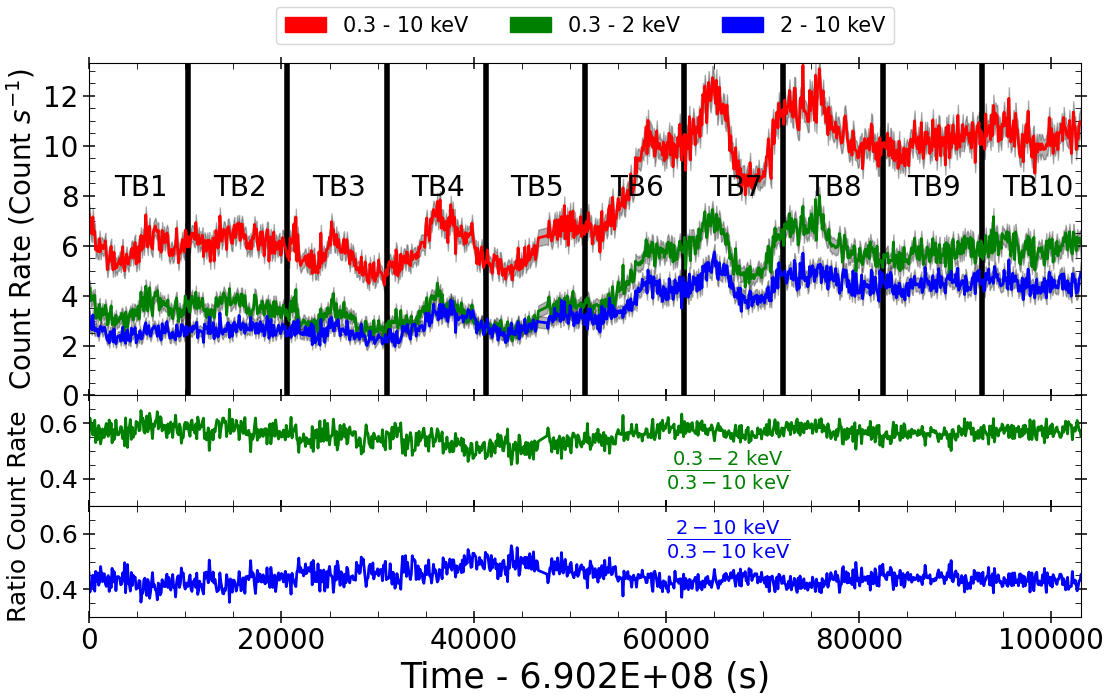}
		\end{subfigure}%
		\begin{subfigure}{0.5\linewidth}
			\includegraphics[width=1\linewidth]{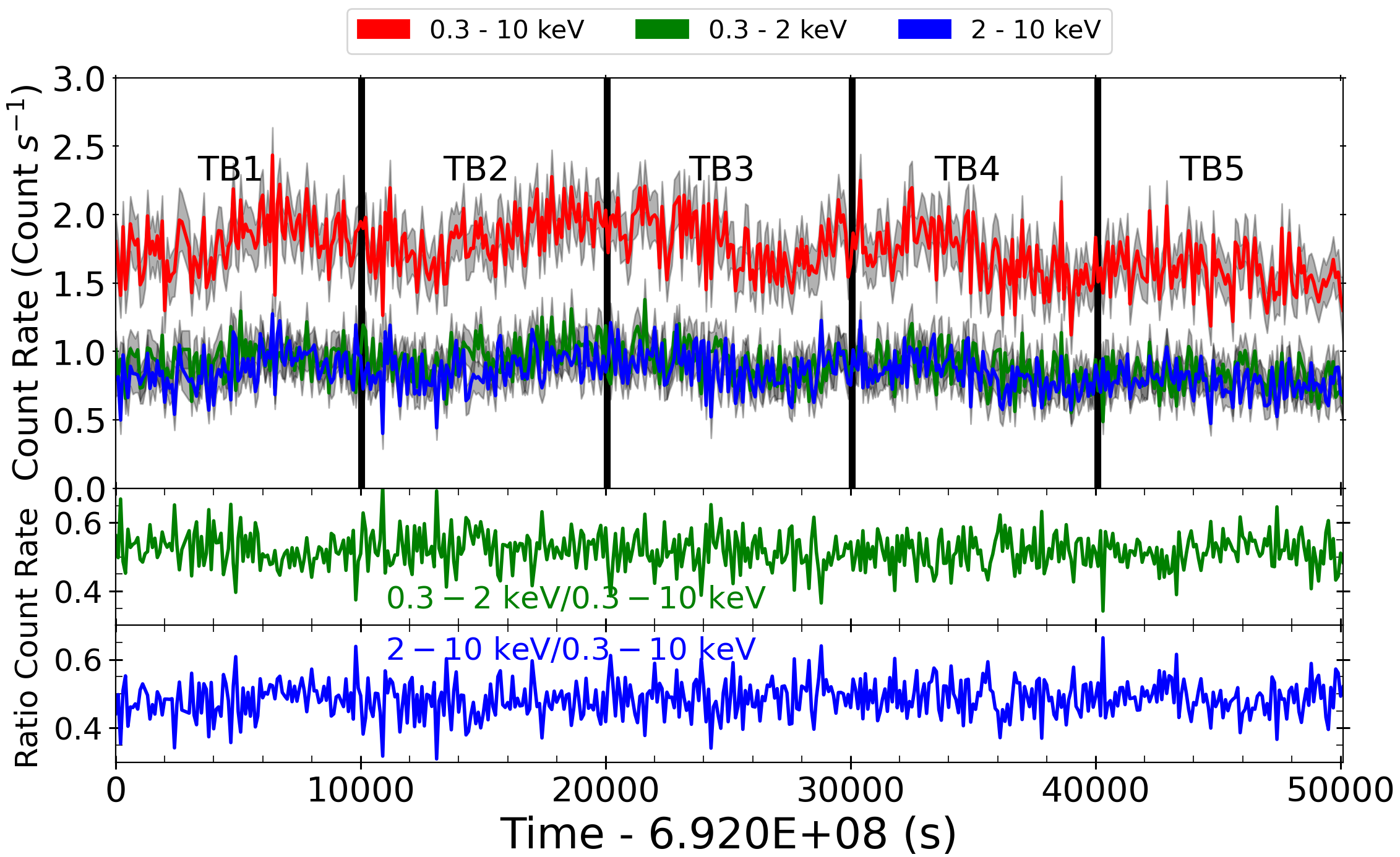}
		\end{subfigure}
		\caption{ Light curves for Obs 1 (left) and Obs 2 (right). \textit{Top panels:} three different energy bands: 0.3 - 10 keV (red), 0.3 - 2 keV (green), and 2 - 10 keV (blue). Also displayed are the timing bins (TB) from which we extracted EPIC-PN spectra for our analysis (see Sect. \ref{Sec:Timing_Bins} for details). The light curves were binned at 100 s. The errors on the energy bin fluxes are shown by the shaded regions.
			\textit{Bottom panels:} Ratio light curves between the 0.3 - 2 keV (green) and 2 - 10 keV (blue) energy bands in the top panel divided by the full 0.3 - 10 keV light curve (red line). The Obs 1 ratios are on the left hand side and the Obs 2 ratios are on the right.}
		\label{Fig:Light_Curves}
	\end{figure*}
	
	\section{Data reduction}
	\label{Sec:Data_Reduction}
	\textit{XMM-Newton} observed NGC 3227 twice during an X-ray obscured state at the end of 2019. The two observations (Obs 1 and Obs 2, hereafter) lasted 103 and 50 ks, respectively, with the details shown in the observation log (Table \ref{Table:Obs_Log}). Here we report on the analysis of the European Photon Imaging Camera pn-CCD \citep[EPIC-PN;][]{Struder2001} spectrum, taken in small-window mode, modelled between 0.35 - 10 keV. The EPIC-PN data were reduced using the \texttt{EPPROC} command in the \texttt{SAS (v18.0.0)} pipeline, where we removed background flaring with counts greater than 0.4 counts s\textsuperscript{-1} in 10 - 12 keV. 
	We extracted the source spectrum using a circle of radius 40 arcseconds and the background spectra were extracted with two circular regions of 30 arcseconds in radius from a source free region on the same CCD; both with quality photons only (\texttt{FLAG} == 0) and using the \texttt{\#XMMEA\_EP} filter. We included both single and double events ($\texttt{PATTERN} <= 4$). Finally, the instrumental response matrices were generated with the \texttt{rmfgen} and \texttt{arfgen} commands. The data were binned using the \cite{KB2016} optimal binning method and we fitted the spectra using \texttt{SPEX} \citep[\texttt{v3.05.00;}][]{Kaastra1996, SPEX305}. We use the Cash statistic \citep[C-statistic, hereafter;][]{Cash1979, Kaastra2017} for statistical significance, with all errors in this paper quoted at $1\sigma$ confidence.
	
	\begin{figure*}
		\centering
		\begin{subfigure}{0.5\linewidth}
			\includegraphics[width=1\linewidth]{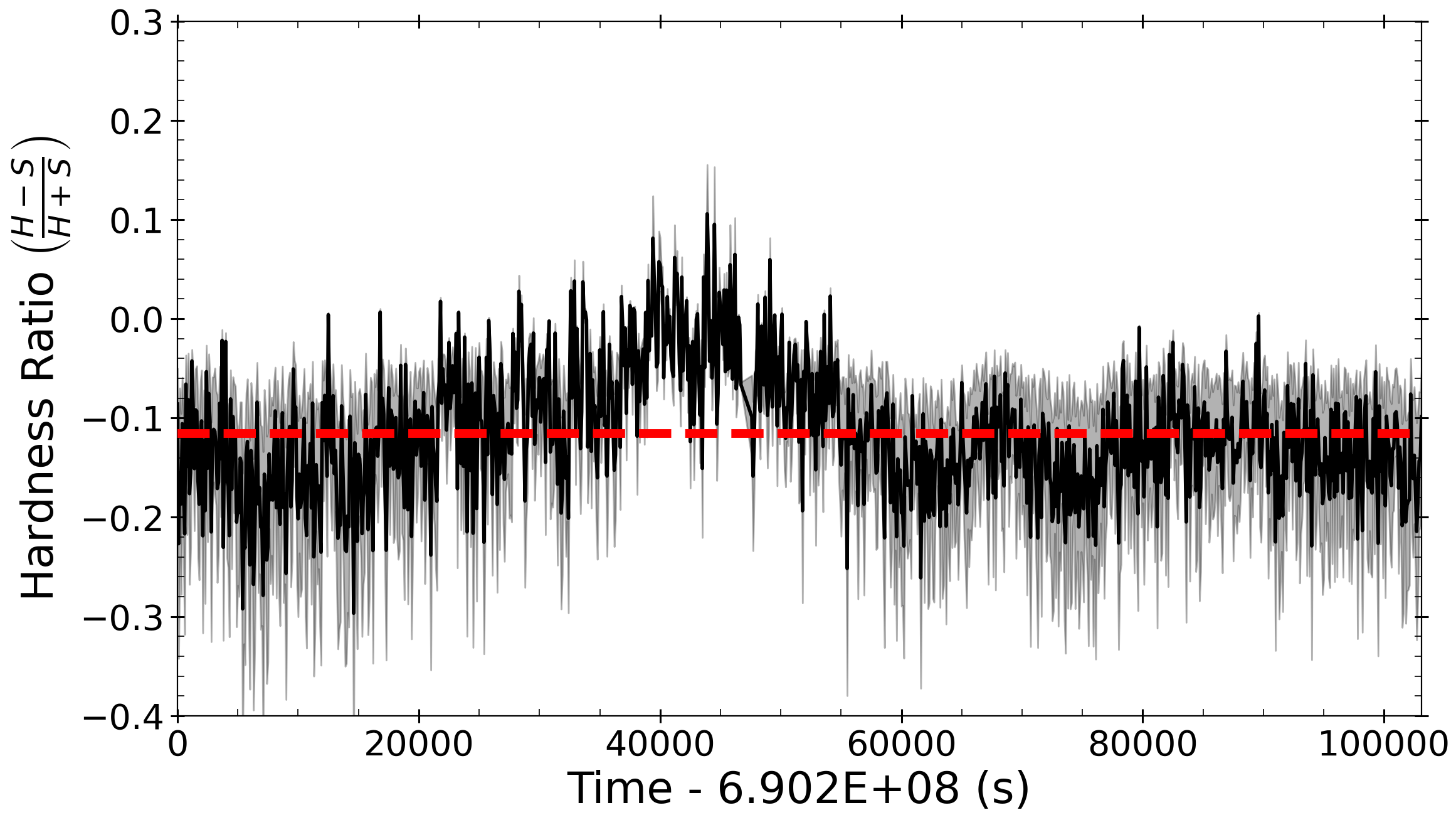}
		\end{subfigure}%
		\begin{subfigure}{0.5\linewidth}
			\includegraphics[width=1\linewidth]{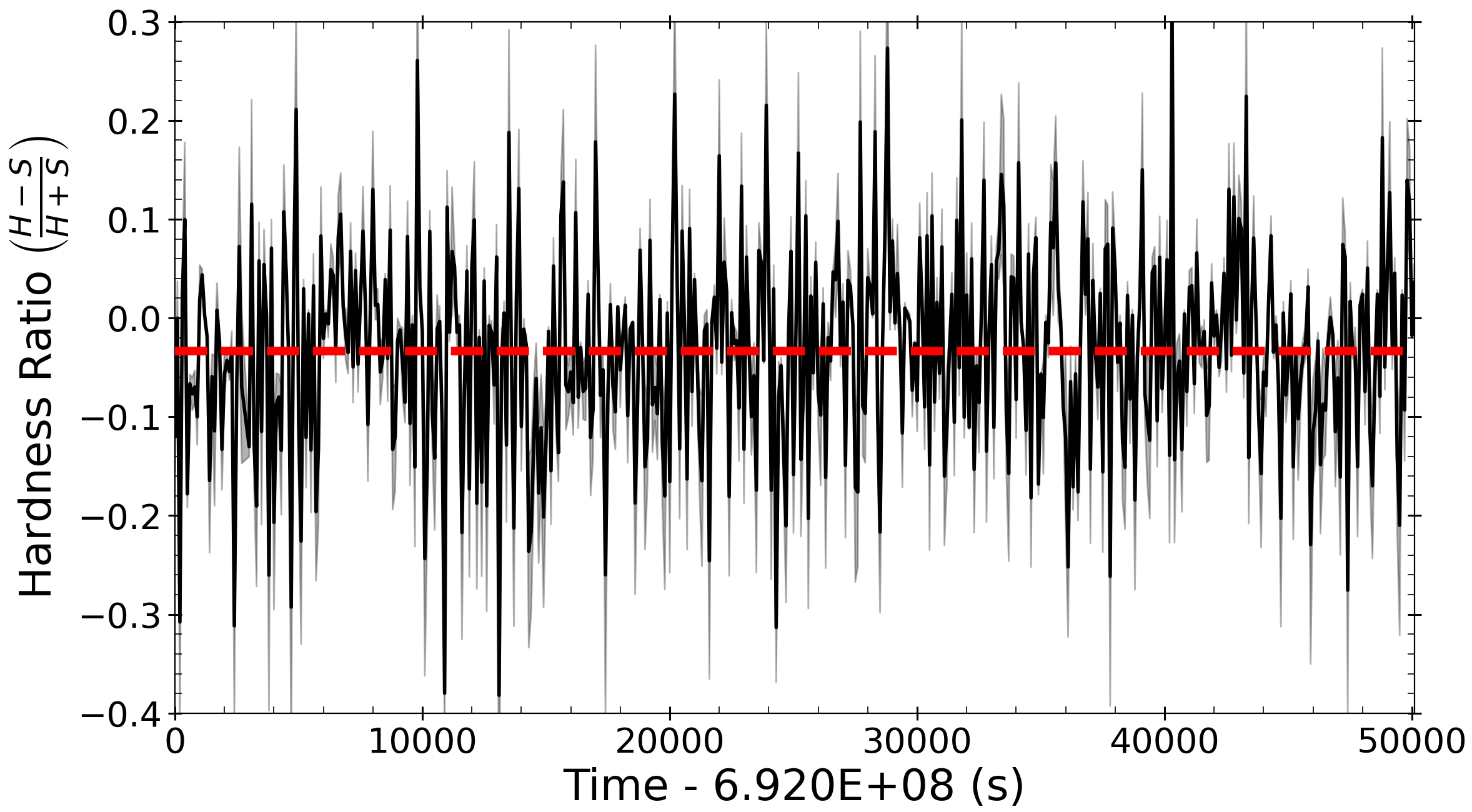}
		\end{subfigure}
		\caption{Hardness ratio between the 2 - 10 keV and 0.3 - 2 keV energy bands, defined as (H-S)/(H+S), for Obs 1 (left) and Obs 2 (right), binned at 100 s. The errors on the hardness ratios are shown by the shaded regions. The red dashed lines show the average hardness ratio ($HR_{avg}$) across each observation. For Obs 1 $HR_{avg} = -0.115$, and for Obs 2 $HR_{avg} = -0.033$.}
		\label{Fig:Hardness_Ratios}
	\end{figure*}
	
	\section{Light curves and timing bins}
	\label{Sec:Observation_analysis}
	\subsection{Light curves}
	\label{Sec:Light_Curves}
	
	The three different energy band light curves for each observation are displayed in Fig. \ref{Fig:Light_Curves}: 0.3 - 10 keV (red; full X-ray range), 0.3 - 2 keV (green; soft X-ray band), and 2 - 10 keV (blue; hard X-ray band). They were extracted using the \texttt{EPICLCCORR} command with \texttt{SAS}, and chosen to cover large enough energy ranges with sufficient counts within the full 0.3 - 10 keV band. Obs 1 (left side of Fig. \ref{Fig:Light_Curves}) shows large variability over the course of the observation, whereas Obs  2 (right side) has far fewer counts and is less variable. To determine how the soft and hard X-ray energy bands varied with respect to each other, we took the ratios between the soft or hard X-ray light curves and the full 0.3 - 10 keV range (shown within the bottom two panels in Fig. \ref{Fig:Light_Curves}). The only strong feature in Fig. \ref{Fig:Light_Curves} (bottom two panels) is during Obs 1 where, between 30 and 60 ks, the 0.3 - 2 keV / 0.3 - 10 kev ratio (green line) decreases, but the 2 - 10 keV / 0.3 - 10 kev (blue) ratio increases. No changes are seen in Obs 2.
	
	We also obtained the hardness ratio (HR) for each observation, which is defined as (H - S)/(H + S), where H and S are the 2 - 10 keV hard band and 0.3 - 2 keV soft band count rates, respectively. For Obs 2 (right side of Fig. \ref{Fig:Hardness_Ratios}), the ratio stays constant over the observing window. However, for Obs 1 (left side of Fig. \ref{Fig:Hardness_Ratios}), there is a strong spectral hardening between 30 and 60 ks, corresponding to the change in soft (0.3 - 2 keV) and hard (2 - 10 keV) flux ratios over this period (see left panel of Fig. \ref{Fig:Light_Curves}). This is either caused by variations in the underlying continuum, or the possible variability of the obscurer, which is what we are investigating in this paper. The rest of the observation shows that NGC 3227 is in a relatively softer state (negative HR), and is consistent with the average value (red dashed line).
	
	\begin{table}
		\centering
		\caption{Observation log for NGC 3227 in 2019.}
		\label{Table:Obs_Log}
		\begin{tabular}{c | c c c c }
			\hline
			\hline
			\multirow{2}{*}{Obs}    &  Obs    &  Start   &  Duration  & Avg. CR \Tstrut\Bstrut \\
			& ID & Date & (ks) & (count s\textsuperscript{-1})\Tstrut\Bstrut \\
			\hline
			1    &  	0844341301   &  2019/11/15   &  103 & 8.01 \Tstrut\Bstrut \\
			2   &  0844341401   &  2019/12/05    &  50 & 1.74  \Tstrut\Bstrut \\
			\hline
		\end{tabular}
	\end{table}
	
	\begin{figure}
		\centering
		\begin{subfigure}{1\linewidth}
			\includegraphics[width=0.999\linewidth]{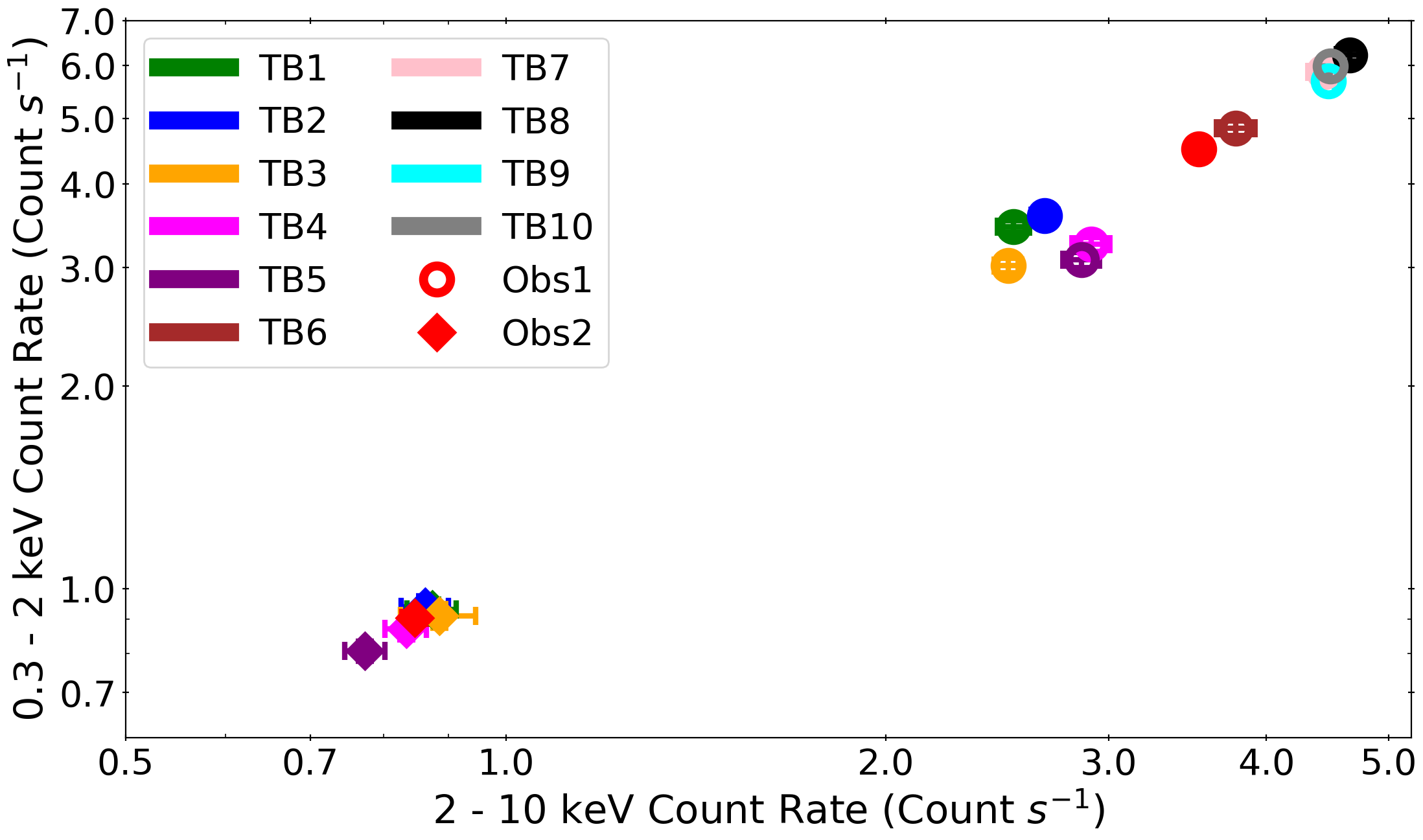}
		\end{subfigure}
		\begin{subfigure}{1\linewidth}
			\includegraphics[width=1.009\linewidth]{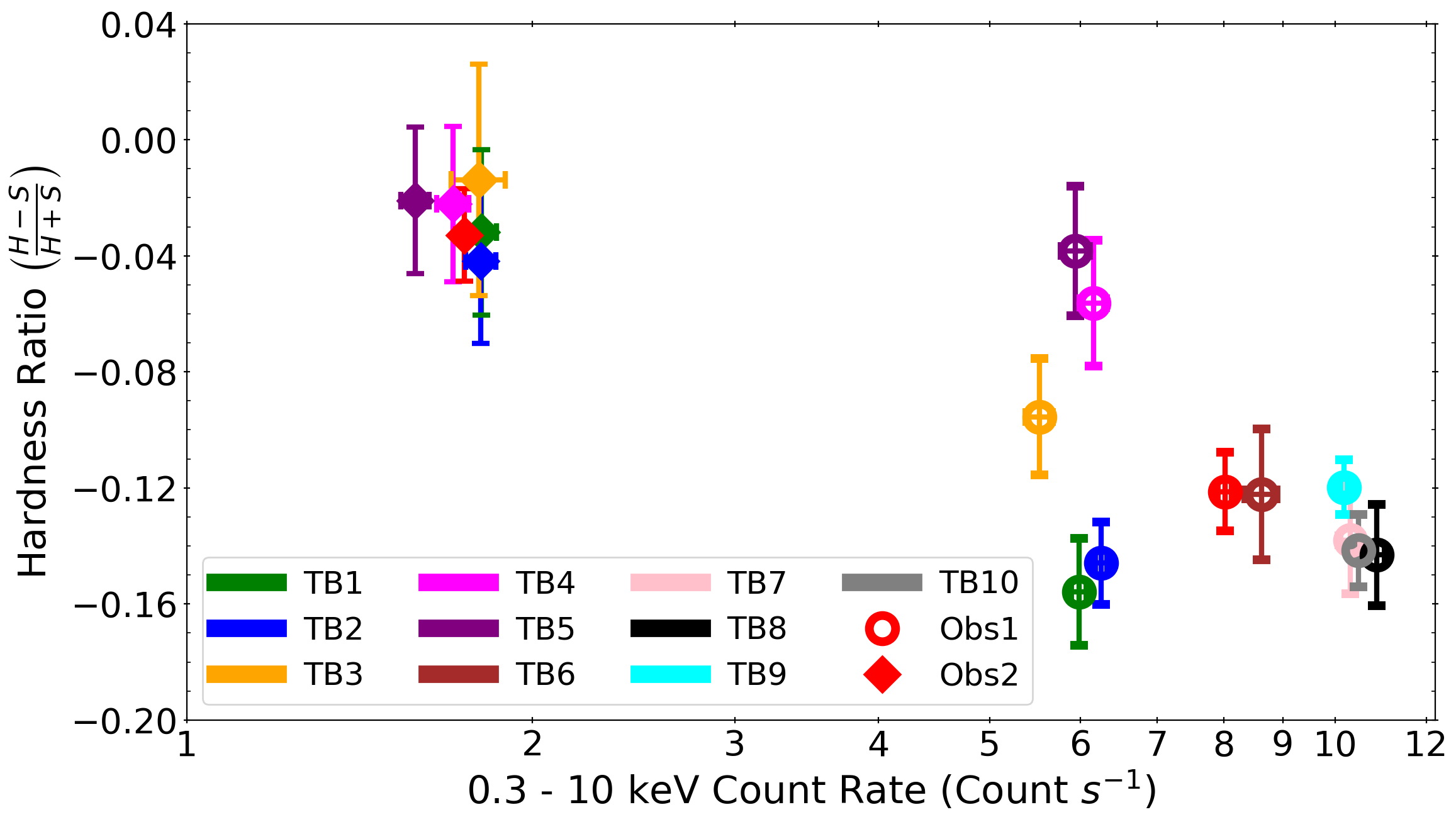}
		\end{subfigure}
		\caption{\textit{Top:} Relation between the averaged 0.3 - 2 keV (soft) and 2 - 10 keV (hard) count rates for each timing bin. Obs 1 (hollow circles) and Obs 2 (diamonds) are compared with the time-averaged values shown in red. \textit{Bottom:} Hardness ratio (H-S / H+S) as a function of the average 0.3 - 10 keV count rates. NGC 3227 shows a softer-when-brighter trend on timescales of weeks as Obs 1 has a larger flux but lower hardness ratio compared to Obs 2.}
		\label{Fig:Counts_Comp_Obs}
	\end{figure}
	
	\subsection{Timing bins}
	\label{Sec:Timing_Bins}
	
	To measure any spectral changes over the duration of the observations, we split each observation into equal length timing bins (TBs) and obtained the full 0.35 - 10 keV spectrum from each. Obs 1 was split up into 10 TBs of length 10.3 ks, while Obs 2 was split into 5 bins with a duration of 10 ks (see Fig. \ref{Fig:Light_Curves}). The top panel of Fig. \ref{Fig:Counts_Comp_Obs} exhibits a strong correlation between the averaged 0.3 - 2 and 2 - 10 keV count rates of each TB in the two observations. The bottom panel of Fig. \ref{Fig:Counts_Comp_Obs} presents the average HR for each TB as a function of the averaged 0.3 - 10 keV count rates. The average HR is lower for each TB in Obs 1 compared to the TBs in Obs 2, but the average count rates are larger. This agrees with the overall flux change in the light curves for each observation, whereby both panels in Fig. \ref{Fig:Counts_Comp_Obs} imply that NGC 3227 shows a softer-when-brighter correlation (similar to, e.g. \citealt{Lobban2020}; \citetalias{Wang_Prep}). This relation is typically associated with intrinsic variability in the continuum, either from steepening of the power law or an increased strength in the soft excess. The increased hardening in Obs 1 from TB1 - TB2 to TB3 - TB5, before decreasing again for TB6 - TB10, is related to the Figs. \ref{Fig:Light_Curves} and \ref{Fig:Hardness_Ratios} where the hard X-ray flux ratio (2 - 10 keV / 0.3 - 10 keV) and the HR increased between 30 and 60 ks.
	
	We also obtained the 0.35 - 10 keV time-averaged spectrum from each observation. Fig. \ref{Fig:Obs_Spec_Bins} shows each TB spectrum for Obs 1 (left) and Obs 2 (right), with the time-averaged spectra shown in red as a comparison. A clear gap can be seen in Obs 1 between the spectra of TB1 - TB5 and TB7 - TB10, with TB6 somewhere in between, relating to the flux increase half way through the observation (see Fig. \ref{Fig:Light_Curves}).
	
	We initially modelled the time-averaged spectrum from each observation, corresponding to the base model, before fitting each TB spectrum. This allowed us to study the obscurer and continuum parameters in order to explain the origin and nature of the X-ray variability within NGC 3227 during obscuration in 2019.
	
	\section{X-ray spectral fitting}
	\label{Sec:Spectral_Fitting}
	
	\subsection{Setting up the model}
	
	In order to analyse the spectral variations over the course of the two observations, we required the models from the previous work on this campaign. The intrinsic (unobscured) SED \citepalias{Mehdipour_Prep} consisted of a power law (\texttt{POW}), a neutral reflection component that takes into account the neutral \ion{Fe}{K$\alpha$} line at 6.4 keV \citep[\texttt{REFL};][]{Magdziarz1995}, a Comptonisation component \citep[\texttt{COMT};][]{Titarchuk1994} for the soft excess, and a disk blackbody component (\texttt{DBB}) for the optical to UV band. Here, we modelled the \texttt{REFL} component without an accretion disk profile, similar to the previous series papers, and do not assume any blurred ionised reflection that could partly explain the soft X-ray excess. Although this could be a plausible alternative explanation for the origin of the soft excess, the \texttt{COMT} component is sufficient for our purposes in the modelling of a broadband SED for NGC 3227 (see e.g. \citetalias{Mehdipour_Prep}, in addition to \citealt{Mehdipour2011, Mehdipour2017, Mehdipour2018, Petrucci2013}). We are therefore neglecting ionised disk reflected emission.
	
	In our modelling, the intrinsic continuum was absorbed by the obscurer, so most of the SED parameters were fixed to the values from \citetalias{Mehdipour_Prep} (for example, the seed photon and plasma temperatures, $T_{\rm seed} = 10.2$ eV and $T_{\rm c} = 60$ eV , respectively, in the \texttt{COMT} component). However, the normalisations of the \texttt{DBB}, \texttt{POW} and \texttt{COMT} components ($N_{\rm dbb}$, $N_{\rm pow}$, $N_{\rm comt}$, respectively), the reflection scaling parameter\footnote{The reflection scaling parameter is the fraction of the observed power law that is reflected, given by $s = \frac{F_{\rm refl}}{F_{\rm pow}}$, where $F_{\rm refl}$ and $F_{\rm pow}$ are the fluxes of the reflected and power law continua, respectively. It is identical to reflection scaling factor of the \texttt{PEXRAV} model in \texttt{XSPEC}, and represents a multiplicative scaling of the reflection once $N_{\rm refl}$ is coupled to $N_{\rm pow}$.} (\textit{s}), and photon index ($\Gamma_{\rm pow}$), were measured for the two 2019 observations during the obscured state in \citetalias{Mao_Prep}; this is the obscured SED.

	\begin{table}
		\centering
		\caption{Parameter values for the obscurer and warm emitter, taken from Tables 2 and 3 from \citetalias{Mao_Prep}. The warm emitter parameters are fixed throughout the modelling, while the obscurer parameters are fitted (see Sect. \ref{Sec:Spectral_Fitting}).}
		\label{Table:Obsc_EM_Pars}
		\begin{tabular}{c c c c }
			\hline
			\multirow{2}{*}{Parameter} & \multicolumn{2}{c}{Obscurer}  & Warm emitter  \Tstrut\Bstrut \\
			& (Obs 1) & (Obs 2) & (Obs 1) \Tstrut\Bstrut \\
			\hline
			$N_{\rm H}$ ($10^{26}$ m\textsuperscript{-2}) & 4.00 & 10.50 & 0.96 \Tstrut\Bstrut \\
			$\log \xi$ ($10^{-9}$ Wm) & 0.50 & 2.60 & 1.37 \Tstrut\Bstrut \\
			CF $^*$ & 0.61 & 0.72 & 0.01\Tstrut\Bstrut \\
			\hline
			
		\end{tabular}
		\tablefoot{
			$^*$ The covering fraction (CF) for the obscurer components is $f_{\rm cov}$ and for the warm emitter is $C_{\rm cov} = \Omega/4\pi$, where $\Omega$ is the solid angle subtended by the warm emitter with respect to the SMBH.
		}
	\end{table}
	
	\begin{figure*}
		\centering
		\begin{subfigure}{0.5\linewidth}
			\includegraphics[width=1\linewidth]{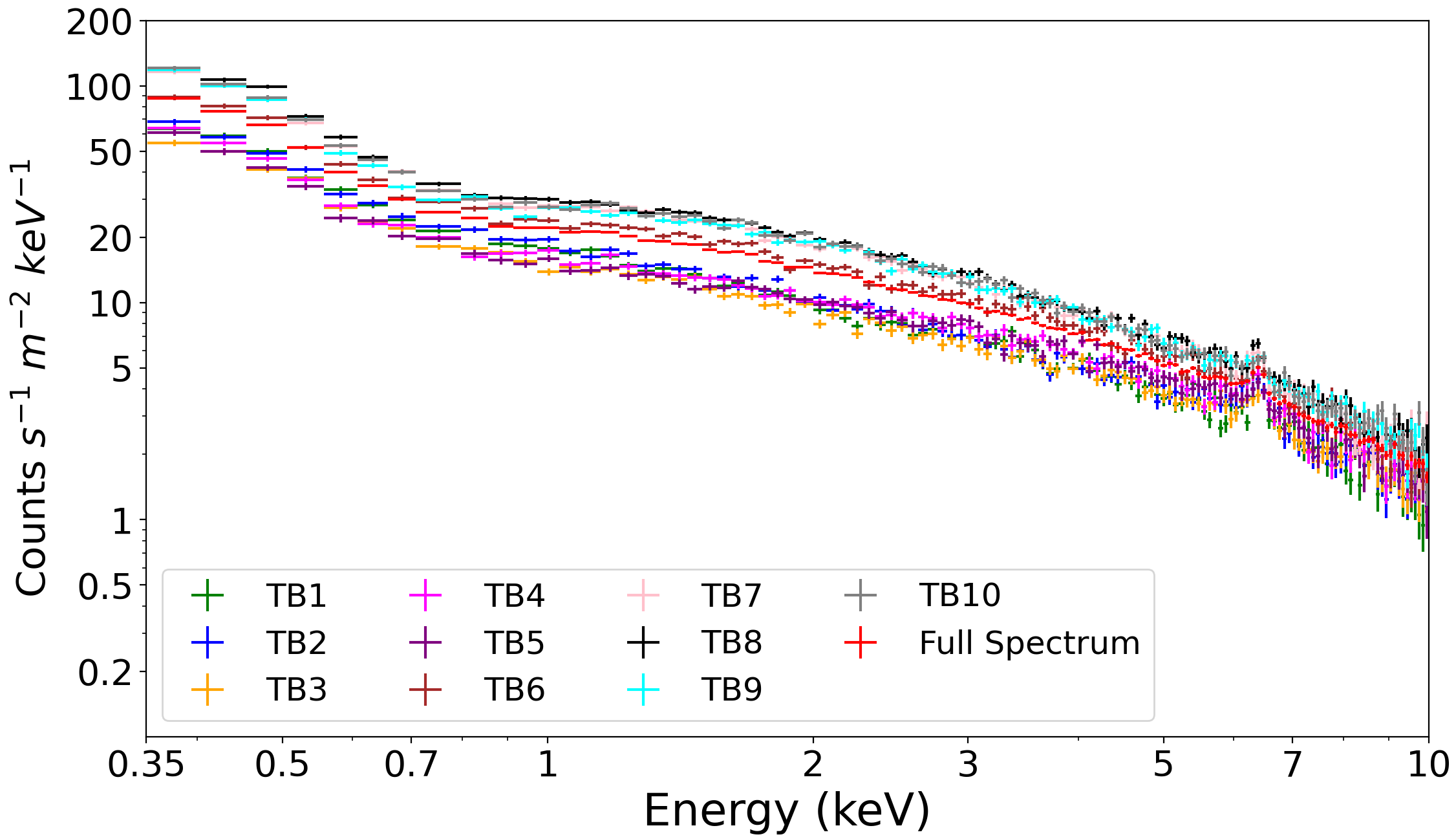}
		\end{subfigure}%
		\begin{subfigure}{0.5\linewidth}
			\includegraphics[width=1\linewidth]{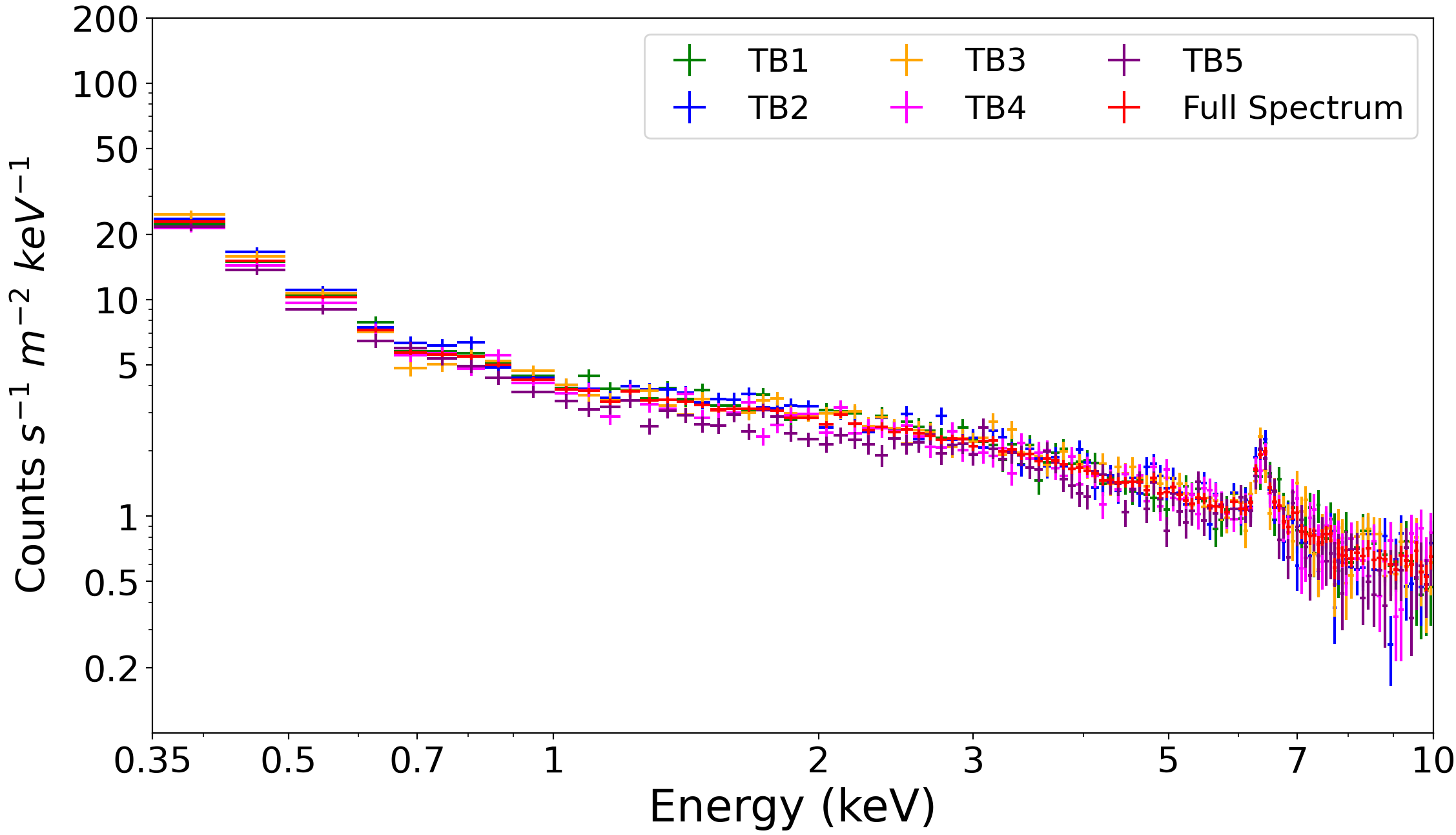}
		\end{subfigure}
		\caption{EPIC-PN spectra for Obs 1 (left) and Obs 2 (right). The time-averaged spectra are shown in red and the TB numbers are presented in the legends by their respective colours. }
		\label{Fig:Obs_Spec_Bins}
	\end{figure*}
	
	The initial model examined in this paper was model M1 from \citetalias{Mao_Prep} (see Tables 2 and 3 for Obs 1 and Obs 2, respectively), where the \texttt{REFL} normalisation ($N_{\rm refl}$) and photon index ($\Gamma_{\rm refl}$) values were coupled to the \texttt{POW} normalisation ($N_{\rm pow}$) and photon index ($\Gamma_{\rm pow}$), for the two 2019 observations. The other models studied in \citetalias{Mao_Prep} were also fitted and are compared in Appendix \ref{Sec:Alt_Mod}: for M2 the $N_{\rm refl}$ is fixed to the 2016 $N_{\rm pow}$ value from \citetalias{Mehdipour_Prep}, and M3 is the same as M1, but models the obscurer with two \texttt{PION} components. For the power law component, we applied upper \citep[309 keV;][]{Turner2018} and lower energy (equal to the disk energy, in this case 10.2 eV from \citetalias{Mehdipour_Prep}) cutoffs. The high energy cutoff was also applied to the reflection component and fixed at 309 keV. The fitted continuum parameters in this paper were $N_{\rm pow}$, $\Gamma_{\rm pow}$, $N_{\rm comt}$, $s$. In addition, we modelled the neutral gas in our Galaxy using the \texttt{HOT} model in \texttt{SPEX}; we fixed the column density and temperature at $N_{\rm H}^{\rm HOT} = 2.07 \times 10^{24}$ m\textsuperscript{-2} \citep{Murphy1996} and $T_{\rm Gal} = 0.5$ eV, respectively.
	
	We also included the four WA components from \citetalias{Wang_Prep}, modelled with \texttt{PION} \citep{Mehdipour2016}, but accounted for their less ionised nature due to the obscurer, which was also accounted for in the previous papers of this series. The de-ionised nature of the WA means only the ionisation parameters of the WAs are lowered as they receive less photons from the central source due to the presence of the obscurer. The other parameters (e.g. $N_{\rm H}$, $v_{\rm out}$) were assumed to be constant over time. This was also seen in NGC 5548 \citep{Kaastra2014} and NGC 3783 \citep{Mehdipour2017}. One \texttt{PION} component was used to model the obscurer in both observations with the initial parameters from Table \ref{Table:Obsc_EM_Pars} for Obs 1 and Obs 2 (taken from Tables 2 and 3 in \citetalias{Mao_Prep}, respectively). The ionisation parameter of the obscurer in Obs 2 is significantly larger compared to the value in Obs 1 which is indicative of there being two completely separate obscuring components that are observed in Obs 1 and Obs 2, respectively, as stated in \citetalias{Mao_Prep}. The fitted obscurer parameters in this paper were the column density ($N_{\rm H}$), ionisation parameter ($\xi$), and covering fraction ($C_f$). The flux changes between the 2016 and 2019 observations \citepalias[see Fig. 2 in][]{Mehdipour_Prep} is a result of the obscurer. The variations we observe are unlikely to be due to the known WA components \citepalias{Wang_Prep} as the ionising continuum is absorbed by the obscurer before it reaches them.
	
	Finally, we accounted for the emission lines, produced by the warm emitter, with a single \texttt{PION} component in both observations. However, the ionising continuum was different to the one used for the obscurer and WA, so the intrinsic 2016 SED (same components as the obscured SED but with different parameter values) was used to ionise the warm emitter (see \citetalias{Mao_Prep} for details of the modelling and Table 2 in \citetalias{Mehdipour_Prep} for the values). The warm emitter values were fixed to those from Table \ref{Table:Obsc_EM_Pars} \citepalias[see also Table 2 in][]{Mao_Prep}.

	\subsection{Fitting the spectra}
	\label{Sec:Fit_Spec}
\begin{figure*}
		\centering
		\begin{subfigure}{0.5\linewidth}
			\includegraphics[width=1\linewidth]{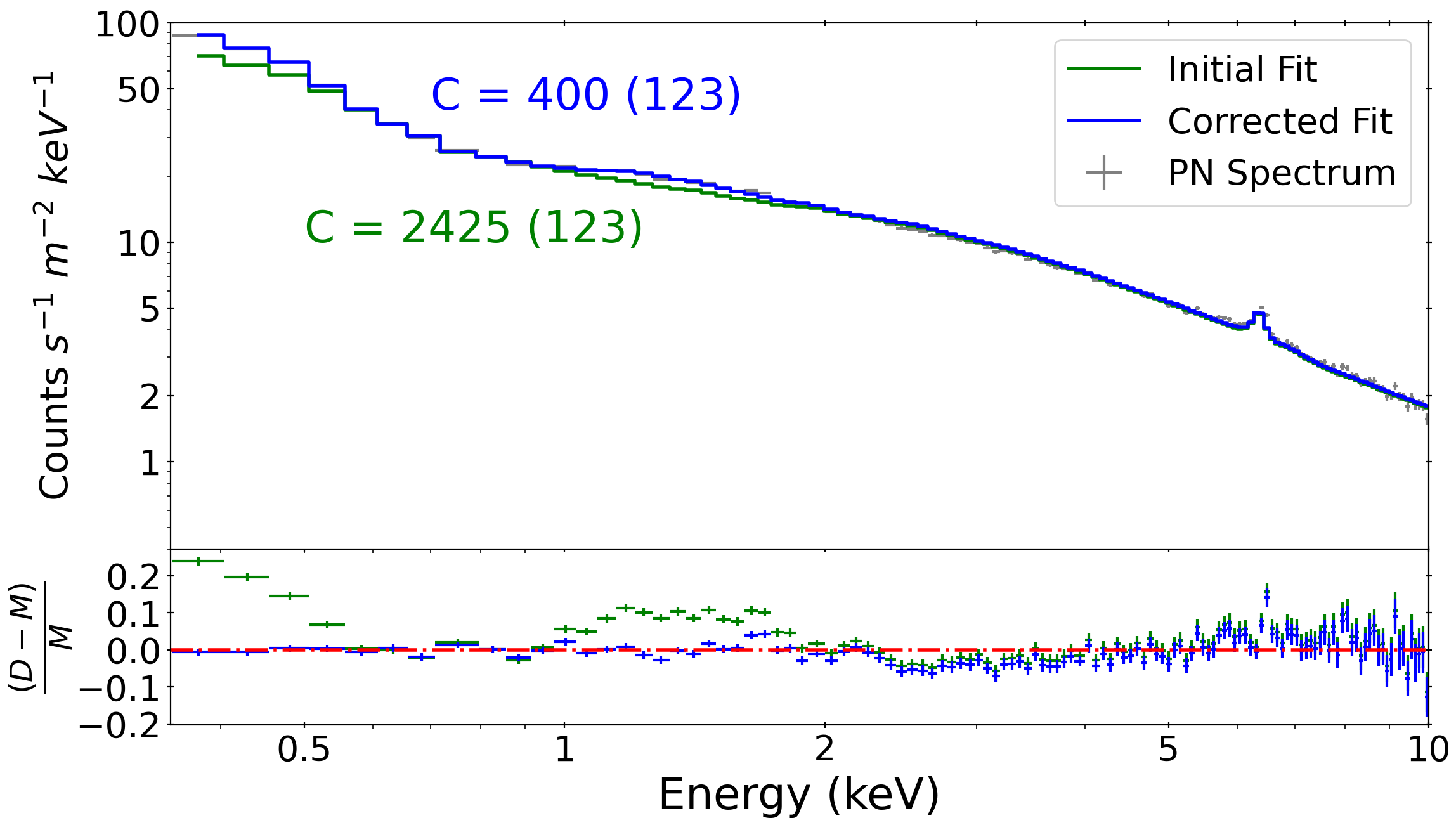}
		\end{subfigure}%
		\begin{subfigure}{0.5\linewidth}
			\includegraphics[width=1\linewidth]{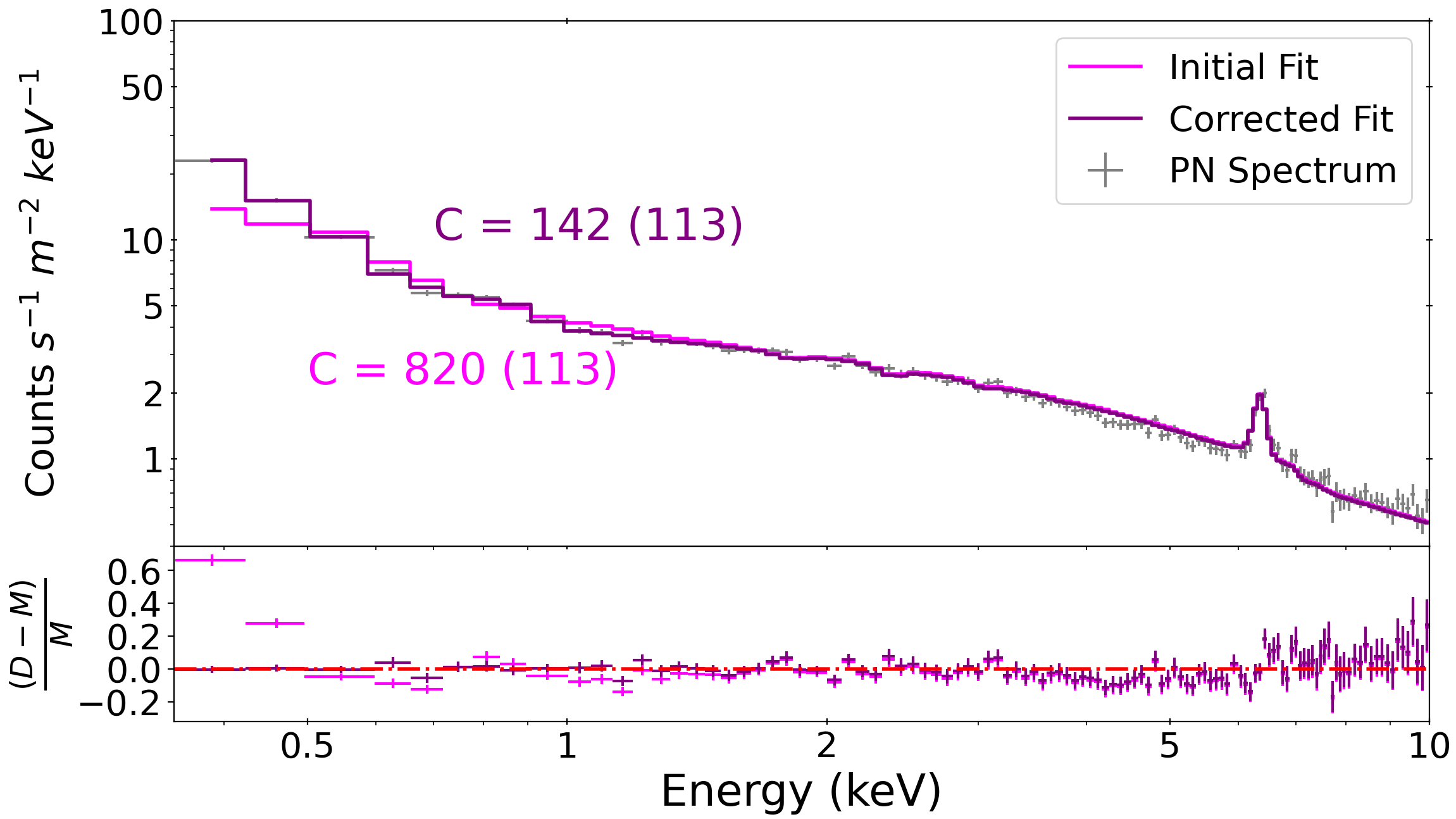}
		\end{subfigure}
		\caption{Initial and \texttt{KNAK} corrected models (see Appendix \ref{Appen:KNAK_Corr}) for Obs 1 (left) and Obs 2 (right). The initial models were from \citetalias{Mao_Prep} (green and pink), while the corrected models are in blue and purple, respectively. The C-statistic values are displayed, along with their respective degrees of freedom in the parentheses. The bottom panels show the residuals between the data and model. Appendix \ref{Appen:KNAK_Corr} explains the steps required to correct the models, shown in Fig. \ref{Fig:Ratio_Compare}.}
		\label{Fig:Model_Compare}
	\end{figure*}

	We folded the model described above to the time-averaged 0.35 - 10 keV EPIC-PN spectrum for each observation. The initial C-statistics were C = 2425 (for 123 degrees of freedom, d.o.f hereafter) and C = 820 (for 113 d.o.f) for Obs 1 and Obs 2, respectively. However, we found that there were strong residuals between the data and model at softer energies in both observations (see the initial fit in Fig. \ref{Fig:Model_Compare}). This was most likely due to the cross calibration issue between the reflection grating spectrometer \citep[RGS;][]{denHerder2001} and EPIC-PN instruments. \citetalias{Mao_Prep} modelled the RGS for the soft X-rays and EPIC-PN data for the hard X-rays, while here we model the full X-ray band using EPIC-PN data only. Therefore, to correct for the cross calibration issue we forced the EPIC-PN data to agree with RGS. We do it this way because RGS is able to resolve the O-edge, while EPIC-PN cannot. The process of how we corrected for this cross calibration effect is explained in Appendix \ref{Appen:KNAK_Corr}. The corrected C-statistics once we accounted for any instrumental effects (Appendix \ref{Appen:KNAK_Corr}) were C = 400 (123 d.o.f) and C = 142 (113 d.o.f) for Obs 1 and Obs2, respectively (see the corrected fit in Fig. \ref{Fig:Model_Compare}).
	
\begin{figure}
		\centering
		\includegraphics[width=1\linewidth]{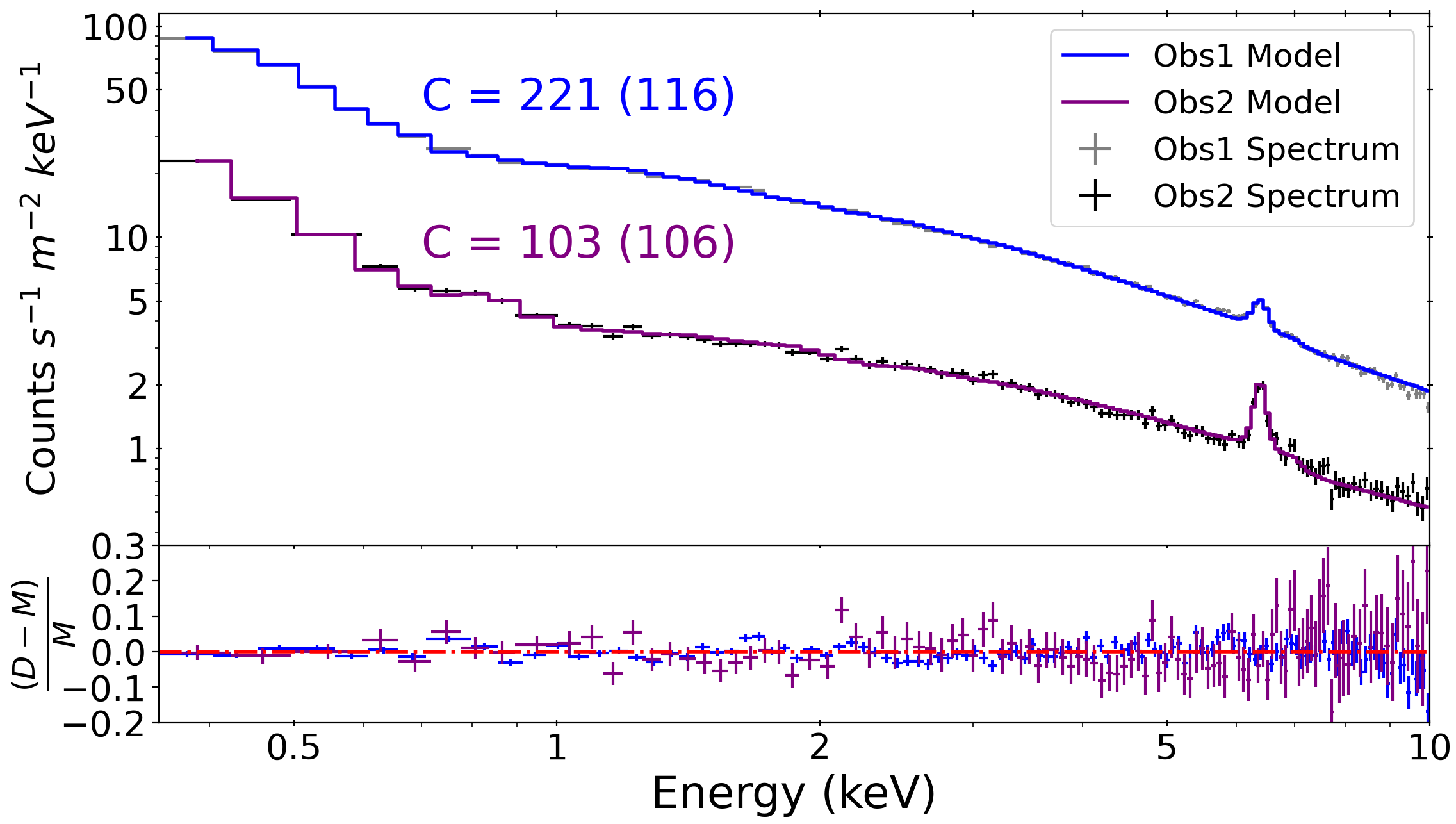}
		\caption{Final best fit models fitted to the time-averaged spectra: Obs 1 grey crosses and blue model; Obs 2 black crosses and purple model. Also shown are the C-statistics and degrees of freedom. The best fit parameter values are displayed in Table \ref{Table:Full_Spec_Results}. The bottom panel shows the residuals between the data and model for Obs 1 (blue) and Obs 2 (purple).}
		\label{Fig:Final_Best_Fits}
	\end{figure}

	For the time-averaged spectrum of each observation, we then fitted $N_{\rm pow}$, $\Gamma$, $N_{\rm comt}$, \textit{s} of the 2019 continuum (obscured SED), and the column density ($N_{\rm H}$), ionisation parameter ($\xi$), and covering fraction ($C_f$) of the obscurer component. Fitting these parameters improved the best fit to C = 221 (116 d.o.f) in Obs 1, and C = 103 (106 d.o.f) for Obs 2. We note that the best fit in Obs 1 is not formally acceptable (C/d.o.f = 1.91), while in Obs 2 the fit is very good (C/d.o.f = 0.97). However, for Obs 1 there are no clearly structured residuals left despite the formally unacceptable fit. One explanation is that the obscurer and continuum parameters for TB1-TB5 are so different from TB6-TB10 that a single model can't explain the average spectrum. Furthermore, this is similar to the analysis of NGC 1068, where \cite{Grafton-Waters2021} found that the 2014 observation, owing to having more than 2.5 times the exposure time compared to 2000 epoch, had a significantly worse fit. The final best fit models of each observation are displayed in Fig. \ref{Fig:Final_Best_Fits}. Now that we obtained a good fit to the EPIC-PN data only, we used this as the baseline model to fit the spectra of each TB. The best fit parameter results are displayed in Table \ref{Table:Full_Spec_Results}.

	\begin{figure*}[!tbp]
		\centering
		\begin{subfigure}{0.5\linewidth}
			\includegraphics[width=1\linewidth]{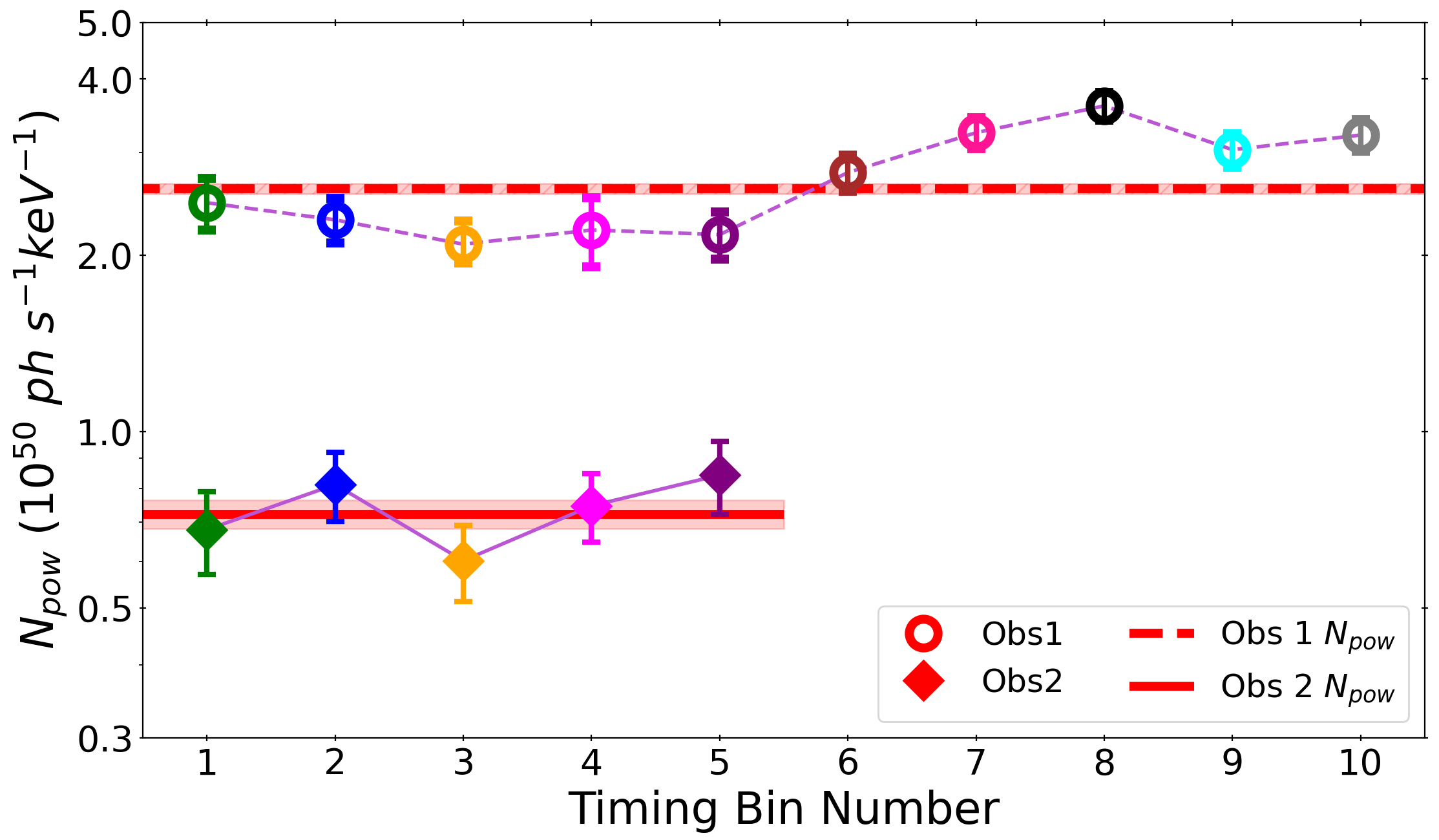}
		\end{subfigure}%
		\begin{subfigure}{0.5\linewidth}
			\includegraphics[width=1\linewidth]{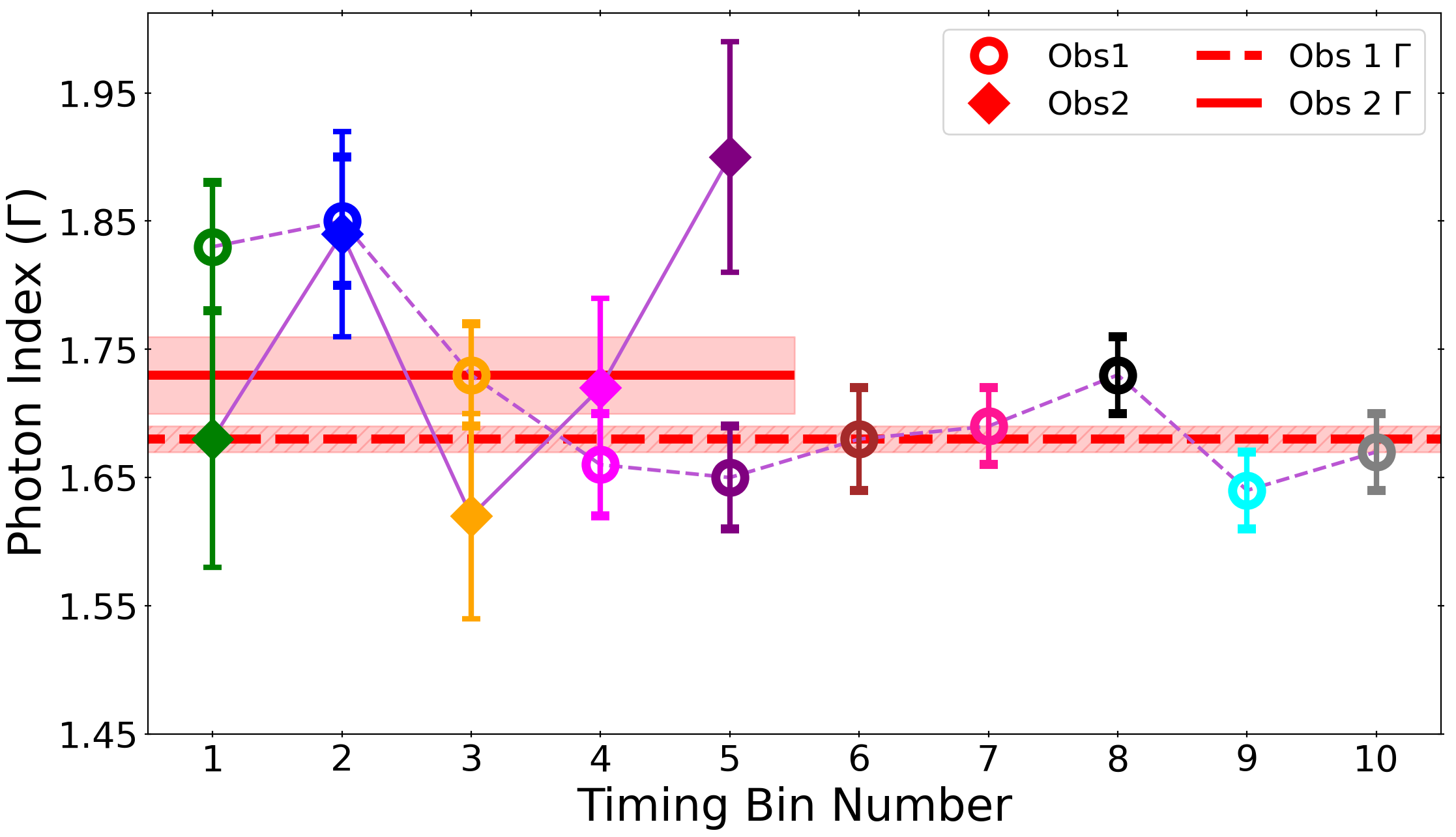}
		\end{subfigure}
		\begin{subfigure}{0.5\linewidth}
			\includegraphics[width=1\linewidth]{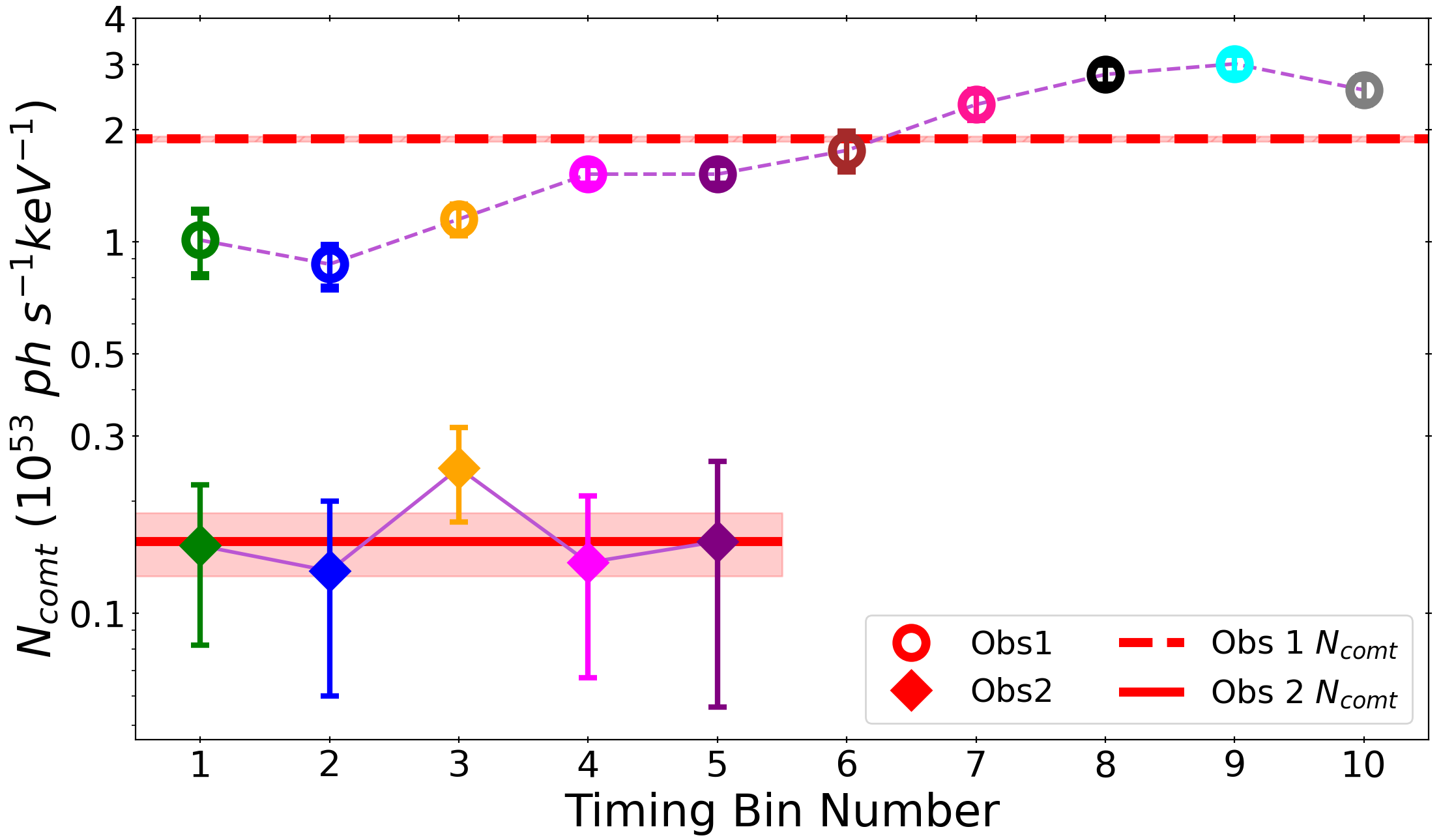}
		\end{subfigure}%
		\begin{subfigure}{0.5\linewidth}
			\includegraphics[width=1\linewidth]{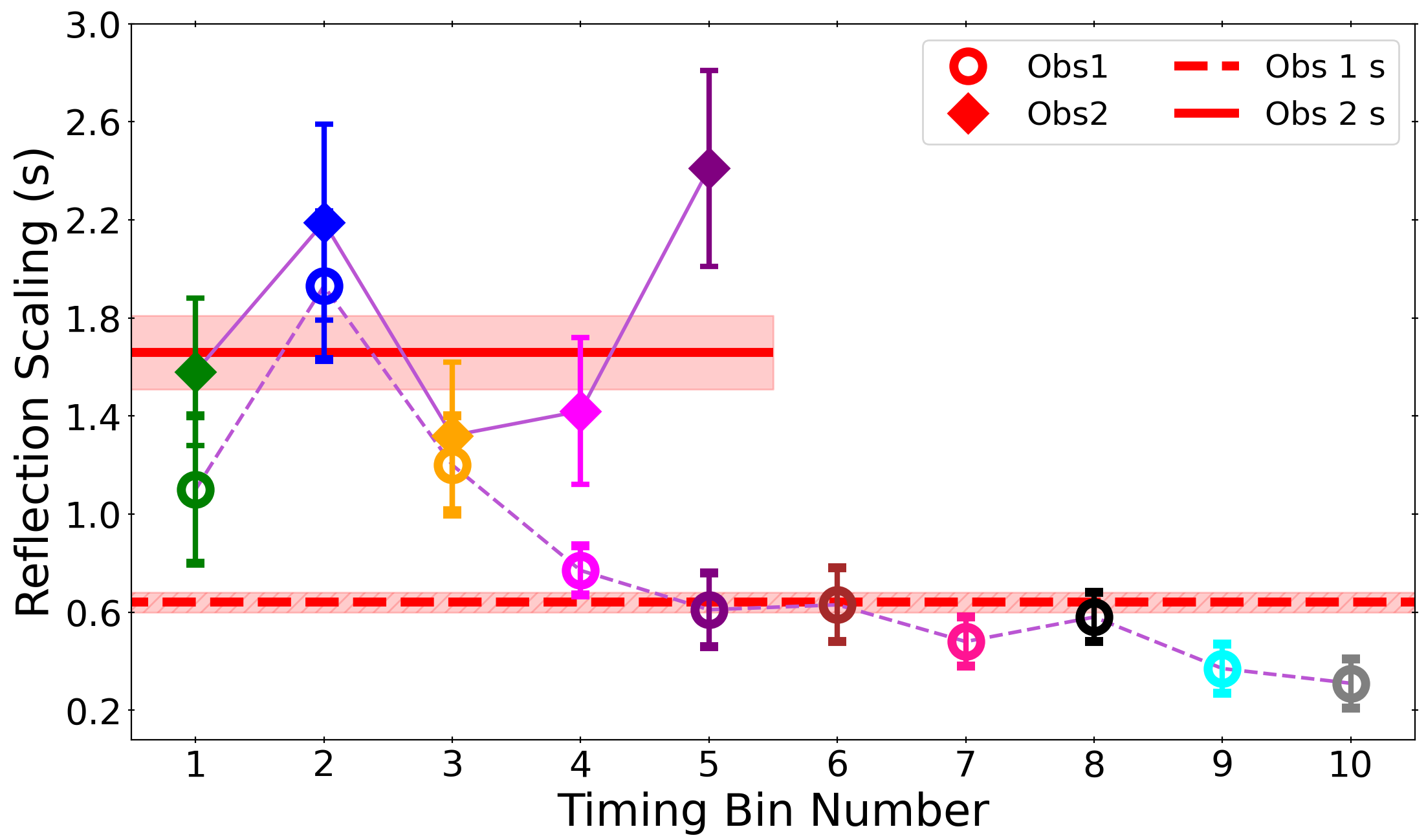}
		\end{subfigure}
		\begin{subfigure}{0.5\linewidth}
			\includegraphics[width=1\linewidth]{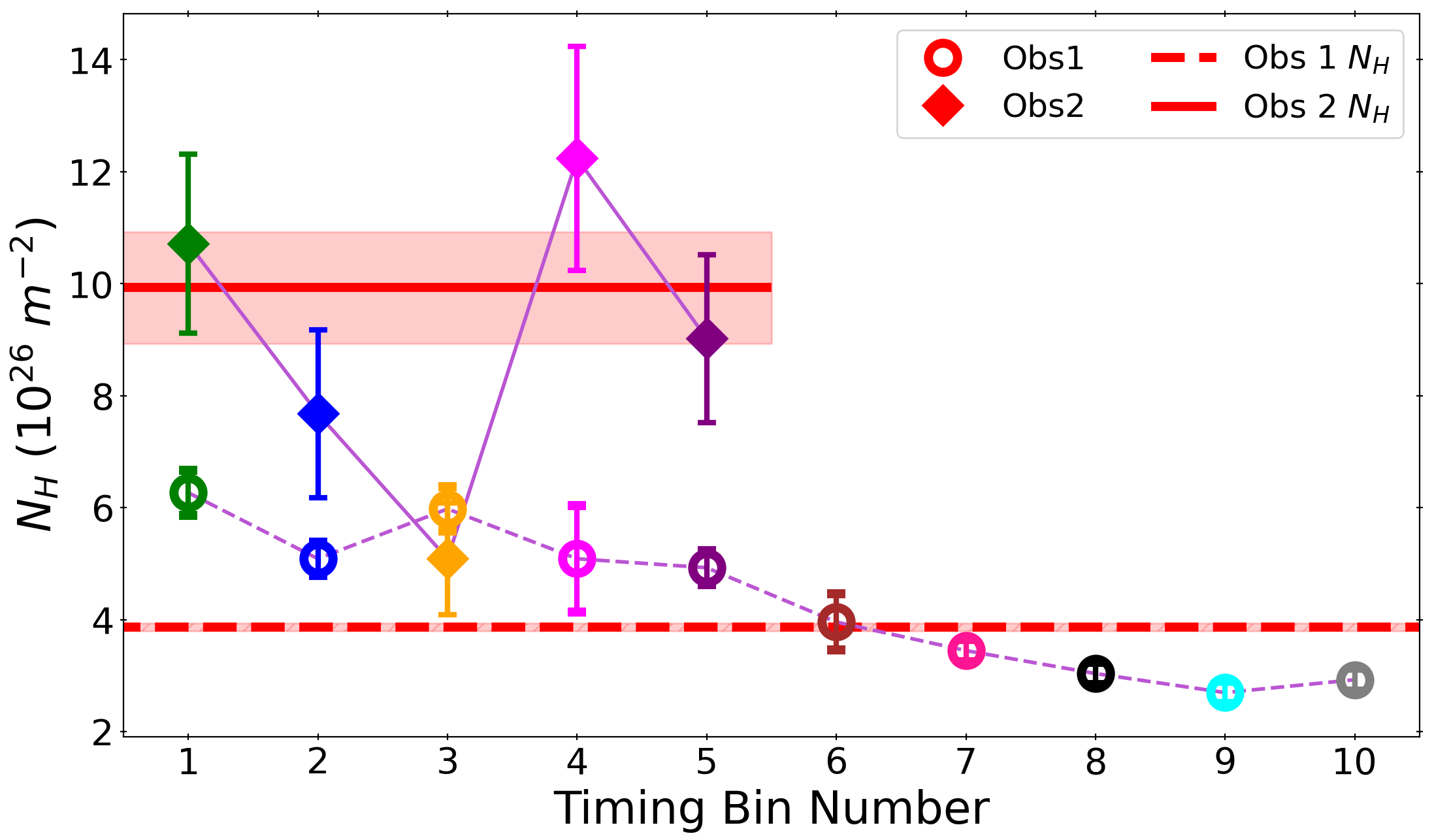}
		\end{subfigure}%
		\begin{subfigure}{0.5\linewidth}
			\includegraphics[width=1\linewidth]{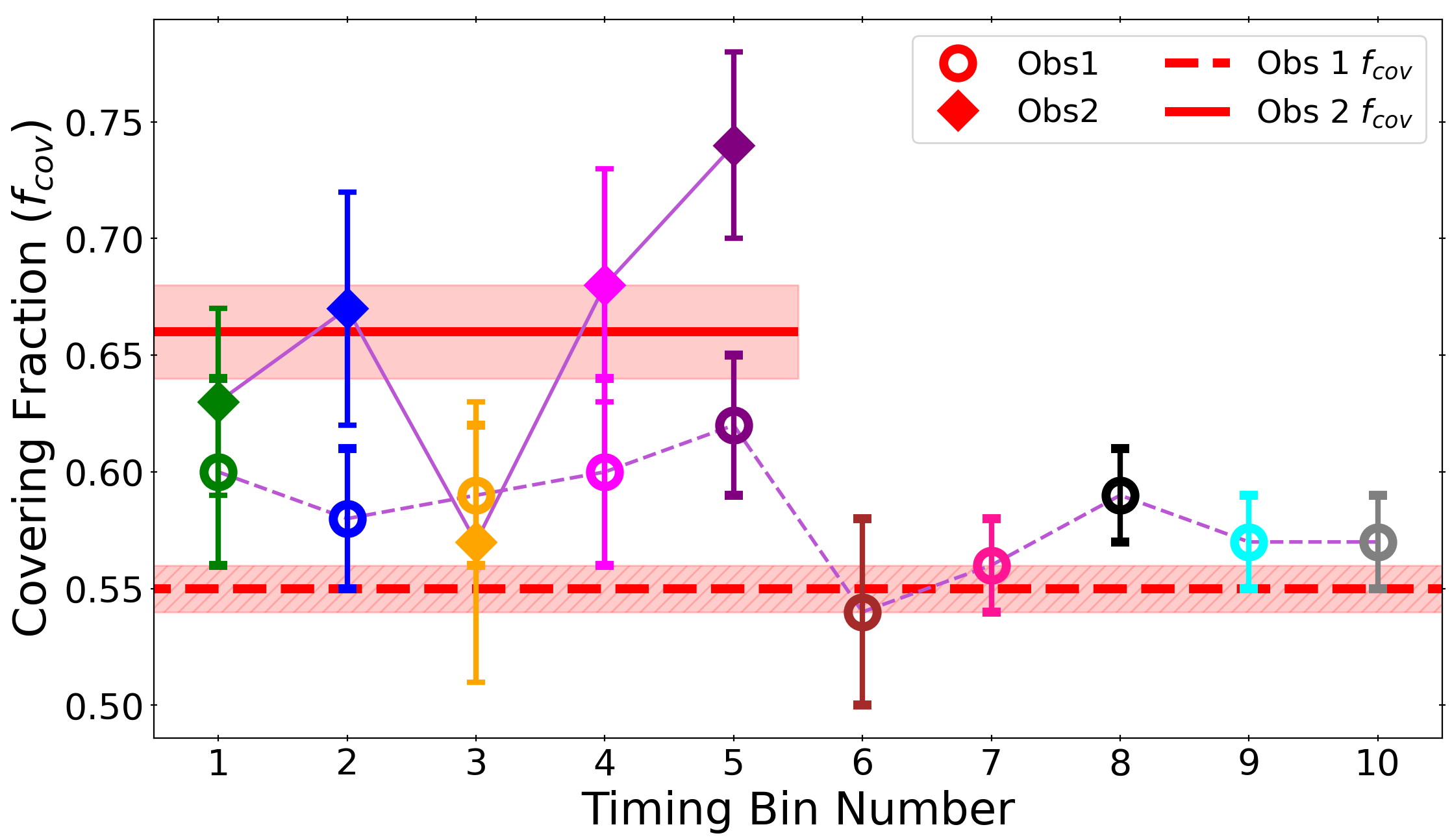}
		\end{subfigure}
		\begin{subfigure}{0.5\linewidth}
			\includegraphics[width=1\linewidth]{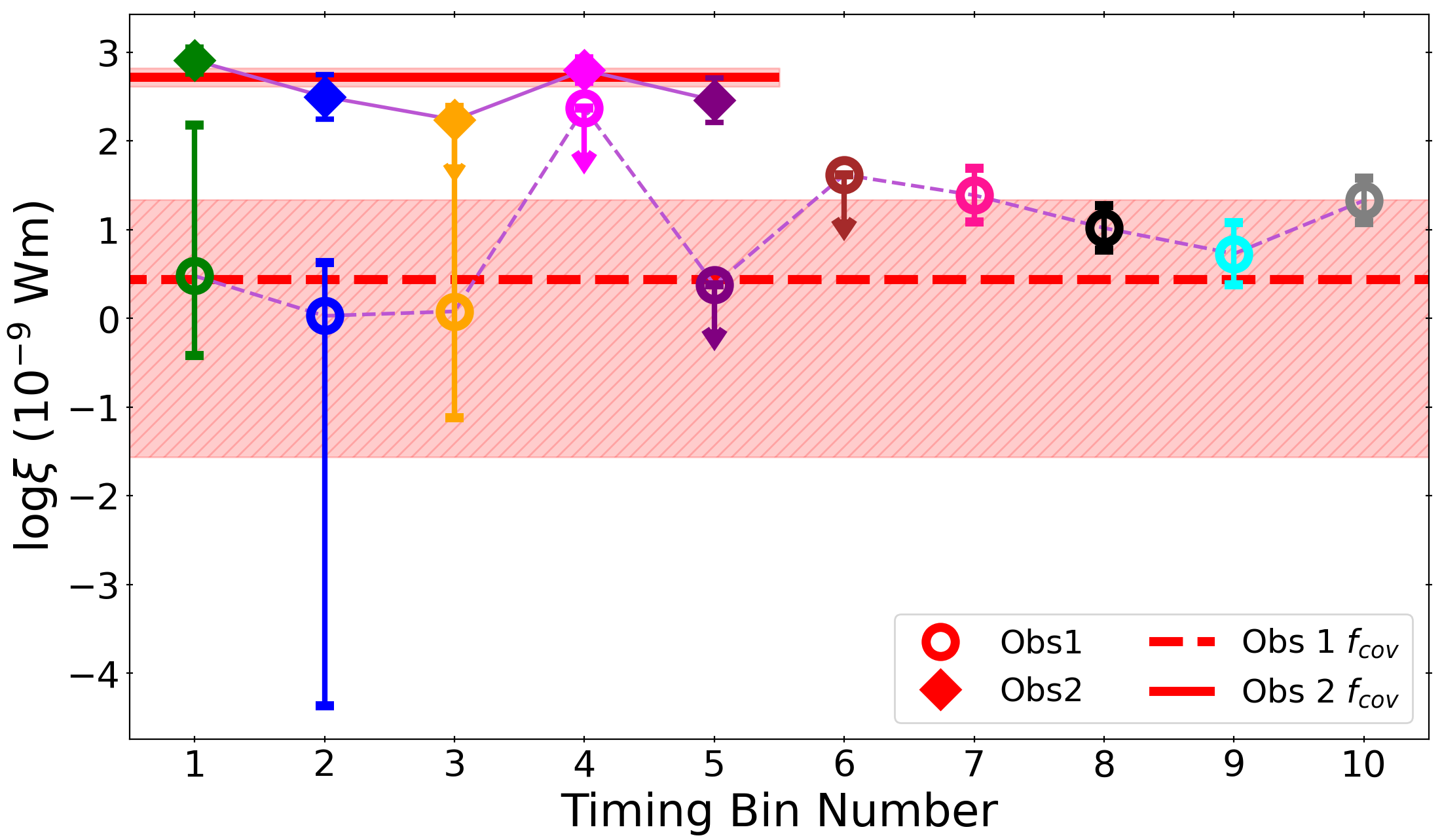}
		\end{subfigure}
		\caption{Model parameter comparisons for each timing bin (TB) in Obs 1 (hollow circles) and Obs 2 (diamonds). The red horizontal dashed and solid lines represent the time-averaged parameter values for Obs 1 and Obs 2, respectively, and the shaded areas show the uncertainties. The TB colours are consistent with their respective spectra in Fig. \ref{Fig:Obs_Spec_Bins}. The model parameter is shown on the y-axis of each panel. The light purple lines (dashed for Obs 1 and solid for Obs 2) connect the TB values for each parameter in the two observations.}
		\label{Fig:TB_Compare}
	\end{figure*}

	\section{Results and discussion}
	\label{Sec:Results_Discussion}
	
	\subsection{Timing bins comparison}
	\label{Sec:TB_Comp}
	Here we compare the parameter results from each TB for the two observations. Table \ref{Table:Full_Spec_Results} displays the best fit parameter values for each TB in Obs 1 and Obs 2, along with the time-averaged spectrum results (All). Figure \ref{Fig:TB_Compare} displays how these parameters changed over the course of each observation. For each panel in Fig. \ref{Fig:TB_Compare}, the time-averaged values are shown by the red dashed and solid horizontal lines for Obs 1 and Obs 2, respectively.
	
	The top two panels in Fig. \ref{Fig:TB_Compare} display how the power law normalisation ($N_{\rm pow}$; left) and photon index ($\Gamma_{\rm pow}$; right) vary over time. For Obs 1, $N_{\rm pow}$ increases during the observation, jumping in flux between TB5 and TB7, similar to the count rate increase in the light curve of Fig. \ref{Fig:Light_Curves}. For Obs 2, $N_{\rm pow}$ stays constant between TBs, within the uncertainties. The photon index in Obs 2 displays strong changes, although similar to $N_{\rm pow}$, they are consistent with each other within the uncertainties. In Obs 1, $\Gamma_{\rm pow}$ does not vary throughout the observation, except for TB1 and TB2. For both observations, $\Gamma_{\rm pow}$ is consistent with a hard power law, similar to previous observations \citep[$\Gamma = 1.5 - 1.8$; e.g.][]{Gondoin2003, Cappi2006, Rivers2011, Arevalo2014}, although this parameter is overall slightly lower in Obs 1.
	
	The top-middle panels in Fig. \ref{Fig:TB_Compare} show the Comptonisation normalisation ($N_{\rm comt}$; left) and reflection scaling parameter ($s$; right) changes. For Obs 1, $N_{\rm comt}$ has the same, but more significant, trend as $N_{\rm pow}$. For Obs 2, similarly to $N_{\rm pow}$, $N_{\rm comt}$ is consistent between TBs. Although the reflection scaling parameter ($s$) of \texttt{REFL} in both observations appears to correlate well with $\Gamma_{\rm pow}$ (top right panel of Fig. \ref{Fig:TB_Compare}), both parameter values are consistent within uncertainties, suggesting no overall correlation. For Obs 1, the reflection scaling of TB2 is larger compared to the other TBs, but from TB3 to TB10, $s$ decreases over time, showing an anti-correlation between the flux (Fig. \ref{Fig:Light_Curves}) and the amount of reflected continuum. In other words, we are observing more intrinsic flux during the latter half of Obs 1. 
	
	\begin{figure*}[!tbp]
		\centering
		\begin{subfigure}{0.5\linewidth}
			\includegraphics[width=1\linewidth]{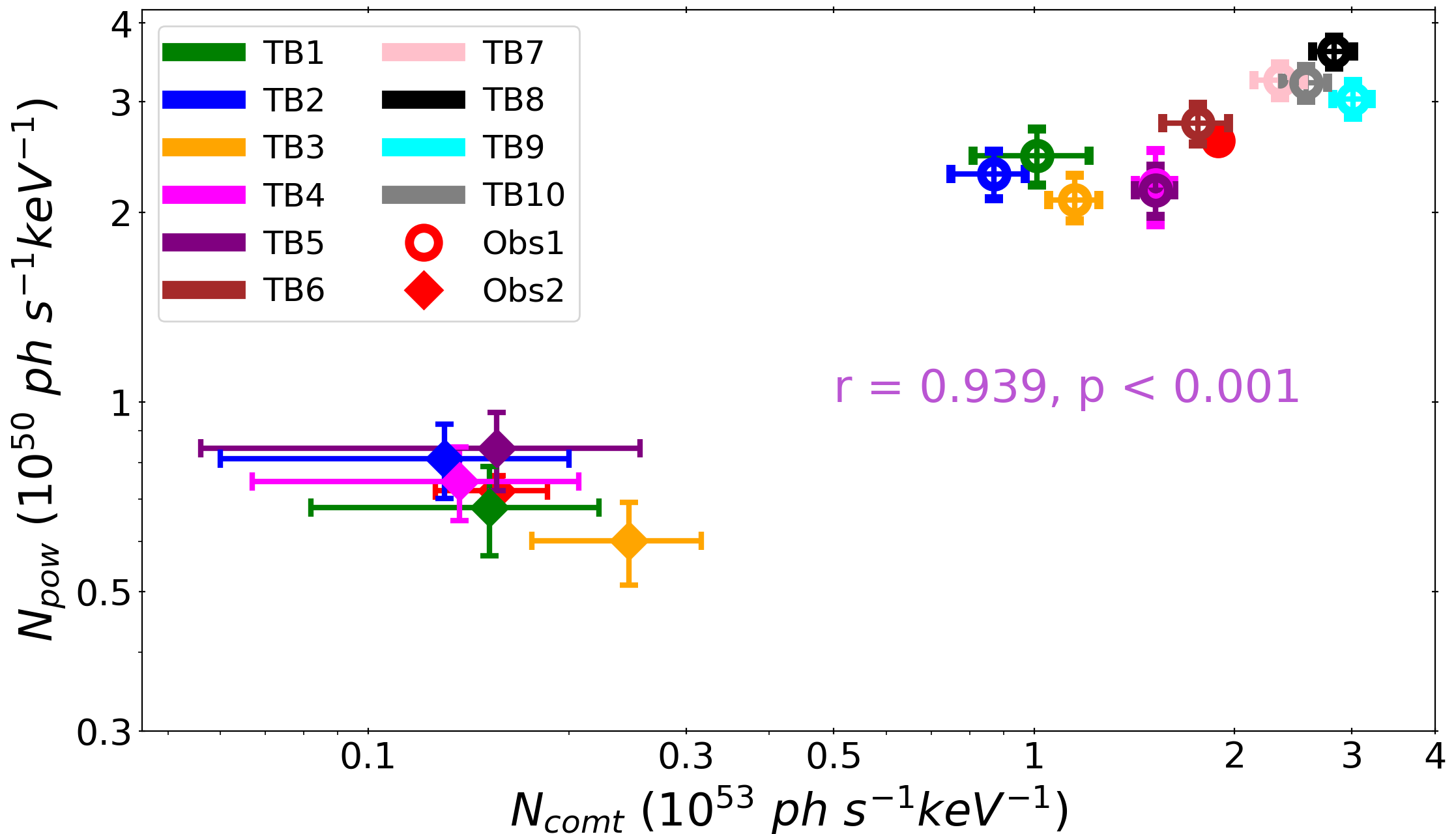}
		\end{subfigure}%
		\begin{subfigure}{0.5\linewidth}
			\includegraphics[width=1\linewidth]{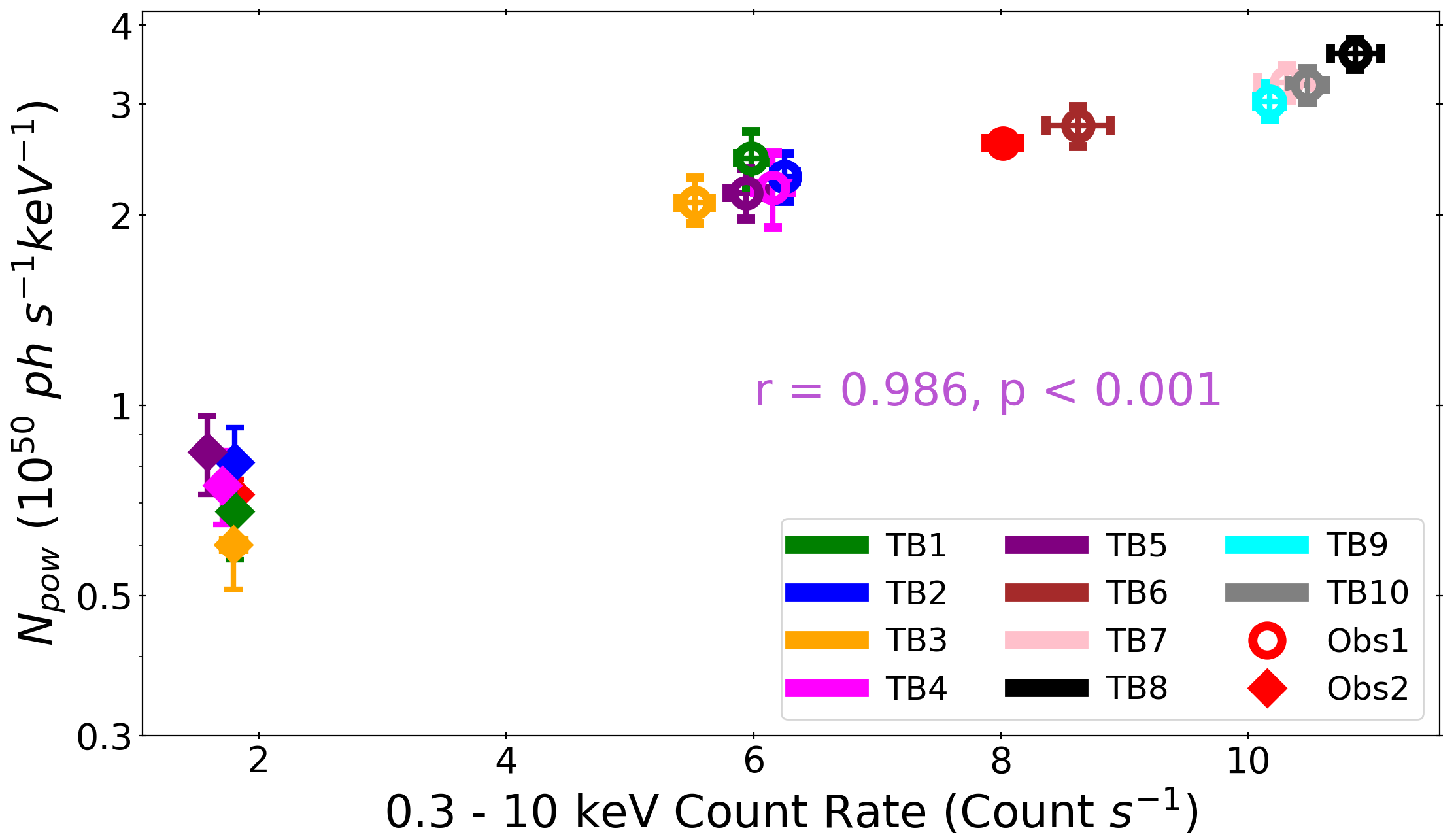}
		\end{subfigure}
		\begin{subfigure}{0.5\linewidth}
			\includegraphics[width=1\linewidth]{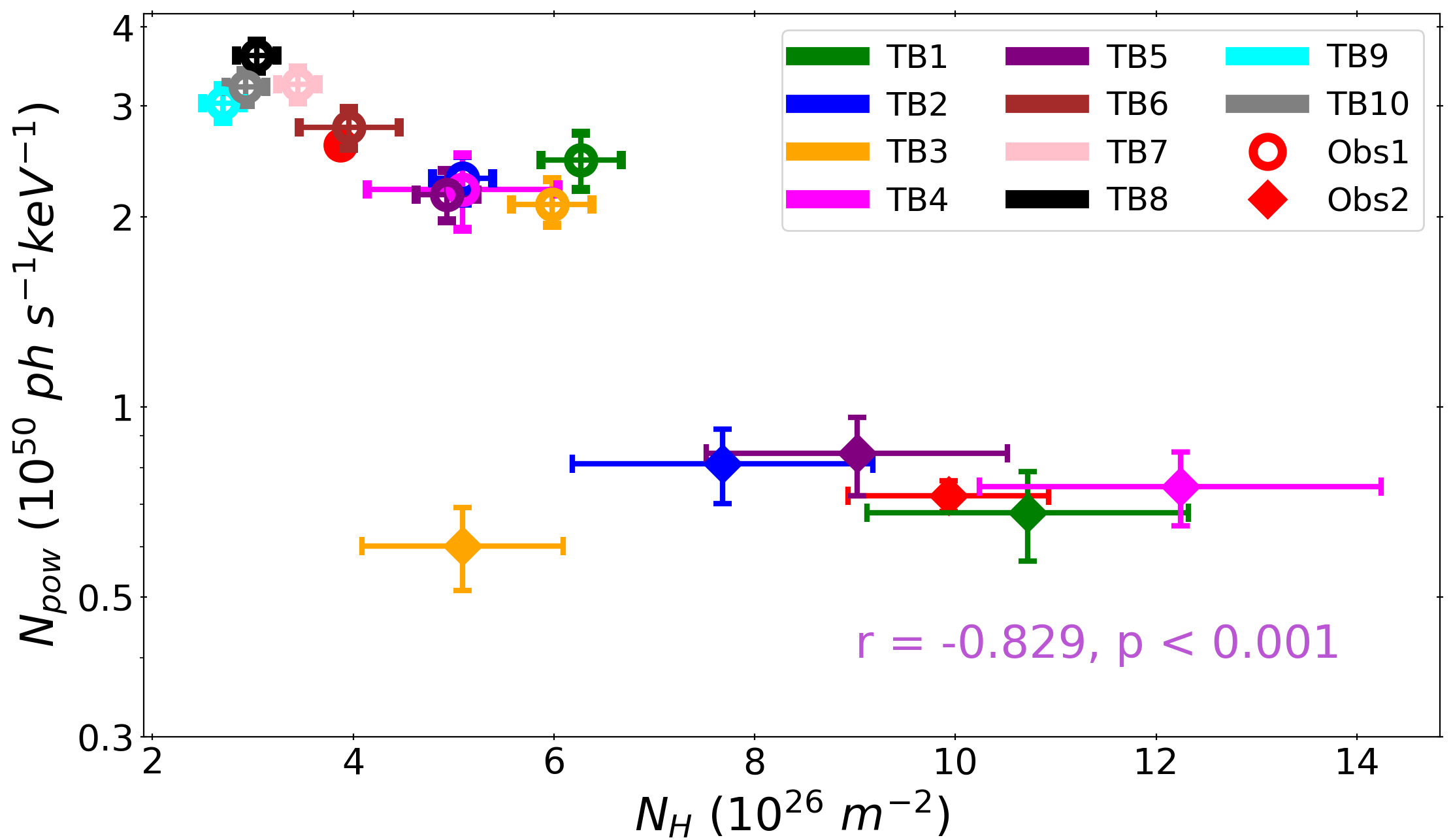}
		\end{subfigure}%
		\begin{subfigure}{0.5\linewidth}
			\includegraphics[width=1\linewidth]{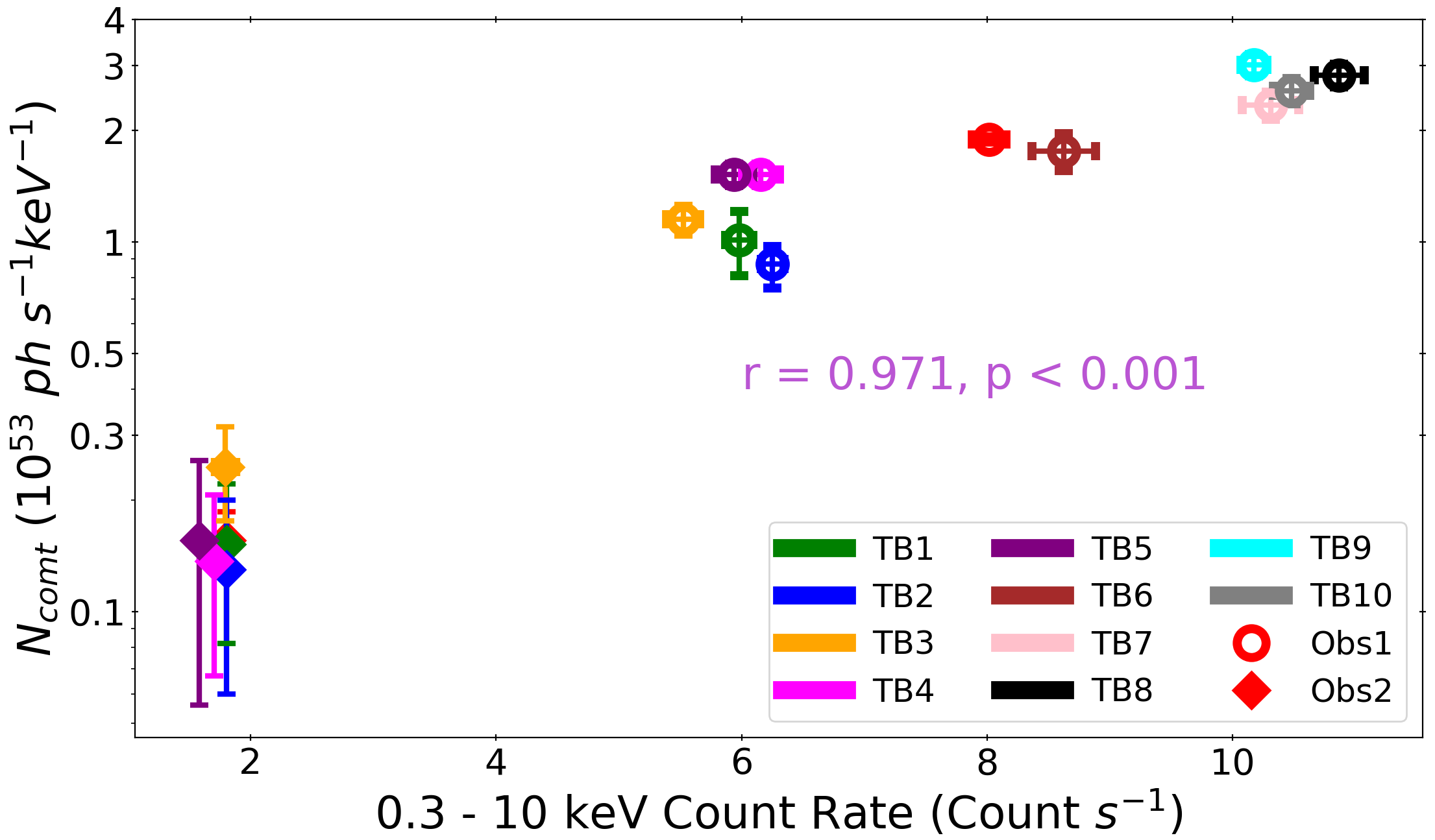}
		\end{subfigure}
		\begin{subfigure}{0.5\linewidth}
			\includegraphics[width=1\linewidth]{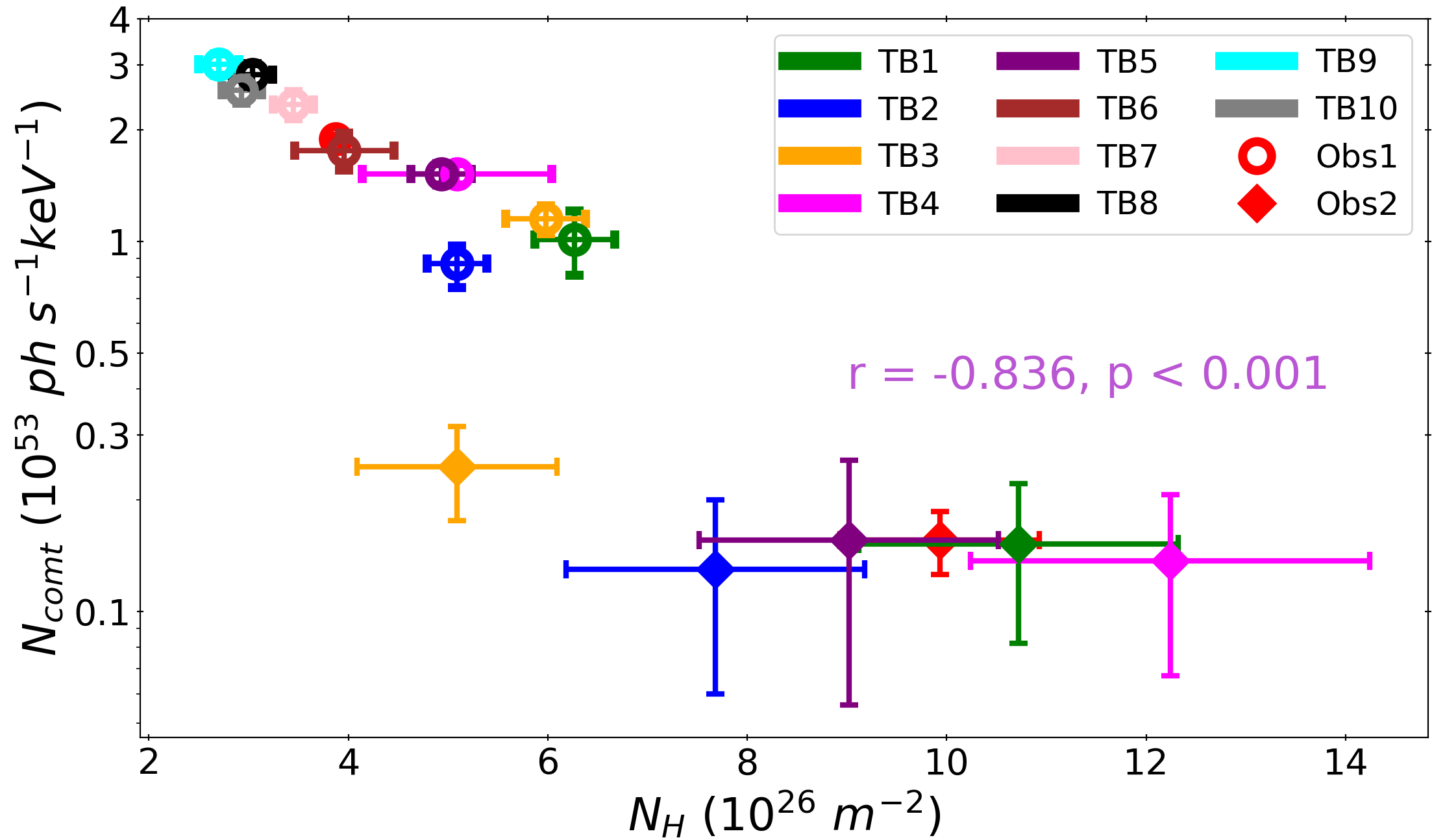}
		\end{subfigure}%
		\begin{subfigure}{0.5\linewidth}
			\includegraphics[width=1\linewidth]{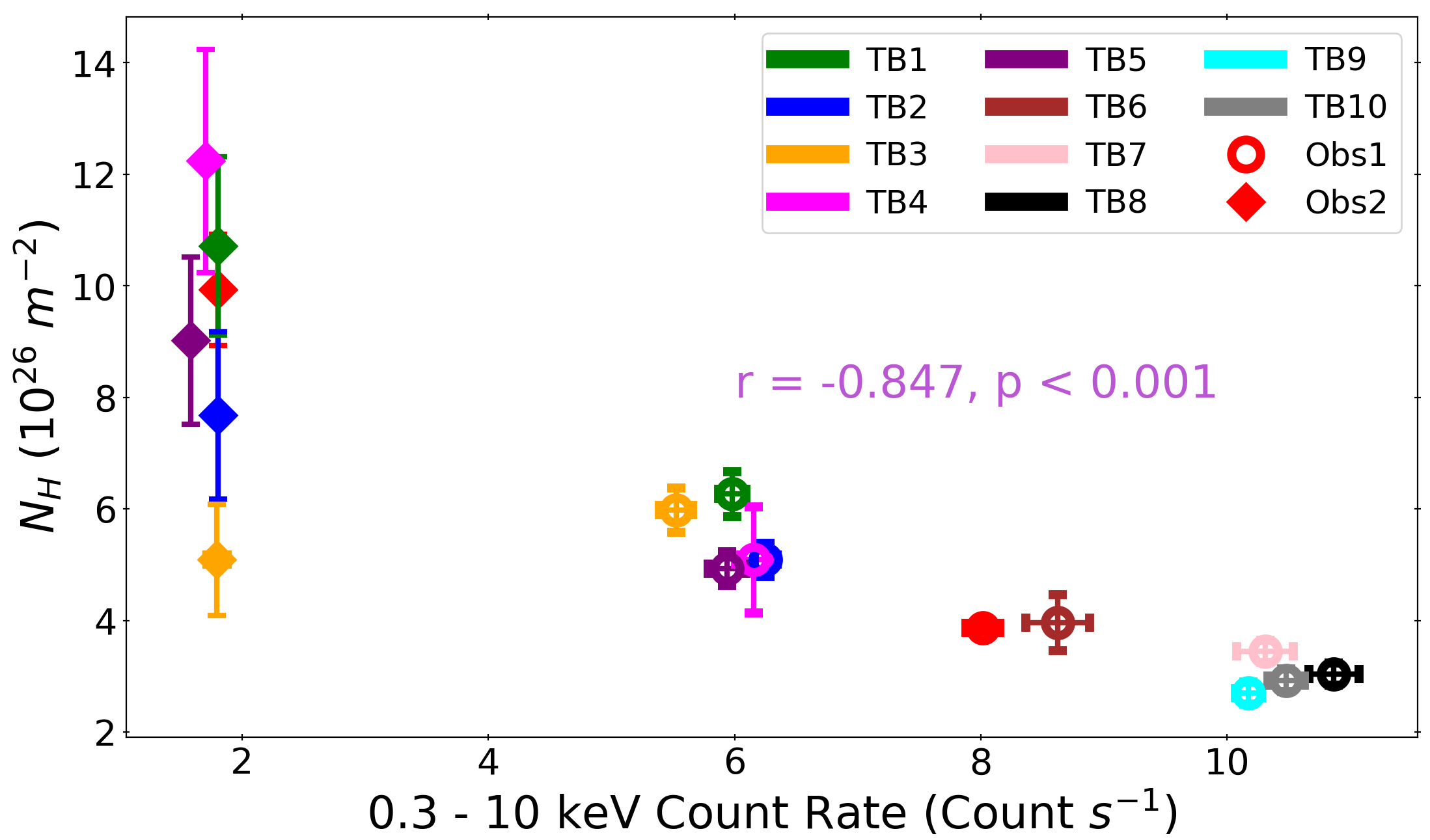}
		\end{subfigure}
		\caption{Correlations between model parameters. \textit{Left:} the normalisations of the power law ($N_{\rm pow}$) and Comptonisation component ($N_{\rm comt}$) are correlated over the two observations (top); $N_{\rm pow}$ against the column density of the obscurer ($N_{\rm H}$) (middle); and $N_{\rm comt}$ against $N_{\rm H}$ (bottom). \textit{Right:} Correlations between $N_{\rm pow}$, $N_{\rm comt}$ and $N_{\rm H}$ against with the 0.3 - 10 keV count rate for top, middle and bottom, respectively. In each panel, the Pearson rank (\textit{r}) and \textit{p}-value is shown to display the significance of each correlation.}
		\label{Fig:Parameter_Compare}
	\end{figure*}
	
	The bottom-middle panels in Fig. \ref{Fig:TB_Compare} present the column density ($N_{\rm H}$; left) and covering fraction ($f_{\rm cov}$; right) of the obscurer component. In Obs 2, it appears that $N_{\rm H}$ fluctuates significantly between TBs. However, with the exception of TB3, the $N_{\rm H}$ values in each TB are constant within the uncertainties of each other and the time-averaged value. Therefore, the given changes in $N_{\rm H}$ are unlikely to occur on timescales between bins, and we suggest that these fluctuations are not real. For Obs 1, on the other hand, $N_{\rm H}$ decreases through each TB, which follows the inverse trend compared to $N_{\rm pow}$ and $N_{\rm comt}$, meaning at the same time there are changes in the obscurer and in the source continuum. We run further tests to explain this inverse trend in Sect. \ref{Sec:Par_Comp}, and discuss the possible causes of this observed variability in Sect. \ref{Sec:Var_Cause} to exclude possible degeneracy between the model parameters. For the obscurer covering fraction, overall, $f_{\rm cov}$ stays constant within the uncertainties throughout each observation. However, $f_{\rm cov}$ is systematically higher in TB1 - TB5 compared to TB6 - TB10. This suggests that the covering fraction is not the parameter driving the observed variability. In Obs 2, TB3 gives apparent outliers for both $N_{\rm H}$ and $f_{\rm cov}$ compared to the other TBs. However, we are unable to constrain $\xi$ (bottom panel of Fig. \ref{Fig:TB_Compare}) in TB3, which, in addition to the lower flux of Obs 2, could explain this drop in obscurer parameter values compared to the other TBs.
	
	Between Obs 1 and Obs 2 ($\sim 3$ weeks), the time-averaged results (Table \ref{Table:Full_Spec_Results}) indicate that $f_{\rm cov}$ and $N_{\rm H}$ increased from 0.55 to 0.66, and from $3.87 \times 10^{26}$ m\textsuperscript{-2} to $9.93 \times 10^{26}$ m\textsuperscript{-2}, respectively. Conversely, the observed $N_{\rm pow}$ and $N_{\rm comp}$  time-averaged parameters decreased from $2.60$ to $0.72 \times 10^{50}$ ph s\textsuperscript{-1} keV\textsuperscript{-1} and from $1.89$ to $0.16 \times 10^{53}$ ph s\textsuperscript{-1} keV\textsuperscript{-1}, respectively. Similar to the obscurer results in this paper, \cite{Beuchert2015} found that the observed variability of NGC 3227 in 2008 could also be produced by an increase in covering fraction (0.7 to 0.9) and column density (5 to $18 \times 10^{26}$ m\textsuperscript{-2}) in an absorption component, in addition to changes caused by the intrinsic continuum. Furthermore, in NGC 5548, the short-term variability (days to weeks) was driven by changes in both the column density and covering fraction of the obscurer \citep{Cappi2016}, whereas for the long-term variability (weeks to months), the covering fraction of the obscurer changed over time \citep{Mehdipour55482016, Mehidpour2022}. Here we discover that over the course of Obs 1 (103 ks) the column density of the obscurer varied significantly in relation to the changes in the observed X-ray flux, despite $\xi$ being free, while $f_{\rm cov}$ stayed constant. However, three weeks later, during Obs 2, both the covering fraction and column density increased, consistent with the strong X-ray flux decrease observed between the two observations. \citetalias{Mao_Prep} attribute this to two separate obscuring clouds occulting the LOS X-ray continuum due to large differences in the ionisation parameters. On the other hand, the continuum normalisations are significantly smaller in Obs 2, compared to Obs 1, and therefore it is not clear what produces the strong flux decrease.
	
	Finally, the bottom panel of Fig. \ref{Fig:TB_Compare} displays the ionisation parameter ($\xi$) of the obscurer for each TB. For Obs 2, the TB values are constant with the time-averaged value, except for TB3 which is an upper limit. For Obs 1, on the other hand, it is very difficult to constrain $\xi$, even for the time-averaged value. For TB1 - TB3, the uncertainties are very large, and for TB4 - TB6 only upper limits can be found. The obscurer ionisation parameter can only be constrained for TB7 to TB10 when the flux is at its highest. Given the size of these uncertainties we cannot conclude one way or the other whether the ionisation state of the obscurer responds to changes in the ionising continuum ($N_{\rm pow}$ and $N_{\rm comt}$). On the other hand, these large $\xi$ errors could indicate an inhomogeneous obscurer that is made up of multiple components, each with a different $N_{\rm H}$, $\xi$, or number density ($n_{\rm e}$), such that any global change would be hard to identify with a single model component.
	
	\begin{table*}
		\centering
		\caption{Best fit parameter results fitted to the 0.35 - 10 keV EPIC-PN time-averaged (`All') and TB spectra. Continuum parameters: power law normalisation and photon index ($N_{\rm pow}$ and $\Gamma_{\rm pow}$, respectively); Comptonisation normalisation ($N_{\rm comt}$); reflection scaling ($s$). Obscurer parameters: column density ($N_{\rm H}$); ionisation parameter ($\xi$); covering fraction ($f_{\rm cov}$). Obs 1 results are shown in the top half and the bottom half displays the best fit Obs 2 results. }
		\label{Table:Full_Spec_Results}
		\begin{tabular}{c c c c c c c c c}
			\hline
			\multicolumn{9}{c}{Obs 1 (d.o.f = 116)}  \Tstrut\Bstrut \\
			\hline
			Time   &  $N_{\rm pow}$   &  \multirow{2}{*}{$\Gamma$}  &  $N_{\rm comt}$ & \multirow{2}{*}{s}& $N_{\rm H}$  & $\log \xi$ & \multirow{2}{*}{$f_{\rm cov}$} & \multirow{2}{*}{C-stat}   \Tstrut\Bstrut \\
			Bin &  (1) & & (2) & & (3) &  (4) & & \Tstrut\Bstrut \\
			\hline
			\hline
			All & $2.60 \pm 0.05$ &  $1.68 \pm 0.01$ &  $1.89 \pm 0.03$ & $0.64 \pm 0.04$ &  $3.87 \pm 0.07$  & $0.44^{+0.90}_{-2.00}$ & $0.55 \pm 0.01$ & 221  \Tstrut\Bstrut \\
			1  & $2.46 \pm 0.25$ & $1.83 \pm 0.05$ & $1.01 \pm 0.20$ & $1.10 \pm 0.30$ &  $6.27 \pm 0.40$ & $0.48^{+1.70}_{-0.90}$ & $0.60 \pm 0.04$ & 158 \Tstrut\Bstrut \\
			2  & $2.40 \pm 0.20$ & $ 1.85 \pm 0.05$ & $0.87 \pm 0.10$& $ 1.93 \pm 0.30$ &  $5.09 \pm 0.30$ & $0.03^{+0.60}_{-4.40}$& $0.58 \pm 0.03$ & 163 \Tstrut\Bstrut \\
			3  & $ 2.09 \pm 0.20$ & $1.73 \pm 0.04$ & $ 1.15 \pm 0.10$ & $ 1.20 \pm 0.20$ & $ 5.98 \pm 0.40$ & $ 0.08^{+2.30}_{-1.20}$ & $ 0.59 \pm 0.03$ & 127  \Tstrut\Bstrut \\
			4  & $ 2.21 \pm 0.30$ & $ 1.66 \pm 0.04$ & $ 1.52 \pm 0.10$ & $ 0.77 \pm 0.15$ & $ 5.09 \pm 0.95$ & $< 2.37$ & $ 0.60 \pm 0.04$ & 133  \Tstrut\Bstrut \\
			5  & $ 2.17 \pm 0.20$ & $ 1.65 \pm 0.04$ & $ 1.52 \pm 0.10$ & $ 0.61 \pm 0.15 $ & $ 4.93 \pm 0.30$ & $ < 0.38 $ & $ 0.62 \pm 0.03$ & 157  \Tstrut\Bstrut \\
			6  & $ 2.7 7\pm 0.20 $ & $ 1.68 \pm 0.04$ & $ 1.76 \pm 0.20$ & $ 0.63 \pm 0.15 $ & $ 3.96 \pm 0.50$ & $ < 1.62 $ & $ 0.54 \pm 0.04$ &  146 \Tstrut\Bstrut \\
			7  & $ 3.24 \pm 0.20$ & $ 1.69 \pm 0.03$ & $ 2.34 \pm 0.20 $ & $ 0.48 \pm 0.10$ & $ 3.45 \pm 0.20$ & $ 1.39 \pm 0.30$ & $ 0.56 \pm 0.02 $ & 101  \Tstrut\Bstrut \\
			8  & $ 3.60 \pm 0.20$ & $ 1.73 \pm 0.03$ & $ 2.82 \pm 0.20 $ & $ 0.58 \pm 0.10$ & $ 3.04 \pm 0.20 $ & $ 1.02 \pm 0.25$ & $ 0.59 \pm 0.02$ &  114 \Tstrut\Bstrut \\
			9  & $ 3.03 \pm 0.20$ & $1.64 \pm 0.03$ & $ 3.01 \pm 0.20$ & $ 0.37 \pm 0.10$ & $2.70 \pm 0.20$ & $ 0.73 \pm 0.35$ & $ 0.57 \pm 0.02$ & 142  \Tstrut\Bstrut \\
			10 & $ 3.21 \pm 0.20$ & $1.67 \pm 0.03$ & $ 2.56 \pm 0.20$ & $ 0.31 \pm 0.10 $ & $2.93 \pm 0.20 $ & $1.33 \pm 0.25$ & $ 0.57 \pm 0.02$ & 136  \Tstrut\Bstrut \\
			\hline
			\hline
			\multicolumn{9}{c}{Obs 2 (d.o.f = 106)}  \Tstrut\Bstrut \\
			\hline
			Time   &  $N_{\rm pow}$   &  \multirow{2}{*}{$\Gamma$}  &  $N_{\rm comt}$ & \multirow{2}{*}{s}&  $N_{\rm H}$  & $\log \xi$ & \multirow{2}{*}{$f_{\rm cov}$} & \multirow{2}{*}{C-stat}   \Tstrut\Bstrut \\
			Bin &  (5) & & (6) & & (3) & (4) & & \Tstrut\Bstrut \\
			\hline
			\hline
			All & $7.23 \pm 0.40$ &  $1.73 \pm 0.03$ & $1.56 \pm 0.30$ & $1.66 \pm 0.15$ &  $9.93 \pm 1.00$  & $2.72 \pm 0.10$ & $0.66 \pm 0.02$ & 102  \Tstrut\Bstrut \\
			1  & $6.80 \pm 1.10$ & $ 1.68 \pm 0.10$ & $ 1.52 \pm 0.70$ & $ 1.58 \pm 0.30 $ & $10.72 \pm 1.60$ & $2.91 \pm 0.15$& $0.63 \pm 0.04$ & 93 \Tstrut\Bstrut \\
			2  & $ 8.12\pm 1.10 $ & $ 1.84 \pm 0.08$ & $1.30 \pm 0.70$ & $ 2.19 \pm 0.40$ & $ 7.68 \pm 1.50$ & $ 2.50 \pm 0.25$ & $ 0.67 \pm 0.05$ & 103  \Tstrut\Bstrut \\
			3  & $ 6.02 \pm 0.90$ & $1.62 \pm 0.08$ & $ 2.46 \pm 0.70$ & $ 1.32 \pm 0.30$ & $ 5.09 \pm 1.00$ & $ < 2.24 $ & $  0.57 \pm 0.06$ & 112   \Tstrut\Bstrut \\
			4  & $7.47 \pm 1.00$ & $1.72 \pm 0.07$ & $1.37 \pm 0.70$ & $ 1.42 \pm 0.30$ & $ 12.24 \pm 2.00$ & $ 2.80 \pm 0.15$ & $ 0.68 \pm 0.05$ &  129 \Tstrut\Bstrut \\
			5 & $ 8.43 \pm 1.10$ & $ 1.90 \pm 0.09$ & $1.56 \pm 1.00$ & $ 2.41 \pm 0.40$ & $9.02 \pm 1.50$ & $2.46 \pm 0.25$ & $0.74 \pm 0.04$ & 107  \Tstrut\Bstrut \\
			\hline
		\end{tabular}
		\tablefoot{
			{(1) $10^{50}$ ph s\textsuperscript{-1} keV\textsuperscript{-1} at 1 keV;}
			{(2) $10^{53}$ ph s\textsuperscript{-1} keV\textsuperscript{-1} at 1 keV;}
			{(3) $10^{26}$ m\textsuperscript{-2};}
			{(4) $10^{-9}$ Wm;}
			{(5) $10^{49}$ ph s\textsuperscript{-1} keV\textsuperscript{-1} at 1 keV;}
			{(6) $10^{52}$ ph s\textsuperscript{-1} keV\textsuperscript{-1} at 1 keV.}
		}
	\end{table*}

	\subsection{Parameter comparison}
	\label{Sec:Par_Comp}
	After fitting the data for each TB, we examined whether there were any correlations between the parameters. In doing so, we find that there are significant trends between $N_{\rm pow}$ and $N_{\rm comt}$, as well as with $N_{\rm H}$. These are displayed in the left side of Fig. \ref{Fig:Parameter_Compare}. For Obs 1, $N_{\rm pow}$ and $N_{\rm comt}$ relate well with Fig. \ref{Fig:Light_Curves}, while an inverse trend was seen with $N_{\rm H}$ (Fig. \ref{Fig:TB_Compare}). In each panel of Fig. \ref{Fig:Parameter_Compare}, the Pearson rank and \textit{p}-values are shown to display how well correlated the parameters are. In all panels, the \textit{p}-values are significantly lower than the quoted value and hence they are shown as $p < 0.001$. The absolute values of $r$ are greater than 0.8 (both in the positive and negative directions) implying that these parameters in Fig. \ref{Fig:Parameter_Compare} are correlated or anti-correlated very well with each other. 
	
	The right side of Fig. \ref{Fig:Parameter_Compare} shows the relation between $N_{\rm pow}$, $N_{\rm comt}$, and $N_{\rm H}$ with the 0.3 - 10 keV count rates, taken from the bottom panel of Fig. \ref{Fig:Counts_Comp_Obs}. Both normalisations are positively correlated with the 0.3 - 10 keV count rates (top right and middle right panels of Fig. \ref{Fig:Parameter_Compare}), while there is an inverse correlation between $N_{\rm H}$ and the count rate (bottom right panel in Fig. \ref{Fig:Parameter_Compare}). This possibly suggests that the observed variability is related to changes in the column density seen in Obs 1 (see Sect. \ref{Sec:Var_Cause}). For Obs 2, the count rates are consistently low for each model parameter, which is a result of high $N_{\rm H}$ and $f_{\rm cov}$ values, and a low continuum, in Obs 2; no strong increase over the observation is seen. 
	
	We further analysed our modelling to test the statistical significance of thawing a parameter from its time-averaged value in the spectral fitting. To do this, we fixed each of the four parameters of interest - $N_{\rm H}$, $N_{\rm pow}$, $N_{\rm comt}$, and $f_{\rm cov}$ - to their time-averaged values (Table \ref{Table:Full_Spec_Results}) in turn, while all other model parameters were free, and we fitted this adjusted model to the ten TB spectra. We repeated this for each TB in Obs 1, before measuring the ratios between the C-statistic values and the d.o.f and summing them ($S_{\rm fixed} = \sum C_{i}/\rm dof_{i}$, where $\rm dof = 117$ and \textit{i} is the TB number). The summed ratios for the four parameters are displayed in Table \ref{Table:FTest}. 
	As a comparison, we also sum the $S_{\rm free} = C_{i}/\rm dof_{i}$ ratios (where $\rm dof = 116$ here) from the best fit TB results in Table \ref{Table:Full_Spec_Results}, which is $S_{\rm free} = 11.87$. We then calculate the difference between these two summations ($\Delta S = S_{\rm fixed} - S_{\rm free}$), which we use to test the statistical significance each parameter has on the model; the differences are also shown in Table \ref{Table:FTest}.
	
	Next, we ran some F-tests to examine whether the variance ($\sigma$) calculated from the $\sigma_{\rm fixed} = C_{i}/\rm dof_{i}$ ratios in each TB when the parameters of interest are fixed, come from the same distribution as the best fit $\sigma_{\rm free} = C_{i}/\rm dof_{i}$ variance when all parameters in the model are free (see Table \ref{Table:Full_Spec_Results}). The ratio between the variances give rise to the F-value, given as $F = \sigma_{\rm fixed} / \sigma_{\rm free}$. The F-value is used, along with the calculated \textit{p}-value, to determine the statistical probability that the two variances come from the same underlying distribution, when compared to the difference in the summed ratios $\Delta S$ for each fixed parameter. If the F-value is larger than $\Delta S$, then it implies that the parameter in question is statistically significant in the best fit model. The results for the F-values and corresponding \textit{p}-values are displayed in Table \ref{Table:FTest}. 
	
	From the results in Table \ref{Table:FTest}, the F-values for $N_{\rm comt}$ and $N_{\rm H}$ are larger than the difference in the summed ratios ($\Delta S$), for the respective parameter. In addition, $N_{\rm comt}$ and $N_{\rm H}$ have sufficiently small \textit{p}-values (less than 0.05) which implies that it is unlikely for the C-statistic variances ($\sigma_{\rm fixed}$) of these two parameters to come from the same distribution as the best fit C-statistic variance ($\sigma_{\rm free}$). These results suggest that $N_{\rm comt}$ and $N_{\rm H}$ have the most statistically significant impact on the best model from Sect. \ref{Sec:Fit_Spec} and Table \ref{Table:Full_Spec_Results}. We also find that the F-value for $f_{\rm cov}$ is larger than the $\Delta S$ measure. However, the \textit{p}-value for $f_{\rm cov}$ is greater than 0.05 (Table \ref{Table:FTest}), which means we cannot conclude whether this parameter has a significant impact on the best fit with certainty, as this result indicates that it is likely for the C-statistic variance from $f_{\rm cov}$ to lie on the same distribution as the best fit model variance. Finally, the F-value for $N_{\rm pow}$ is smaller than the $\Delta S$ value and therefore means it does not have a significant impact on the best fit model in Obs 1, compared to $N_{\rm comt}$ and $N_{\rm H}$.
	
	\begin{table}
		\centering
		\caption{Testing the significance of the parameters on the best fit model (see Sect. \ref{Sec:Par_Comp} for details).} 
		\label{Table:FTest}
		\begin{tabular}{c | c c | c c}
			\hline
			Parameter & $S_{\rm fixed}$  & $\Delta S$ & F-value & \textit{p}-value  \Tstrut\Bstrut \\
			\hline
			$N_{\rm pow}$ & 12.78 & 0.91 & 0.69 & 0.98  \Tstrut\Bstrut \\
			$N_{\rm comt}$ & 14.72 & 2.85 & 4.81 & $3.33 \times 10^{-16}$  \Tstrut\Bstrut \\
			$N_{\rm H}$ & 13.62 & 1.75 & 2.10 & $3.75 \times 10^{-5}$  \Tstrut\Bstrut \\
			$f_{\rm cov}$ & 12.09 & 0.22 & 0.82 & 0.85  \Tstrut\Bstrut \\
			\hline	
		\end{tabular}
	\end{table}
	
	\begin{figure*}
		\centering
		\begin{subfigure}{0.25\linewidth}
			\includegraphics[width=1\linewidth]{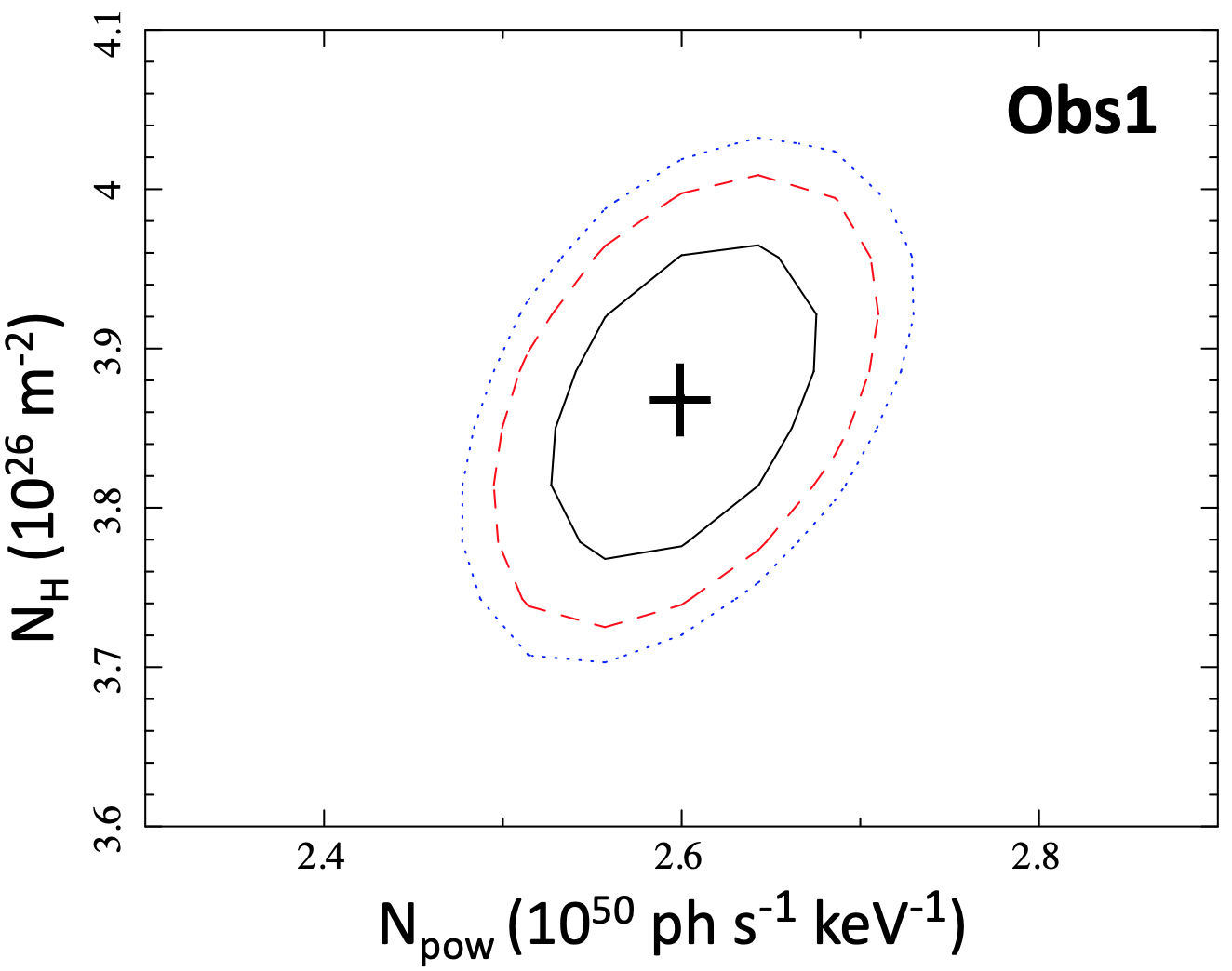}
		\end{subfigure}%
		\begin{subfigure}{0.25\linewidth}
			\includegraphics[width=1\linewidth]{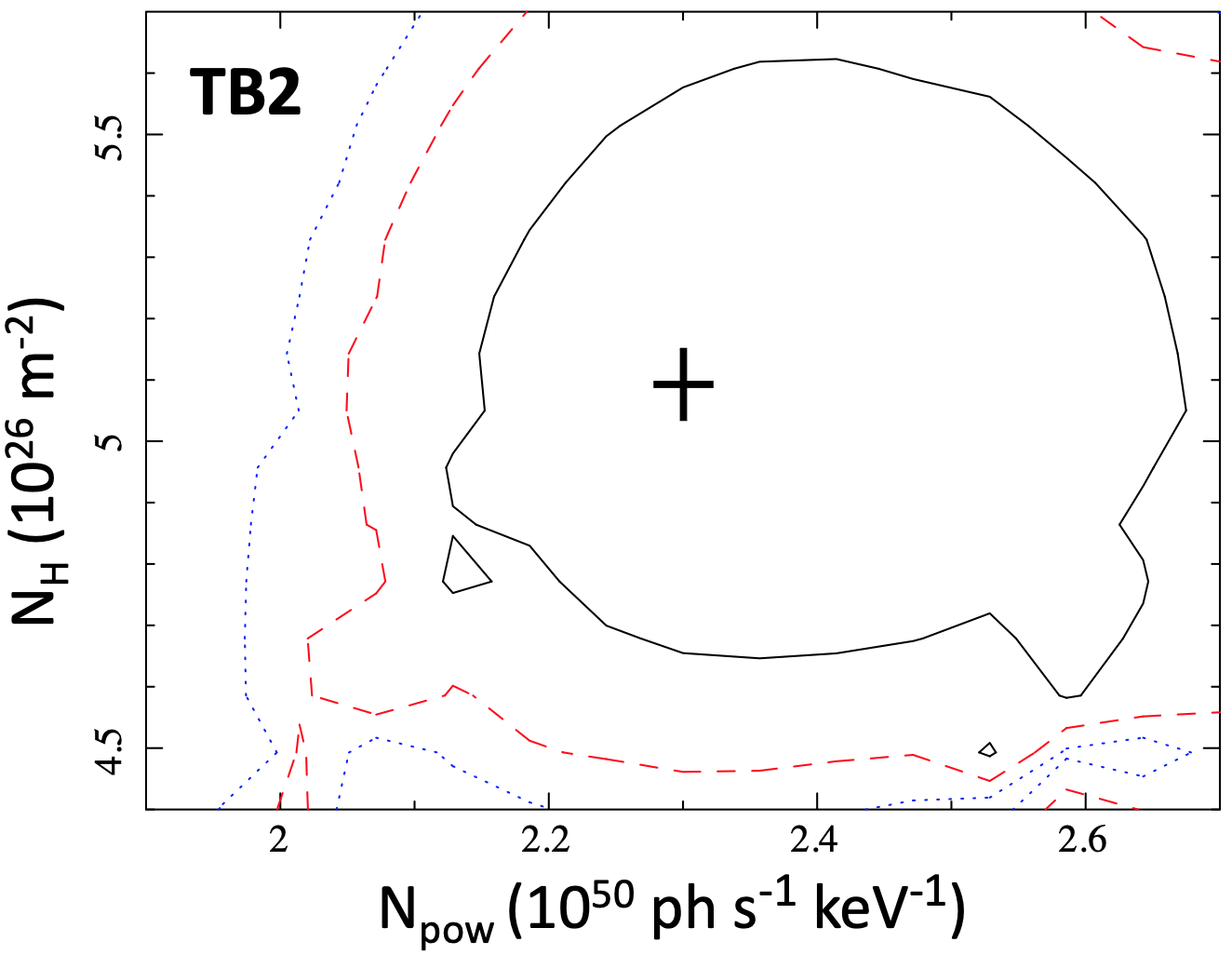}
		\end{subfigure}%
		\begin{subfigure}{0.25\linewidth}
			\includegraphics[width=1\linewidth]{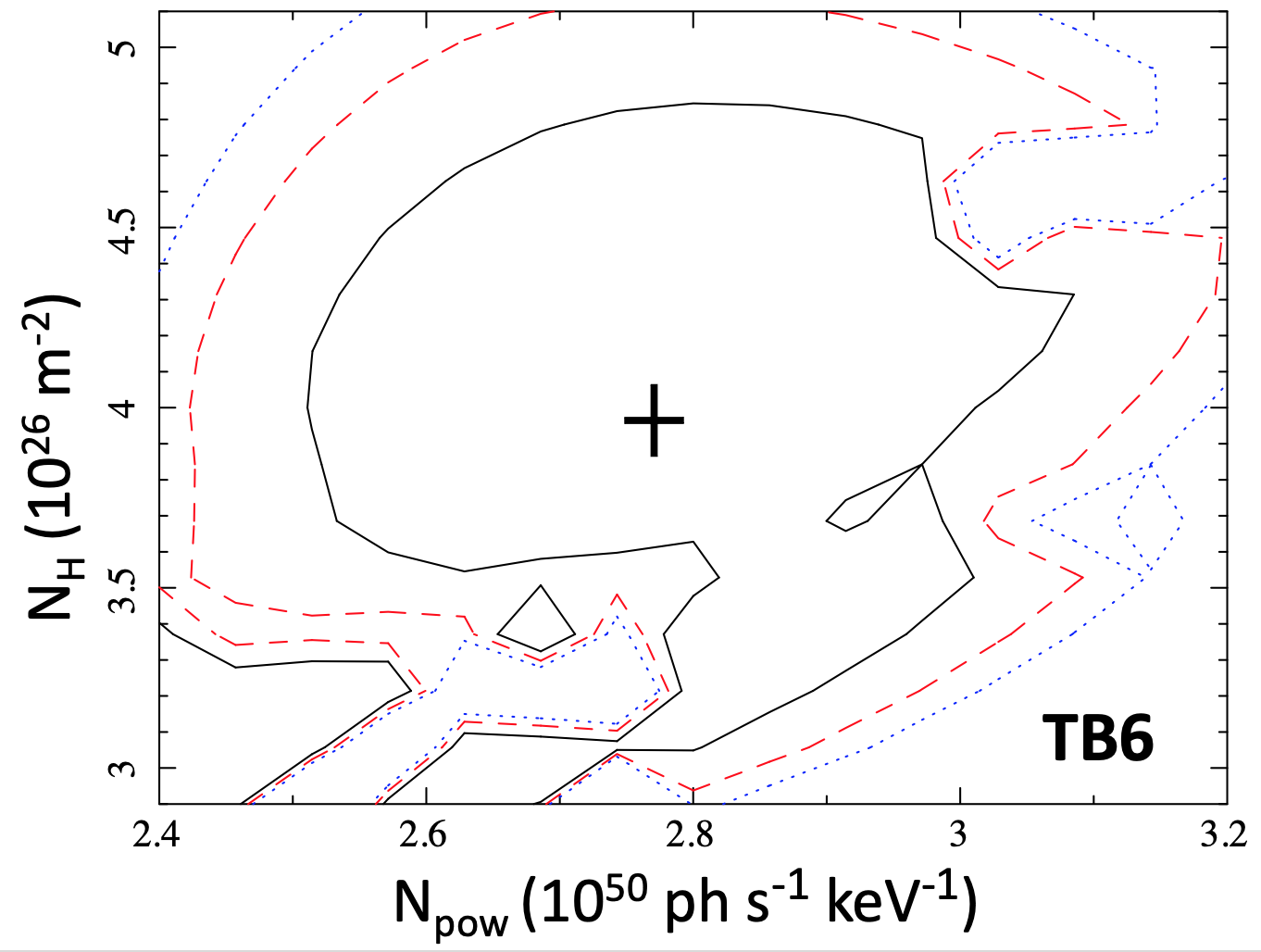}
		\end{subfigure}%
		\begin{subfigure}{0.25\linewidth}
			\includegraphics[width=1\linewidth]{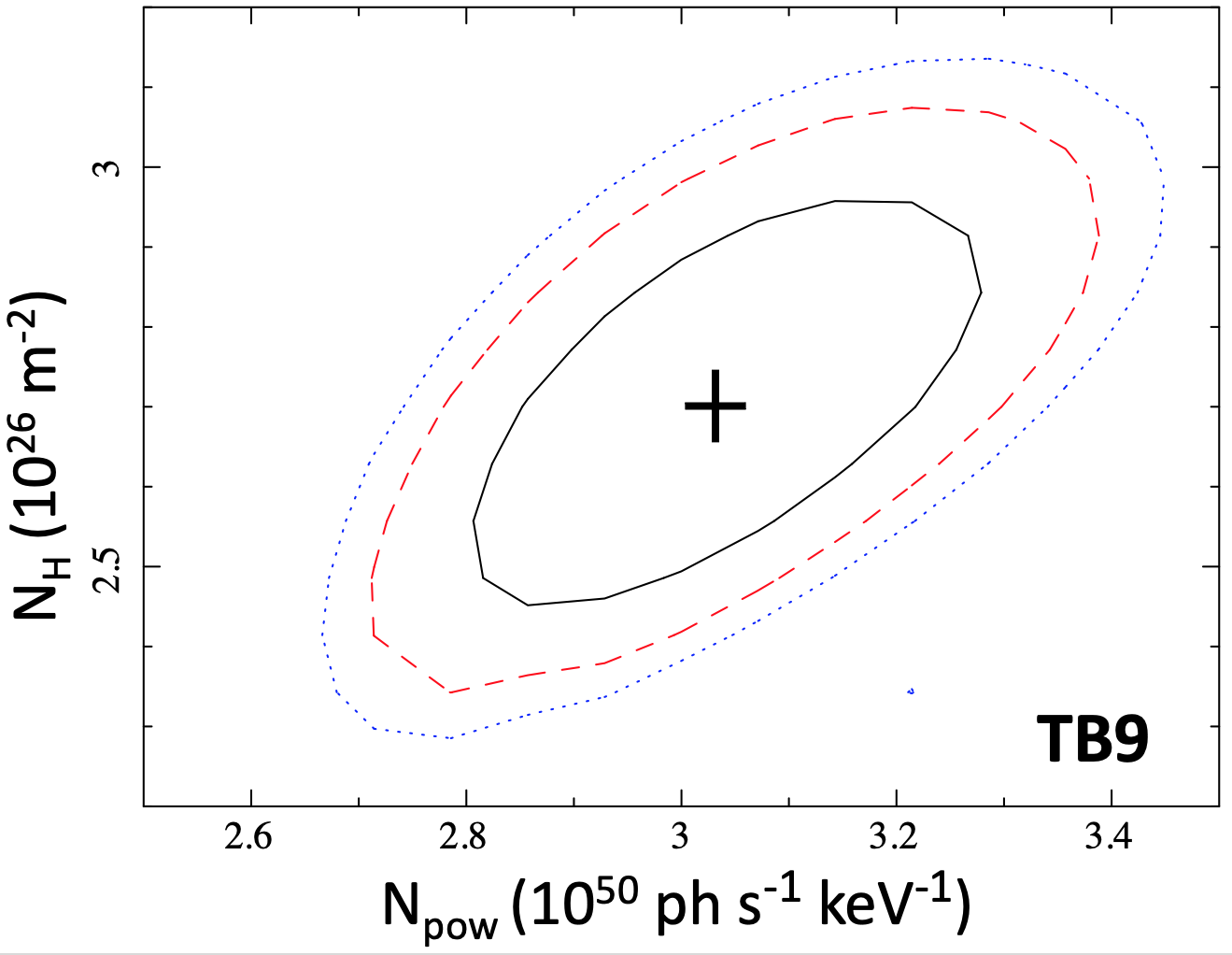}
		\end{subfigure}
		
		\begin{subfigure}{0.25\linewidth}
			\includegraphics[width=1\linewidth]{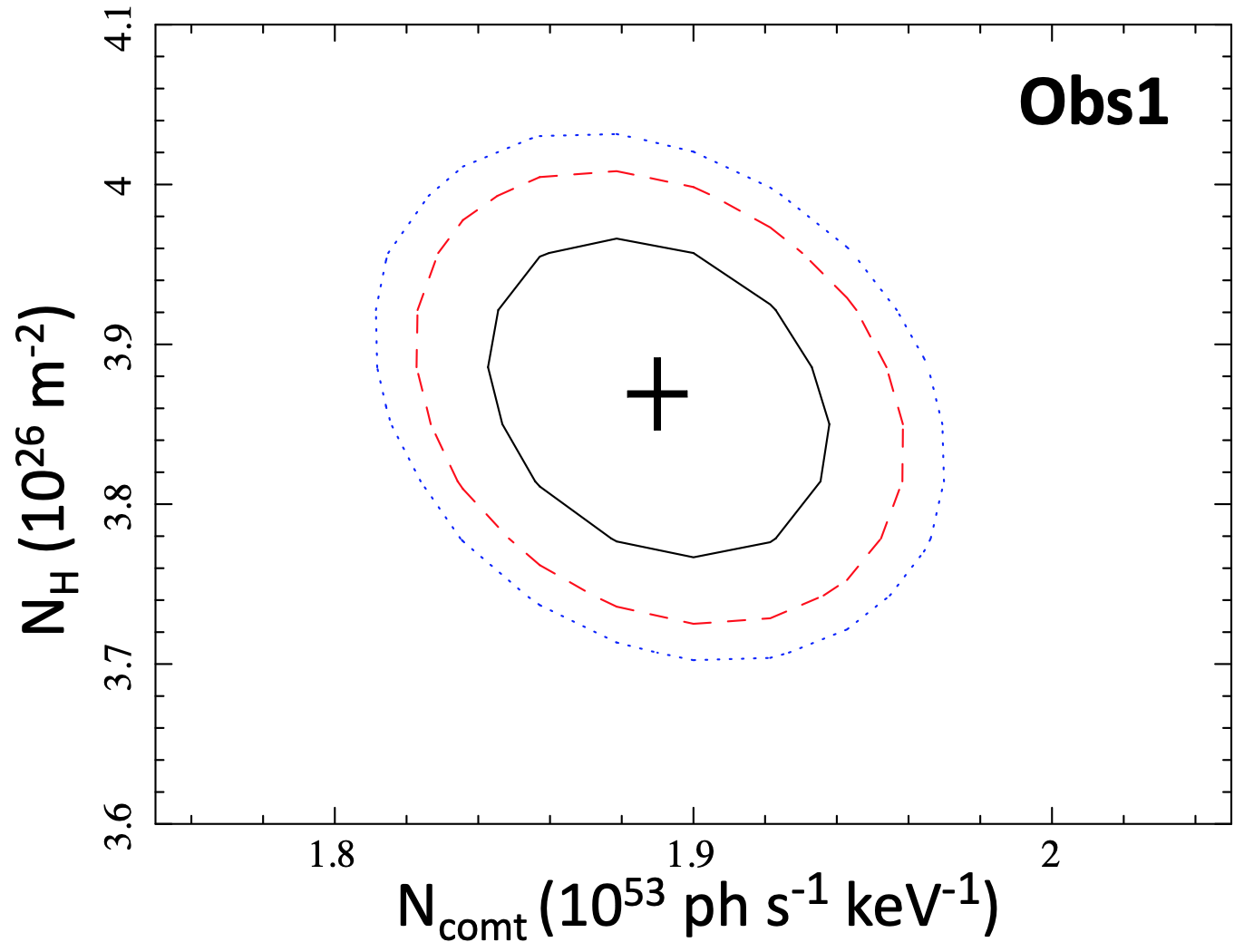}
		\end{subfigure}%
		\begin{subfigure}{0.25\linewidth}
			\includegraphics[width=1\linewidth]{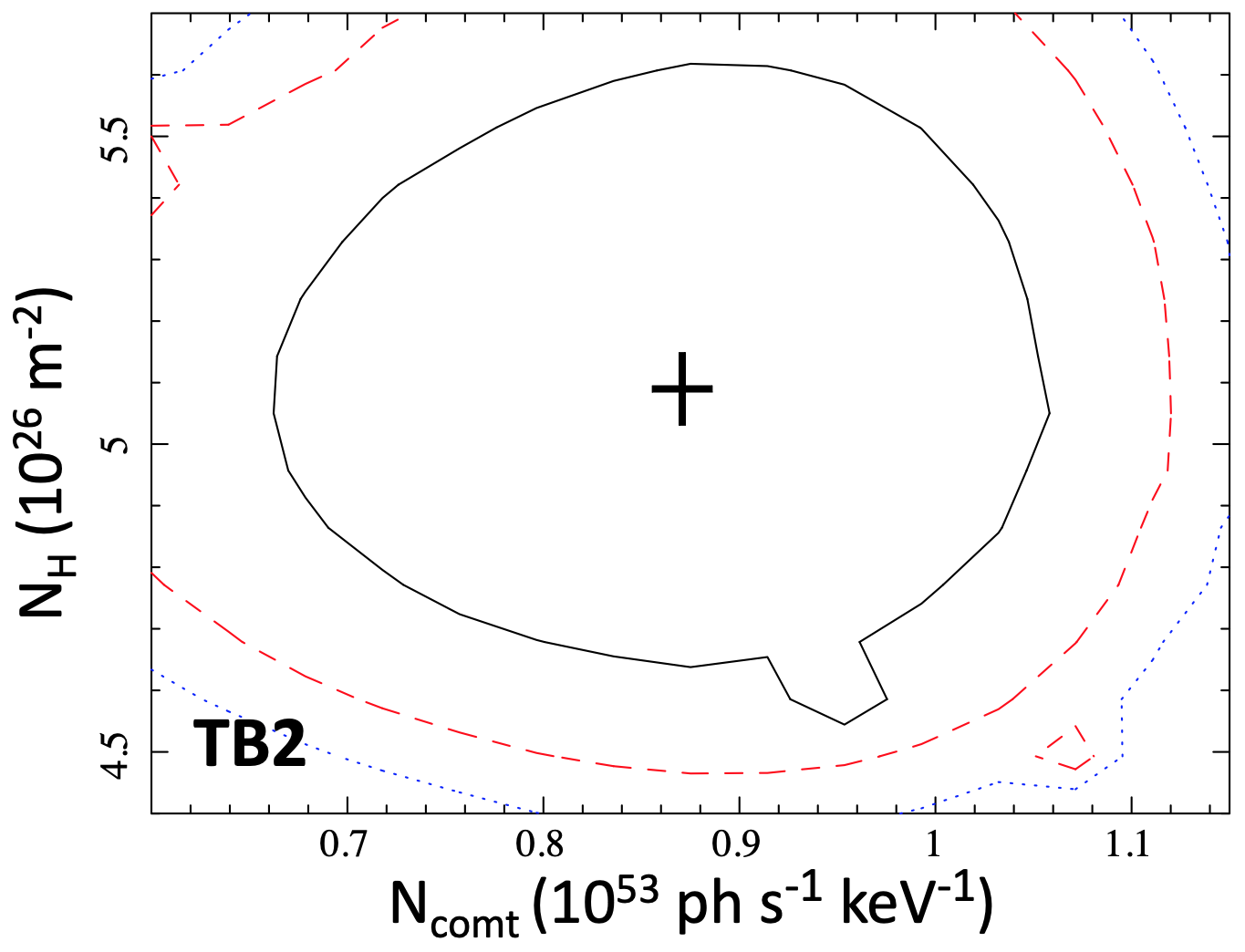}
		\end{subfigure}%
		\begin{subfigure}{0.25\linewidth}
			\includegraphics[width=1\linewidth]{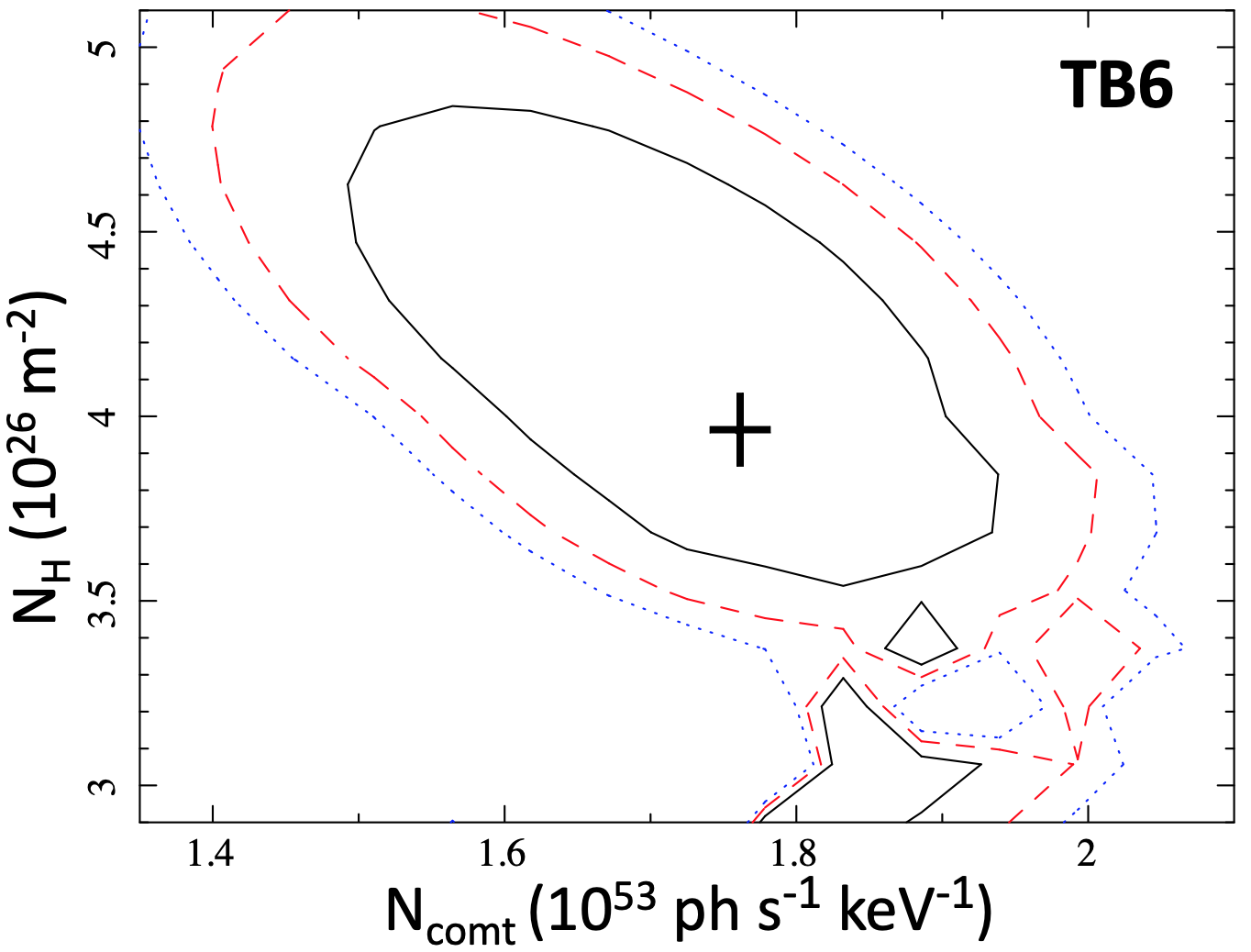}
		\end{subfigure}%
		\begin{subfigure}{0.25\linewidth}
			\includegraphics[width=1\linewidth]{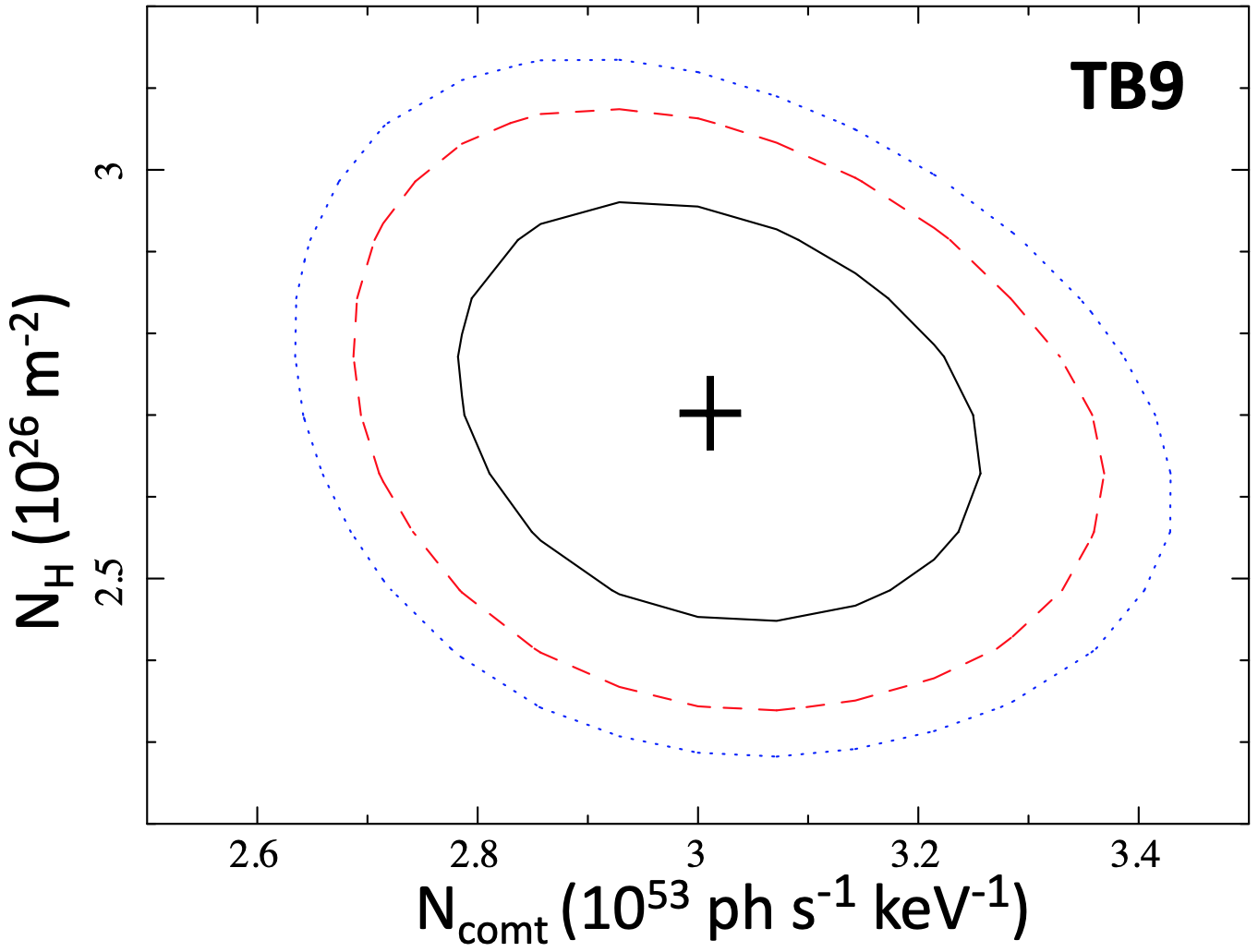}
		\end{subfigure}
		\caption{Contour plots of $N_{\rm H}$ ($10^{26}$ m\textsuperscript{-1}) against $N_{\rm pow}$ ($10^{50}$ ph s\textsuperscript{-1} keV\textsuperscript{-1}; top panels) and $N_{\rm H}$ ($10^{26}$ m\textsuperscript{-1}) against $N_{\rm comt}$ ($10^{53}$ ph s\textsuperscript{-1} keV\textsuperscript{-1}; bottom panels). From left to right, the panels show the contours for the time-averaged model, TB2, TB6, and TB9, respectively, in Obs 1. The levels of confidence presented are at 68 \% ($\Delta C = 2.30$; black solid line), 90 \% ($\Delta C = 4.61$; red dashed line), and 95 \% ($\Delta C = 6.17$; blue dotted line). The black crosses show the parameter values (but not the errors) from the best fit models in Table \ref{Table:Full_Spec_Results}.}
		\label{Fig:Contours}
	\end{figure*}
	
	To explore the anti-correlation between the obscurer and continuum (left side of Fig. \ref{Fig:Parameter_Compare}), we create some contour plots of $N_{\rm H}$ against $N_{\rm pow}$ and $N_{\rm comt}$ (Fig. \ref{Fig:Contours}) for the time-averaged spectrum in Obs 1, and for the spectra of TB2, TB6, and TB9 as they correspond to a low flux bin, a change in flux bin, and a high flux bin, respectively (see Fig. \ref{Fig:Light_Curves}). The far left panels in Figure \ref{Fig:Contours} show that the time-averaged parameters ($N_{\rm pow}$ top, and $N_{\rm comt}$ bottom) are relatively well constrained within the 68 \% confidence level (black solid lines). This suggests that there are no significant degeneracies between these three parameters. For TB2 (middle left panels) and TB6 (middle right panels), the contours are less smooth and round compared to the time-averaged plots. The contours for TB9 (far right panels), however, display a similarity with the time-averaged contours. A possible explanation for this is that TB2 and TB6 have lower fluxes compared to TB9, and therefore a lower signal-to-noise ratio (S/N).
	
	There are slight differences between the widths of the 68\% confidence contours (black lines) of $N_{\rm pow}$ (0.48 to 0.55 $\times 10^{50}$ ph s\textsuperscript{-1} keV\textsuperscript{-1}) and $N_{\rm comt}$ (0.40 to 0.48 $\times 10^{53}$ ph s\textsuperscript{-1} keV\textsuperscript{-1}) for TB2, TB6, and TB9, in Fig. \ref{Fig:Contours}. On the other hand, there is a large difference between the widths of the confidence intervals of $N_{\rm H}$ (0.4 to 1.0 $\times 10^{26}$ m\textsuperscript{-2}). For the time-averaged contours, the intervals covered by the 68\% confidence contours are $\Delta N_{\rm pow} = 0.16 \times 10^{50}$ ph s\textsuperscript{-1} keV\textsuperscript{-1}, $\Delta N_{\rm comt} = 0.10 \times 10^{53}$ ph s\textsuperscript{-1} keV\textsuperscript{-1} and  $\Delta N_{\rm H} = 0.20 \times 10^{26}$ m\textsuperscript{-2}. Therefore, from Fig. \ref{Fig:Contours} there does not appear to be any strong degeneracy between these model parameters, both from the time-averaged and TB spectra, and the apparent anti-correlation between the continuum and obscurer column density seen in Figs. \ref{Fig:TB_Compare} and \ref{Fig:Parameter_Compare} are due to a physical relation discussed in Sect. \ref{Sec:Var_Cause}. 
	
\begin{figure}
		\centering
		\includegraphics[width=\linewidth]{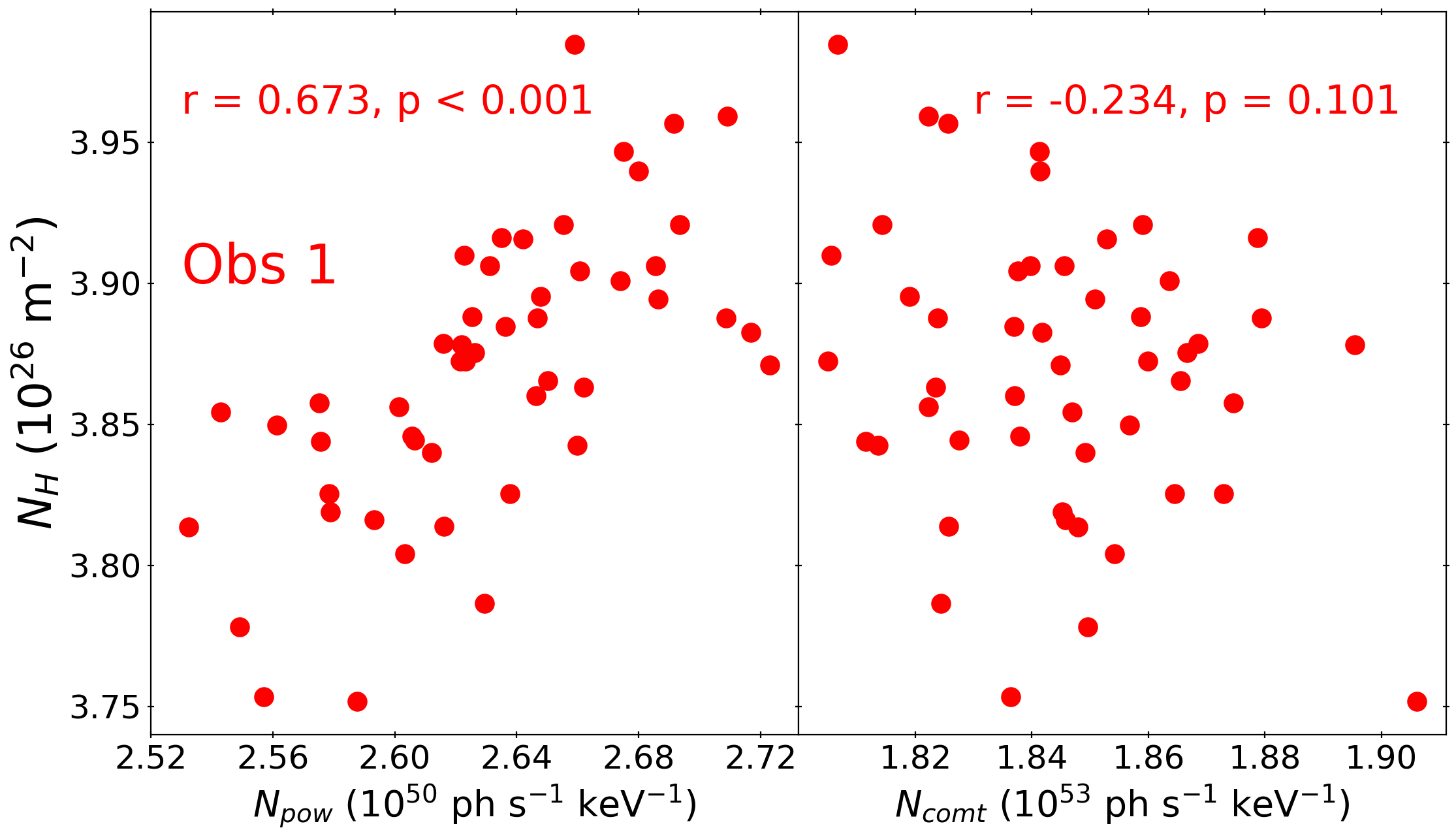}
		\includegraphics[width=\linewidth]{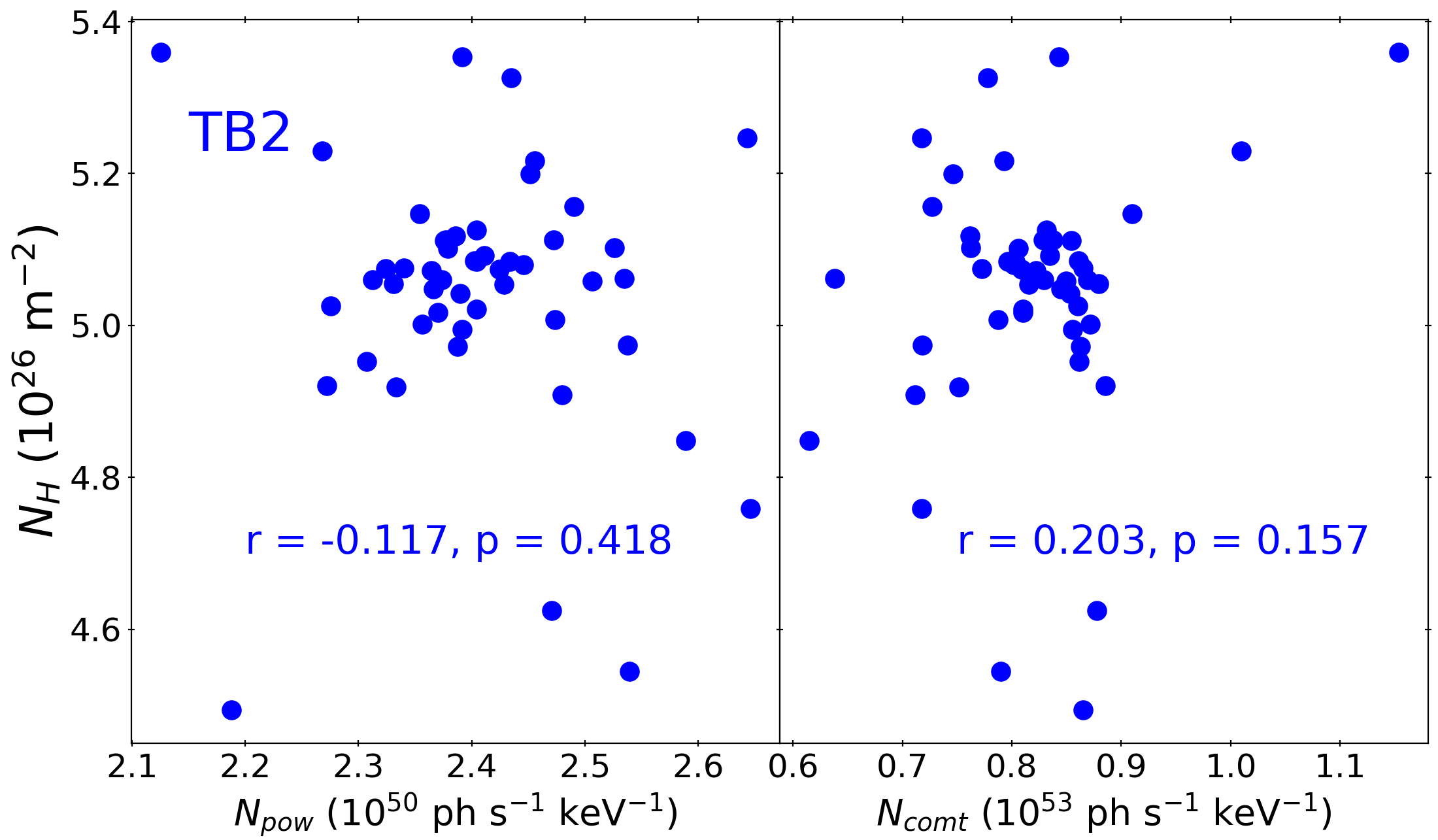}
		\includegraphics[width=\linewidth]{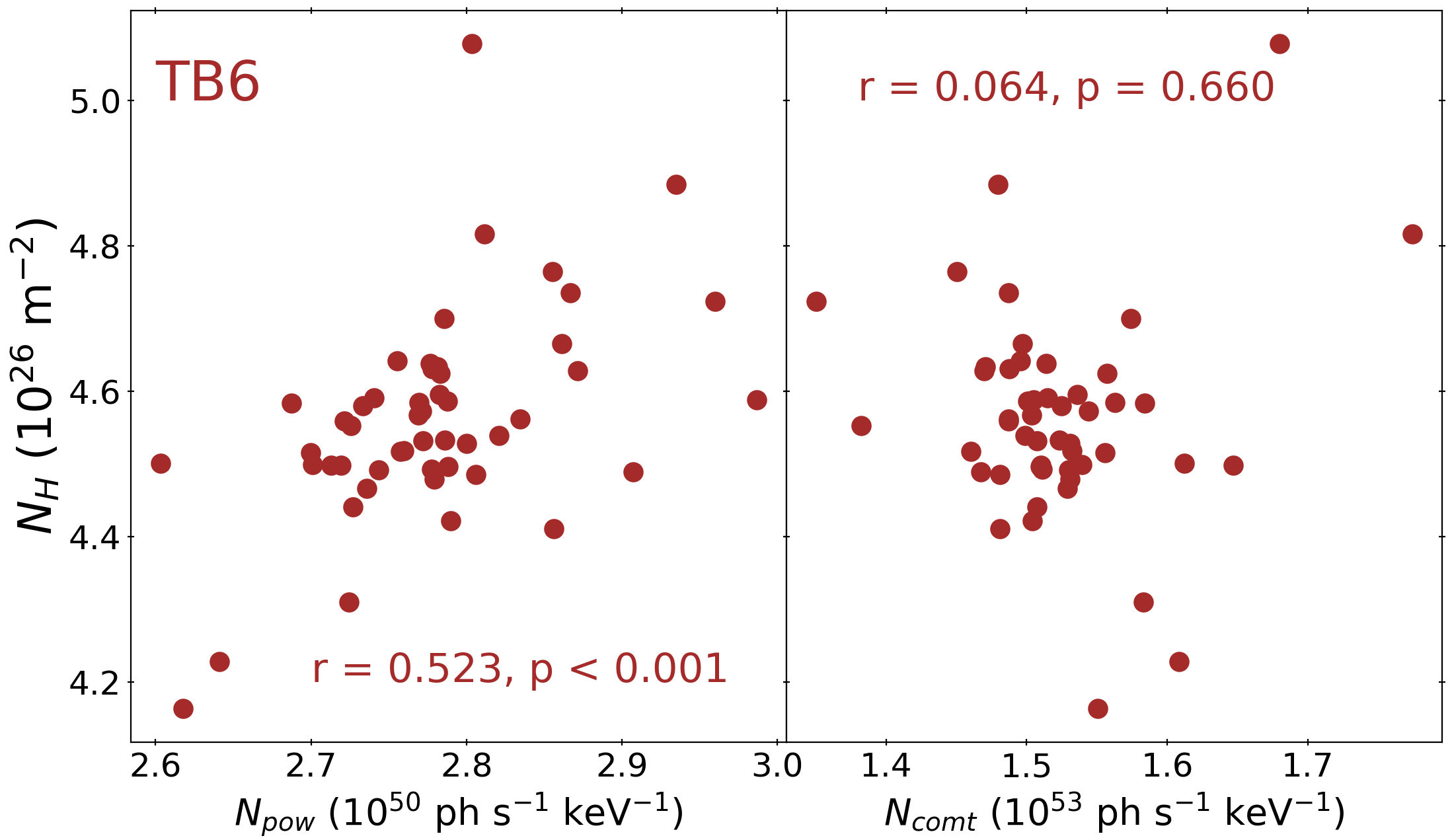}
		\includegraphics[width=\linewidth]{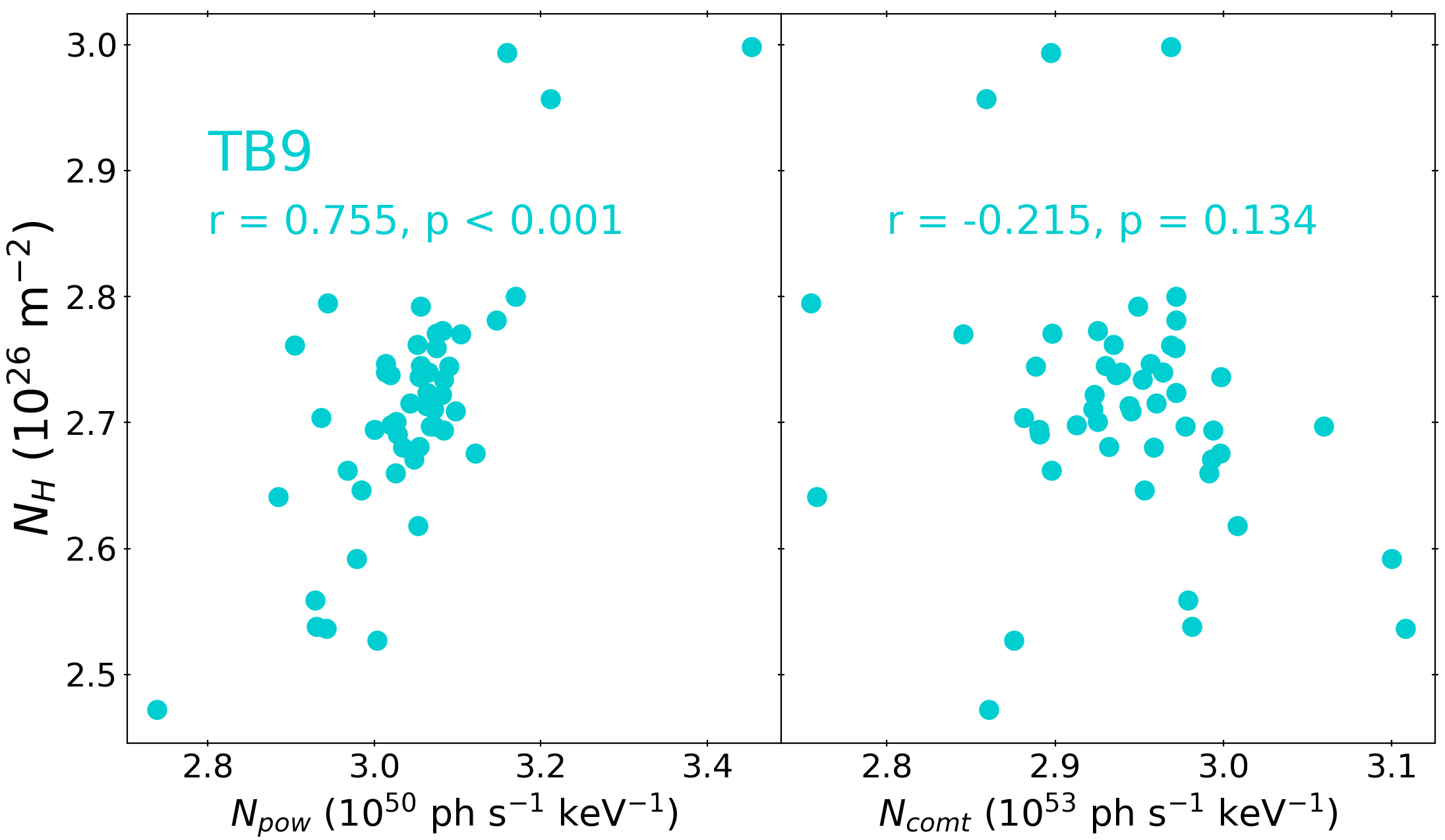}
		\caption{Monte Carlo spectral simulation results for the time-averaged model (top panel; red) and the models from TB2 (top-middle; blue), TB6 (bottom-middle; brown), and TB9 (bottom panel; cyan). The left and right sides of each panel display the results of $N_{\rm H}$ against $N_{\rm pow}$ and $N_{\rm H}$ and $N_{\rm comt}$, respectively, after fitting the best fit models to the simulated spectra 50 times each. Also displayed in each panel are the Pearson rank ($r$) and \textit{p}-values to demonstrate the linear correlation coefficient and corresponding probabilities for the parameters, used to determine parameter degeneracies.}
		\label{Fig:N_Sim_Plot}
	\end{figure}

	Finally, we carry out Monte Carlo spectral simulations to test whether there are any degeneracies between $N_{\rm H}$ and $N_{\rm pow}$ or $N_{\rm comt}$. We do this by simulating spectra in \texttt{SPEX} using the best fit models (Table \ref{Table:Full_Spec_Results}) fitted to the time-averaged spectrum of Obs 1 and the spectra of TB2, TB6, and TB9. We then refit the best fit models (all of the parameters in Table \ref{Table:Full_Spec_Results}) to the simulated spectra (time-averaged, TB2, TB6, or TB9), repeating this process 50 times each. The results of these simulations are shown in Fig. \ref{Fig:N_Sim_Plot}, where we plot $N_{\rm H}$ against $N_{\rm pow}$ in the left side of each panel and $N_{\rm H}$ against $N_{\rm comt}$ in the right side. In addition, we also present the Pearson rank ($r$) and \textit{p}-value in each panel, demonstrating the linear correlation coefficient and corresponding probabilities for the parameters. 
	
	Overall, there does not appear to be any correlation between $N_{\rm H}$ and $N_{\rm comt}$ for any TB or time-averaged simulations (right panels in Fig. \ref{Fig:N_Sim_Plot}), suggesting that there is no degeneracy between these two parameters (the Pearson rank values are between $r= -0.23$ and $0.20$ and the \textit{p}-values are larger than 0.05, suggesting the correlations are not significant). This is a similar result for $N_{\rm H}$ against $N_{\rm pow}$ in TB2, where there is no significant correlation between these parameters (left panels in Fig. \ref{Fig:N_Sim_Plot}). However, for the Obs 1 time-averaged, TB6, and TB9 simulated results, there appears to be some correlation between $N_{\rm H}$ and $N_{\rm pow}$, where the Pearson rank values are $r = 0.673$, $0.523$, and $0.755$, respectively, and $p < 1 \times 10^{-5}$ (displayed as $< 0.001$ in Fig. \ref{Fig:N_Sim_Plot} for visual purposes). These correlations in the left panels of Fig. \ref{Fig:N_Sim_Plot} imply that there could be some degeneracy between $N_{\rm H}$ and $N_{\rm pow}$. The results in Fig. \ref{Fig:N_Sim_Plot} also suggests that if the obscurer was to change with the continuum, it would vary with $N_{\rm pow}$ (not $N_{\rm comt}$). On the other hand, we conclude that there is no degeneracy between $N_{\rm H}$ and $N_{\rm comt}$.

	\subsection{Cause of the observed variability}
	\label{Sec:Var_Cause}
	From Figs. \ref{Fig:TB_Compare} and \ref{Fig:Parameter_Compare}, and the discussion in Sects. \ref{Sec:TB_Comp} and \ref{Sec:Par_Comp}, there is a clear inverse correlation between the continuum ($N_{\rm pow}$ and $N_{\rm comt}$) and the column density of the obscurer ($N_{\rm H}$) in Obs 1, but no clear correlation between $\xi$ and the continuum. 
	We therefore carried out further tests to see whether the observed changes could be caused by the continuum, ionisation parameter, or the column density.
	
\begin{figure*}
		\centering
		\begin{subfigure}{0.5\linewidth}
			\includegraphics[width=1\linewidth]{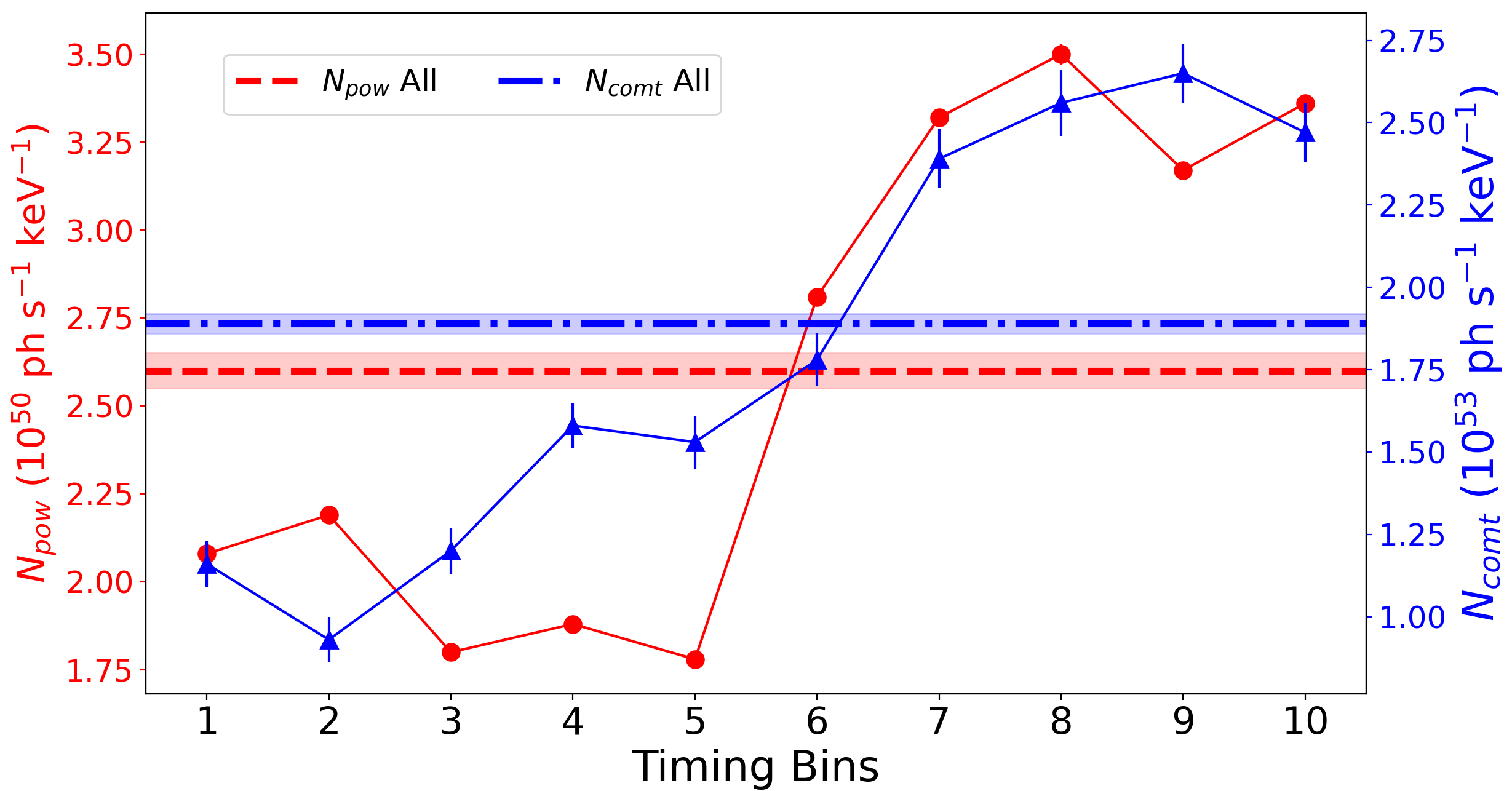}
		\end{subfigure}%
		\begin{subfigure}{0.5\linewidth}
			\includegraphics[width=1\linewidth]{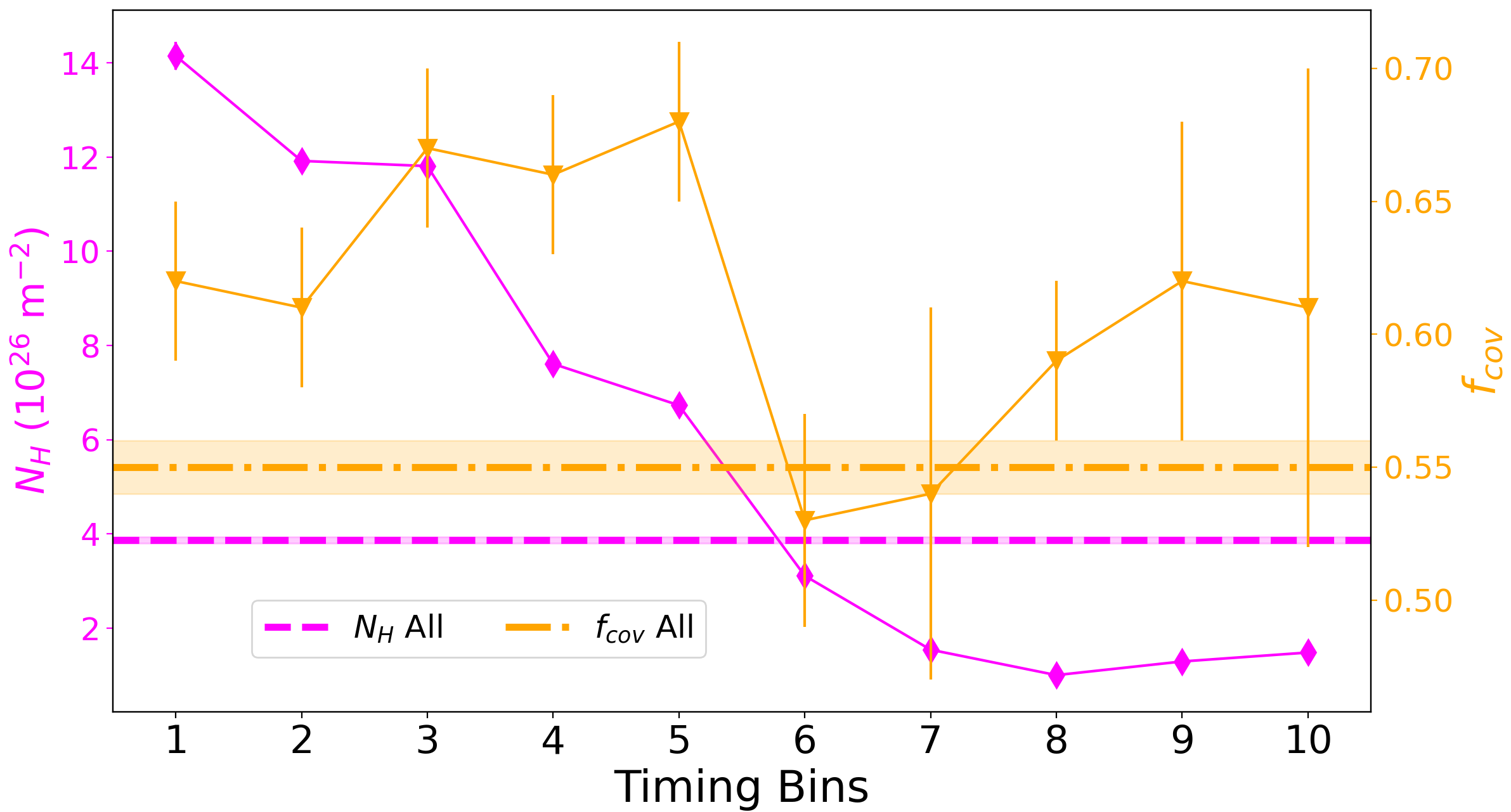}
		\end{subfigure}
		\caption{Further tests to determine the cause of the observed variability. \textit{Left:} Fixing the obscurer parameters to their time-averaged values in Obs 1 (Table \ref{Table:Full_Spec_Results}), and fitting $N_{\rm pow}$ (red circles) and $N_{\rm comt}$ (blue triangles) in each TB. \textit{Right:} Fixing the continuum parameters to their time-averaged values and fitting $N_{\rm H}$ (magenta diamonds) and $f_{\rm cov}$ (orange triangles). In both scenarios the changes in the parameters suggest that they could contribute to the observed variability. See Sect. \ref{Sec:Var_Cause} for details. The uncertainties on $N_{\rm pow}$ (left) and $N_{\rm H}$ (right) are relatively small that they are within the data points. The photon index and reflection scaling parameters from the continuum show the same trends as Fig. \ref{Fig:TB_Compare} when the obscurer parameters are fixed. The ionisation parameter of the obscurer does not change when the continuum parameters are fixed.}
		\label{Fig:Test_NH_Cont_Fixed}
	\end{figure*}

	To do this, we firstly fixed the obscurer parameters in all of the Obs 1 TBs to the time-averaged values of $N_H = 3.87 \times 10^{26}$ m\textsuperscript{-2}, $\log \xi = 0.44$, and $f_{\rm cov} = 0.55$ (see Table \ref{Table:Full_Spec_Results}), and fitted the normalisations of \texttt{POW} and \texttt{COMT}, in addition to $\Gamma_{\rm pow}$ and $s$. The results for $N_{\rm pow}$ and $N_{\rm comt}$ can be seen in the left side of Fig. \ref{Fig:Test_NH_Cont_Fixed}. We then fixed the continuum parameters to their time-averaged values in each TB, $N_{pow} = 2.60 \times 10^{50}$ ph s\textsuperscript{-1} keV\textsuperscript{-1}, $N_{comt} = 1.89 \times 10^{53}$ ph s\textsuperscript{-1} keV\textsuperscript{-1}, $\Gamma_{\rm pow} = 1.68$ and $s = 0.64$, and fitted $N_{\rm H}$, $\xi$ and $f_{\rm cov}$. The results can be seen in the right side of Fig. \ref{Fig:Test_NH_Cont_Fixed} for $N_{\rm H}$ and $f_{\rm cov}$.
	
	From the left side of Fig. \ref{Fig:Test_NH_Cont_Fixed}, with the obscurer parameters fixed, both $N_{\rm pow}$ and $N_{\rm comt}$ increase over the course of Obs 1. This strong change suggests that the most of the variability is driven by continuum changes. The photon index and reflection scaling parameters show the same trend as their respective panels in Fig. \ref{Fig:TB_Compare}. On the other hand, from the right side of Fig. \ref{Fig:Test_NH_Cont_Fixed}, when the continuum parameters are fixed, both $N_{\rm H}$ and $f_{\rm cov}$ respond to changes in the spectrum independently to the continuum, which could also suggest that the obscurer parameters drive the variability. Nevertheless, given the large error bars on $f_{\rm cov}$ (Fig. \ref{Fig:Test_NH_Cont_Fixed}) it seems that the covering fraction is consistent throughout Obs 1 and is therefore not changing, similar to Fig. \ref{Fig:TB_Compare}.
	
	Fig. \ref{Fig:Test_NH_Cont_Fixed}, therefore, is not very conclusive in determining the cause of the observed variability in Obs 1. There are two scenarios here: 1) the continuum varies and as a consequence, so does the obscurer; 2) the obscurer varies independently of the continuum changes. For scenario 1), the ionisation parameter $\xi$ does not appear to vary with the continuum in Obs 1 (Fig. \ref{Fig:TB_Compare}, which is supported further by the test in Fig. \ref{Fig:Test_NH_Cont_Fixed}). However, from the left side of Fig. \ref{Fig:Test_NH_Cont_Fixed}, it is clear that the continuum itself can drive the variability we observe, regardless of the obscurer. It is possible that $\xi$ does change, but given the errors in Fig. \ref{Fig:TB_Compare}, we cannot conclude one way or the other with full certainty; either possibility is plausible from these results. However, a likely explanation as to why the errors are so large is because we are fitting a multi-phased cloud, made up of different $N_{\rm H}$, $\xi$ and $n_{\rm e}$ regions, with a single model component. Alternatively for scenario 2), the obscurer varies independently of the continuum, which can be seen in the right side of Fig. \ref{Fig:Test_NH_Cont_Fixed}, where $N_{\rm H}$ and $f_{\rm cov}$ have to change in order to obtain a good fit when the continuum parameters are fixed. However, it is not clear what causes the changes in the column density of the obscurer over the course of Obs 1.
	
	One way in which the column density of the obscurer can change is by moving transversely across our LOS to the X-ray source. If the obscurer was clumpy, for example, then by crossing our LOS it would produce variations in $N_{\rm H}$. The other two possible scenarios are that the obscurer expands, causing the column density to decrease, for an increase in covering fraction (but this requires a large expansion velocity of order 1120 km s\textsuperscript{-1} - see Appendix \ref{Appen:Expansion} for the derivation), or alternatively, the material condenses to become visible dust (very unlikely as the ionising flux increases over the course of Obs 1). As the latter two explanations are less probable, we determine whether the obscurer could cross our LOS, and thus cause changes in $N_{\rm H}$. 
	
	The location of the obscurer in Obs 1 was estimated to be between 0.012 and 0.055 pc \citepalias{Mao_Prep}, assuming a spherical cloud geometry. Using $v_{\rm cross} = \sqrt{\frac{GM_{BH}}{r}}$, where $G$ is the gravitational constant, $M_{BH} = 5.96 \times 10^6$ M$_{\odot}$ is the black hole mass, and $r$ is the distance of the obscurer from the SMBH, the crossing velocity of the obscurer, assuming it follows a Keplerian orbit \citepalias{Mao_Prep}, is between 680 and 1470 km s\textsuperscript{-1}, depending on which distance is used. For this range in velocities, the obscurer would travel a distance of $6.80 \times 10^{10} - 1.51 \times 10^{11}$ m ($8 - 17$ R\textsubscript{g}) during Obs 1 ($\sim 103$ ks). Assuming that the radius of the X-ray corona is between 7 and 15 R\textsubscript{g} (e.g. \citealt{Chainakun2019}; a fiducial value of 10 R\textsubscript{g} was adopted in NGC 3783 by \citealt{Mehdipour2017}), then the diameter of the X-ray source would be within the range of $1.24 - 2.65 \times 10^{11}$ m. Here, there is clearly an overlap in the size of the corona and the crossing distance travelled by the obscurer during Obs 1. Therefore, changes in the obscurer column density for Obs 1 (Fig. \ref{Fig:TB_Compare}) could be explained with the obscurer moving transversely across our LOS towards the X-ray corona.
	
	In comparison, the length of time between observations was 1810 ks (three weeks), such that the distance travelled by the obscurer, assuming $v_{cross} = 680 - 1470$ km s\textsuperscript{-1}, is $1.23 - 2.65 \times 10^{12}$ m ($140 - 300$ R\textsubscript{g}). This is larger than the size of the ionising source, suggesting that the increase in $N_{\rm H}$ between Obs 1 and Obs 2 could be a result of the same obscurer moving transversely into our LOS for the two observations. This would explain the large drop in flux in the spectra and light curves between Obs 1 and Obs 2. However, the more likely argument to explain the difference in obscuration and continuum properties is that we are viewing two completely different components in the two observations. The main evidence is due to the ionisation parameter in Obs 2 being larger compared to Obs 1, even though the flux decreases greatly, as discussed in \citetalias{Mao_Prep}.
	
	Therefore, from these arguments and different tests (in Sects. \ref{Sec:Par_Comp} and \ref{Sec:Var_Cause}), we conclude that the observed variability in Obs 1 is likely to be driven by the continuum, reflecting changes in the intrinsic brightness of the central engine in NGC 3227. Some explanations for the continuum driving the observed changes could include: (1) changes in the accretion rate of NGC 3227; (2) fluctuations in the heating and cooling rates in the corona \citep[e.g.][studied how a warm corona responds to the accretion disk, in addition to the heating and cooling effects]{Ballantyne2020}; (3) variations in the corona size or location from the black hole \citep[this was an argument to explain the observed changes in NGC 4151;][]{Couto2016}. However, we cannot rule out scenario 2), whereby the possible changes could be caused by $N_{\rm H}$ as a result of the obscurer moving transversely over the X-ray source within our LOS during the observation time; the crossing distance is comparable to the X-ray corona size.
	
	One final possible solution to explain the trend between the continuum and obscurer column density could be established if the obscurer is a wind. In NGC 1365, \cite{Connolly2014} found that the absorbing column density decreased for an increase in luminosity, with no change in the ionisation state of the material; there is a similar trend here and can be seen in Figs. \ref{Fig:TB_Compare} and \ref{Fig:Parameter_Compare} for NGC 3227. In NGC 1365, this trend was interpreted through a disk wind model, where if the X-ray luminosity increased, then the launching radius of the wind became larger, causing the amount of material in the LOS to decrease \citep{Connolly2014}. Therefore, if we view the X-ray source through the inner edge of the obscurer within NGC 3227, and the continuum flux increases, then the obscurer would move outwards from within our LOS, causing the NH to decrease. A schematic can be seen in Figure 5 of \citep{Connolly2015}. However, this solution is speculative in NGC 3227, because there is no evidence that the obscurer is a wind due to the lack of UV absorption lines \citepalias{Mehdipour_Prep, Mao_Prep}, while in NGC 5548 \citep{Kaastra2014} and NGC 3783 \citep{Mehdipour2017, Kriss2019} the outflow velocities of the obscurers were measured with UV absorption features. In addition, we note that a leading model for X-ray continuum variability across a range of timescales is the model of propagating fluctuations in the local mass accretion rate \citep[e.g.][]{Lyubarskii1997, Arevalo2006, Ingram2013}, that could also explain the changes we observe in NGC 3227 during the obscuration event at the end of 2019.

	\subsection{Excess variance}
	\label{Sec:Fvar}
	Finally, we further explore the variability of each observation using the variance and fractional root mean square (rms) variability amplitude \citep[$F_{\rm var}$;][]{Vaughan2003, Ponti2004}, similar to \cite{Lobban2020} for NGC 3227 and \cite{Mehdipour55482016} for NGC 5548. We calculate $F_{\rm var}$ following the steps from \cite{Vaughan2003}, where the difference between the source variance ($S^2$) and the measured errors ($\overline{\sigma_{\rm err}^2}$) give rise to the excess variance $\sigma_{\rm XS}^2$. Here, $S^2 = \sum\limits_{i=1}^{N} \frac{(x_i - \overline{x})^2}{(N - 1)}$ and $\overline{\sigma_{\rm err}^2} = \frac{1}{N} \sum\limits_{i=1}^{N} \sigma_{\rm err, i}^2$, where $x_i$ is the measured value (flux), $\overline{x}$ is the mean, $N$ is the number of measurements, and $\sigma_{\rm err, i}^2$ is the uncertainty, or mean square error, on each measurement. The fractional rms variability amplitude is related to the excess variance via
	\begin{equation}
	F_{var} = \sqrt{\frac{\sigma_{\rm XS}^2}{\overline{x}^2}} = \sqrt{\frac{S^2 - \overline{\sigma_{\rm err}^2}}{\overline{x}^2}}.
	\label{Eq_Fvar}
	\end{equation}
	To compute the uncertainty on $F_{\rm var}$, we used Eq. B2 in Appendix B from \cite{Vaughan2003}. 
	
\begin{figure}
		\centering
		\includegraphics[width=1\linewidth]{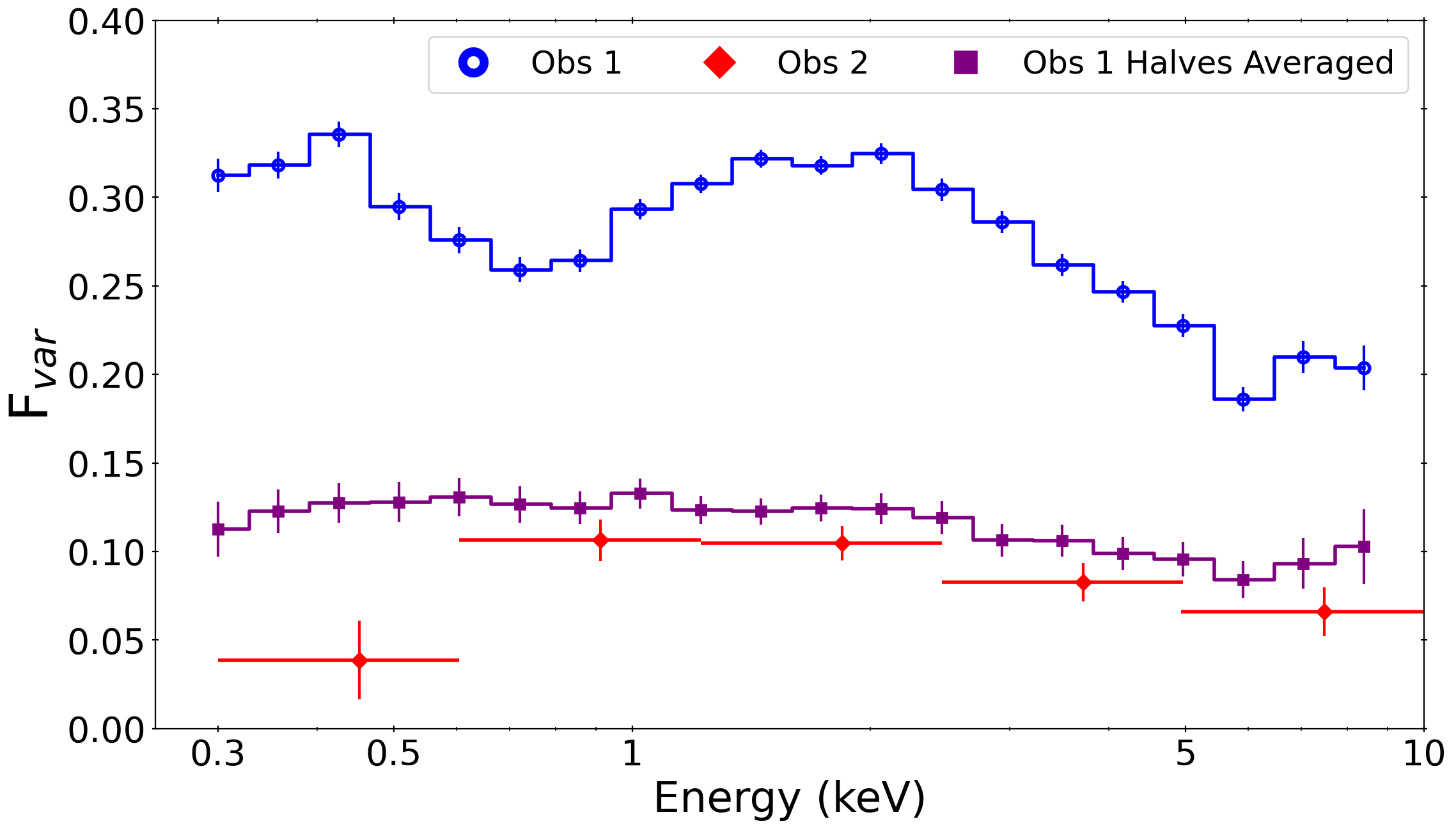}
		\includegraphics[width=1\linewidth]{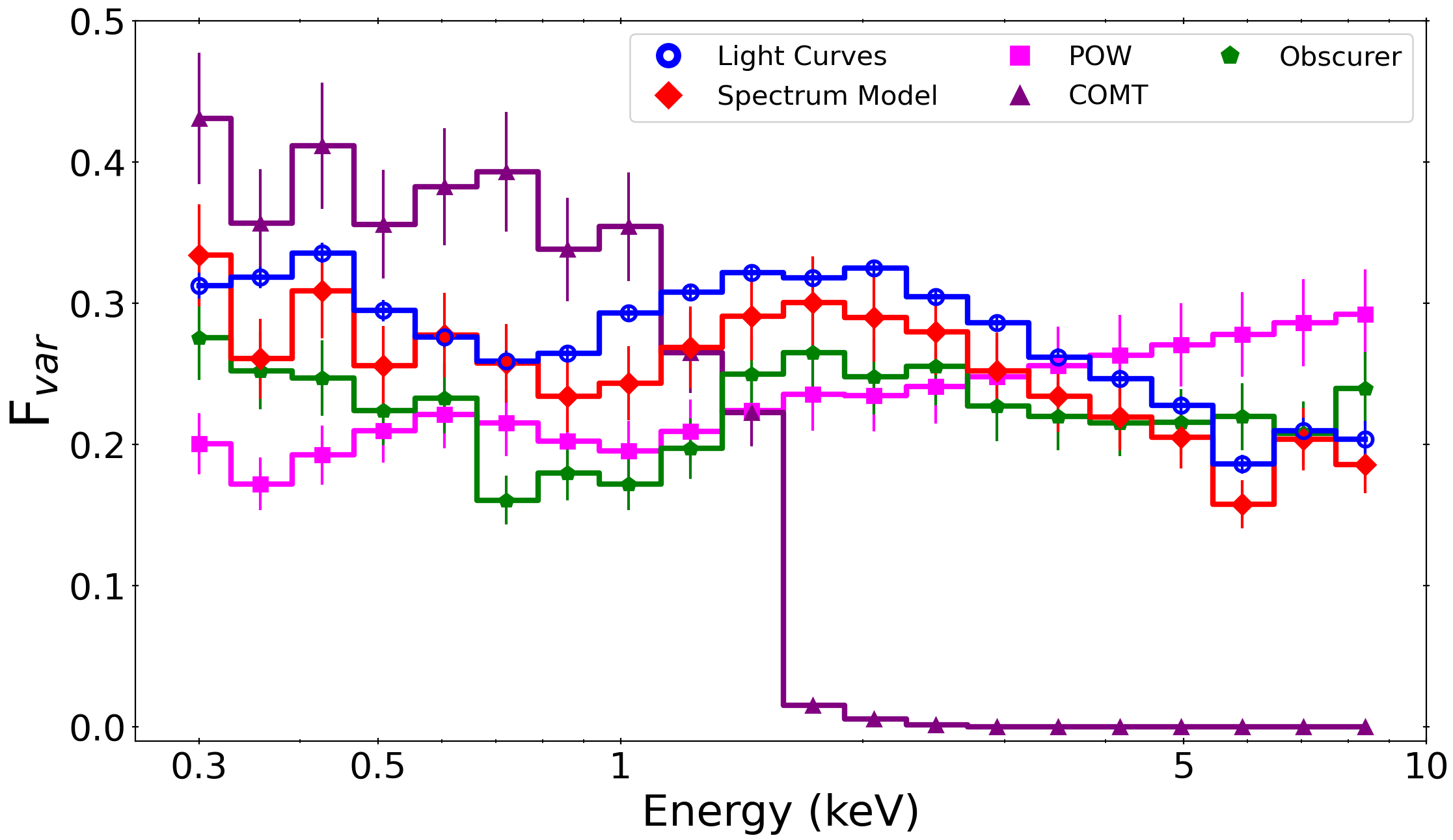}
		\includegraphics[width=1\linewidth]{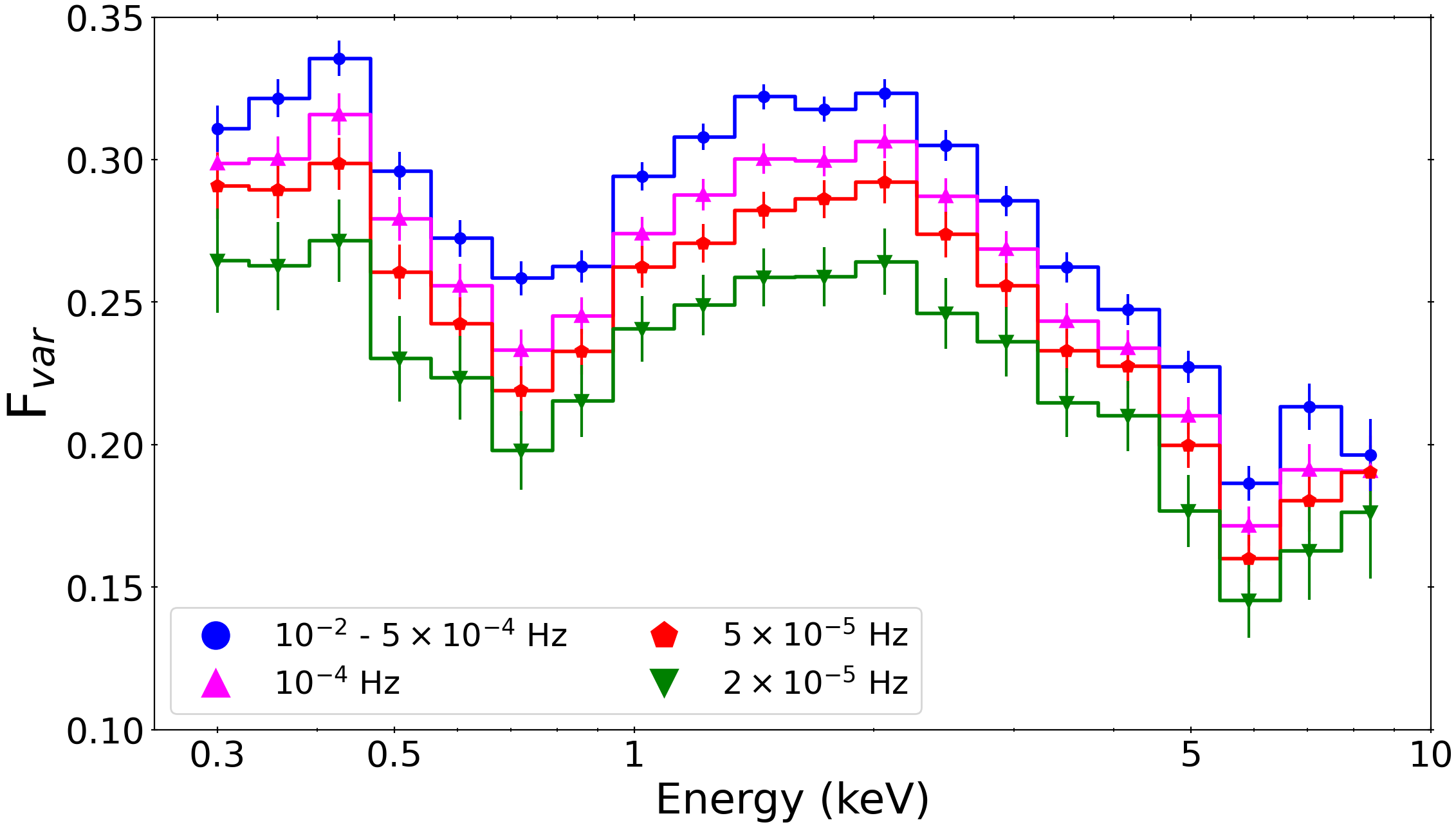}
		\caption{Fractional root mean square (rms) variability amplitude ($F_{\rm var}$) spectra as a function of energy. \textit{Top panel:} Obs 1 (blue circles) and Obs 2 (red diamonds), where the light curves were binned at 1000 s. For Obs 1, the energies are binned logarithmically such that the 20 data points are evenly spaced. For Obs 2, large energy bins are used to remove negative excess variance ($\sigma_{\rm XS}^2$) values caused by the significant drop in flux compared to Obs 1; the bin sizes are displayed by the \textit{x}-axis error bars. The purple $F_{\rm var}$ spectrum shows the results from averaging the two halves of Obs 1. See Sect. \ref{Sec:Fvar} for details.
			\textit{Middle panel:} Fractional rms variability amplitude spectra for the best fit spectral model (red line) and the three main model components: \texttt{POW} (magenta line), \texttt{COMT} (purple line), and the obscurer (\texttt{PION}; green line). The blue line shows the spectrum from the light curves (see top panel). From this plot, it appears that the \texttt{COMT} component drives the fractional rms variability below 1 keV, while the obscurer drives the $F_{\rm var}$ spectrum between 1 and 5 keV; above 5 keV \texttt{POW} dominates.
			\textit{Bottom panel:} Fractional rms variability amplitude spectra comparing different timing bins for the EPIC-PN light curves from Obs 1. The frequencies are $10^{-2}$ - $5 \times 10^{-4}$ Hz ($\Delta t = 100, 500, 1000$ and $2000$ s; blue), $10^{-4}$ Hz ($\Delta t = 10$ ks; magenta), $5 \times 10^{-5}$ Hz ($\Delta t = 20$ ks; red), and $2 \times 10^{-5}$ Hz ($\Delta t = 50$ ks; green). The shape of each curve stays consistent between frequency bands, suggesting that the model components vary for all light curve timing bins.}
		\label{Fig:Fvar_Plots}
	\end{figure}

	The top panel of Fig. \ref{Fig:Fvar_Plots} shows the fractional rms variance spectrum for the full duration of Obs 1 (103 ks) and Obs 2 (50 ks), calculated using Eq. \ref{Eq_Fvar}, with the EPIC-PN light curves binned at 1000 s for both observations. For Obs 1 (blue circles) the energies are binned logarithmically, such that the 20 data points are evenly spaced out. For Obs 2 (red diamonds in top panel of Fig. \ref{Fig:Fvar_Plots}), we have used large energy bins, such that there is sufficient S/N to prevent negative excess variance ($\sigma_{\rm XS}^2$) values resulting from the much weaker signal in Obs 2, compared to Obs 1. 
	
	The comparison between Obs 1 and Obs 2 does not take into account red-noise variability, as the duration of Obs 2 is half of Obs 1, and hence would infer lower variability for the Obs 2. Therefore, to test this, we split Obs 1 into two halves and found $F_{\rm var}$ in each energy bin (same 20 bins in Fig. \ref{Fig:Fvar_Plots}), averaging the values between the two halves. This can be seen by the purple $F_{\rm var}$ spectrum in the top panel of Fig. \ref{Fig:Fvar_Plots}. The result produces a fractional rms variability spectrum that is slightly above Obs 2 when judged on the same timescales (although the difference is less than implied by the original comparison), and the spectrum is more constant across the X-ray energy range compared to the blue line. 
	
	To determine what causes the features in the $F_{\rm var}$ curve of Obs 1 (blue line in the top panel of Fig. \ref{Fig:Fvar_Plots}), we obtained the $F_{\rm var}$ values for the best fit model and the \texttt{POW}, \texttt{COMT}, and obscurer components \citep[similar to][in their analysis of NGC 5548]{Mehdipour55482016, Cappi2016}. This is displayed in the middle panel of Fig. \ref{Fig:Fvar_Plots}. To obtain these $F_{\rm var}$ spectra, we binned each TB spectrum (Fig. \ref{Fig:Obs_Spec_Bins}) into 20 energy bins and obtained the $F_{\rm var}$ values for each energy bin from all ten TB model spectra. We did this for the best fit model fitted to each TB, which is the red line in Fig. \ref{Fig:Fvar_Plots}. The shape of the fractional rms variability amplitude spectrum for the model follows a similar shape as the $F_{\rm var}$ spectrum from the EPIC-PN light curves (blue line in the middle panel). The $F_{\rm var}$ spectra for the three main components in the best fit models are also displayed in the middle panel - \texttt{POW} (magenta line), \texttt{COMT} (purple line), and the obscurer \texttt{PION} (green line). In order to obtain the $F_{\rm var}$ values for the obscurer, we took the difference between the absorbed best fit model (red line in Fig. \ref{Fig:Fvar_Plots}) and the model when $N_{\rm H} = 0$ m\textsuperscript{-2}. The uncertainties on the measured values for the TBs ($\sigma_{\rm err, i}^2$) are the standard error on the mean for each energy bin \citep[similar to][]{Lobban2020}.
	
	From plotting the $F_{\rm var}$ spectra of the model components, we can see that the \texttt{COMT} component accounts for the variability at energies below 1 keV, while the obscurer causes changes between 1 and 5 keV; \texttt{POW} dominates above 5 keV. This could give additional evidence that the obscurer contributes significantly to the observed variability, and therefore this scenario should not be ruled out compared to the continuum, as discussed in Sect. \ref{Sec:Var_Cause}. Moreover, Fig. \ref{Fig:Fvar_Plots} further confirms the results from the C-statistic summation test in Sect. \ref{Sec:Par_Comp}, which shows that $N_{\rm comt}$ and $N_{\rm H}$ are statistically significant parameters in the best fit model. In the middle panel of Fig. \ref{Fig:Fvar_Plots}, the curves for the \texttt{COMT} component (purple line) and the obscurer (green line) follow the best fit model $F_{\rm var}$ spectrum (red line), suggesting that they likely contribute significantly to the fractional rms variability values within the specific energy ranges. Finally, there is a drop in $F_{\rm var}$ at 6 keV which likely corresponds to the Fe K emission energy band, from material far from the SMBH, and therefore varies less compared to the underlying continuum. This was also found in the 2016 observations by \cite{Lobban2020}.

	In addition to obtaining the fractional rms variability amplitude from the spectral model components (middle panel of Fig. \ref{Fig:Fvar_Plots}), we also explored how different timing bins for the EPIC-PN light curves affected the fractional rms variability amplitudes in Obs 1 \citep[similar to, e.g.][for NGC 3227]{Arevalo2014, Lobban2020}. For each energy bin of the EPIC-PN light curves from Obs 1 we constructed $F_{\rm var}$ spectra for seven different timing bin lengths, $\Delta t =$ 0.1, 0.5, 1, 2, 10, 20, and 50 ks. For the four smallest timing bins, we found that the $F_{\rm var}$ values are similar and therefore average these values together. We convert these timing bins to frequency bands of $10^{-2} - 5 \times 10^{-4}$ Hz for the averaged values, and $10^{-4}$, $5 \times 10^{-5}$, and $2 \times 10^{-5}$, for 10, 20 and 50 ks, respectively \citep{Arevalo2014, Lobban2020}. The result is displayed in the bottom panel of Fig. \ref{Fig:Fvar_Plots}. For all frequencies, the energy bins correlate with each other. In other words, if there is an increase in $F_{\rm var}$ for a particular energy range, such as between 1 and 5 keV (caused by the obscurer; middle panel of Fig. \ref{Fig:Fvar_Plots}), all frequency fractional variability amplitudes follow this increase. This result suggests that all model components that vary, vary on all timescales \citep{Arevalo2014}. As the frequency increases, the shape of the curve stays the same but the whole curve rises in $F_{\rm var}$, as shown in the bottom panel of Fig. \ref{Fig:Fvar_Plots}. This result is different from the analysis by \cite{Lobban2020} on NGC 3227 in 2016, whereby they found that as the frequency increased, the steepness of the curve decreased, until the highest frequency band displayed little change in $F_{\rm var}$ over the full energy range (top panel in their Fig. 7).

	\section{Summary and conclusions}
	\label{Sec:Conclusions}
	
	NGC 3227 has undergone many occultation events \citep{Lamer2003, Markowitz2014, Beuchert2015, Turner2018} lasting a variety of time scales. At the end of 2019, we observed this AGN to undergo obscuration again, detected by our \textit{Swift} monitoring programme, where follow-up observations from \textit{XMM-Newton}, \textit{NuSTAR} and HST were obtained. In \citetalias{Mehdipour_Prep}, we studied the unobscured and obscured SEDs from 2016 and 2019, respectively, in order to probe the properties and origin of the obscuring material in the sequential papers in the series. \citetalias{Wang_Prep} analysed the WAs and past occultation periods using the broadband continuum model, before the 2019 observations were studied in great detail in \citetalias{Mao_Prep}.
	
	In this paper, using the initial model parameter values from \citetalias{Mao_Prep}, we investigated the possible origins of variability during the two observations of NGC 3227 taken at the end of 2019. In Obs 1, we found an anti-correlation between $N_{\rm H}$ and the continuum normalisations, $N_{\rm pow}$ and $N_{\rm comt}$ (Figs. \ref{Fig:TB_Compare} and \ref{Fig:Parameter_Compare}). We conclude that the observed variation in Obs 1 is likely to be driven by the continuum. However, we cannot rule out the possible changes could be caused by $N_{\rm H}$ if the obscurer moved transversely over the X-ray source within our LOS during the observation time.
	
	For Obs 2, the count rate was significantly lower compared to Obs 1, due to the combination of a high column density and covering fraction ($f_{\rm cov}$) in the obscurer and strong flux decrease in the continuum. The lack of strong variability suggests that the continuum driving mechanism is not as significant compared to Obs 1. Between Obs 1 and Obs 2, the time-averaged results for $f_{\rm cov}$ and $N_{\rm H}$ increased from 0.55 to 0.66, and from $3.87 \times 10^{26}$ m\textsuperscript{-2} to $9.93 \times 10^{26}$ m\textsuperscript{-2}, respectively. Comparatively, $N_{\rm pow}$ and $N_{\rm comp}$ decreased from $2.60$ to $0.73 \times 10^{50}$ ph s\textsuperscript{-1} keV\textsuperscript{-1}at 1 keV, and $1.89$ to $0.16 \times 10^{53}$ ph s\textsuperscript{-1} keV\textsuperscript{-1}, respectively.
	
	In conclusion, we argue that the continuum is likely to drive the strong variability seen in NGC 3227 during Obs 1, but we cannot rule out the possible effects from $N_{\rm H}$ of the obscurer. There is no apparent change in the ionisation parameter of the obscurer in Obs 1, but given the large uncertainties, we cannot confirm this with certainty. A likely explanation here is that the obscurer is multi-phased, which we fit with only one component. Obs 2, on the other had, shows little change over the course of the observation, and the large increase in $N_{\rm H}$ and $f_{\rm cov}$, causing the significant drop in X-ray flux, is a result of a different component in our LOS compared to the one in Obs 1.

	\begin{acknowledgements}
		This work is based on observations obtained with \textit{XMM-Newton}, an ESA science mission with instruments and contributions directly funded by ESA Member States and NASA. We would like to thank the referee for their constructive and helpful feedback, and thank you to C. Done for suggesting a relevant paper to help explain the observed relationships between obscurer and continuum. SGW acknowledges the support of a PhD studentship awarded by the UK Science \& Technology Facilities Council (STFC). GP acknowledges funding from the European Research Council (ERC) under the European Union’s Horizon 2020 research and innovation programme. BDM acknowledges support from Ramón y Cajal Fellowship RYC2018-025950-I (grant agreement No 865637). SB acknowledges financial support from ASI under grants ASIINAFI/037/12/0 and n. 2017-14-H.O and from PRIN MIUR project `Black Hole winds and the Baryon Life Cycle of Galaxies: the stone-guest at the galaxy evolution supper', contract no. 2017PH3WAT. POP acknowledges financial support from the French High Energy National Programme (PNHE) and from the French space agency (CNES). DJW acknowledges support from STFC in the form of an Ernest Rutherford Fellowship (ST/N004027/1).
	\end{acknowledgements}


\begin{thebibliography}{}
		
	\bibitem[Arévalo et al.(2014)]{Arevalo2014} Arévalo, P. \& Markowitz, A. 2014, ApJ, 783, 82
	\bibitem[Arévalo et al.(2006)]{Arevalo2006} Arévalo, P. \& Uttley, P. 2006, MNRAS, 367, 801
	\bibitem[Ballantyne \& Xiang(2020)]{Ballantyne2020} Ballantyne, D. R. \& Xiang, X. 2020, MNRAS, 496, 4255
	\bibitem[Beuchert et al.(2017)]{Beuchert2017} Beuchert, T., Markowitz, A. G., Dauser, T., et al. 2017, A\&A, 603, A50
	\bibitem[Beuchert et al.(2015)]{Beuchert2015} Beuchert, T., Markowitz, A. G., Krauß, F., et al. 2015, A\&A, 584, A82
	\bibitem[Cappi et al.(2016)]{Cappi2016} Cappi, M., De Marco, B., Ponti, G., et al. 2016, A\&A, 592, A27
	\bibitem[Cappi et al.(2006)]{Cappi2006} Cappi, M., Panessa, F., Bassani, L., et al. 2006, A\&A, 446, 459
	\bibitem[Cash (1979)]{Cash1979} Cash, W. 1979, ApJ, 228, 939
	\bibitem[Chainakun et al.(2019)]{Chainakun2019} Chainakun, P.,Watcharangkool, A., Young, A. J., \& Hancock, S. 2019, MNRAS, 487, 667
	\bibitem[Connolly et al.(2014)]{Connolly2014} Connolly, S. D., McHardy, I. M., \& Dwelly, T. 2014, MNRAS, 440, 3503
	\bibitem[Connolly et al.(2015)]{Connolly2015} Connolly, S. D., McHardy, I. M., \& Dwelly, T. 2015, arXiv e-prints, arXiv:1502.07501
	\bibitem[Couto et al.(2016)]{Couto2016} Couto, J. D., Kraemer, S. B., Turner, T. J., \& Crenshaw, D. M. 2016, ApJ, 833, 191
	\bibitem[Crenshaw et al.(2001)]{Crenshaw2001} Crenshaw, D. M., Kraemer, S. B., Bruhweiler, F. C., \& Ruiz, J. R. 2001, ApJ, 555, 633
	\bibitem[de Vaucouleurs et al.(1991)]{deVaucouleurs1991} de Vaucouleurs, G., de Vaucouleurs, A., Corwin, Herold G., J., et al. 1991, Third Reference Catalogue of Bright Galaxies
	\bibitem[den Herder et al.(2001)]{denHerder2001} den Herder, J. W., Brinkman, A. C., Kahn, S. M., et al. 2001, A\&A, 365, L7
	\bibitem[Detmers et al.(2010)]{Detmers2010} Detmers, R. G., Kaastra, J. S., Costantini, E., et al. 2010, A\&A, 516, A61
	\bibitem[Ebrero et al.(2016)]{Ebrero2016} Ebrero, J., Kriss, G. A., Kaastra, J. S., \& Ely, J. C. 2016, A\&A, 586, A72
	\bibitem[Gandhi et al.(2015)]{Gandhi2015} Gandhi, P., Hönig, S. F., \& Kishimoto, M. 2015, ApJ, 812, 113
	\bibitem[Gehrels et al.(2004)]{Gehrels2004} Gehrels, N., Chincarini, G., Giommi, P., et al. 2004, ApJ, 611, 1005
	\bibitem[George et al.(1998)]{George1998} George, I. M., Mushotzky, R., Turner, T. J., et al. 1998, ApJ, 509, 146
	\bibitem[Gondoin et al.(2003)]{Gondoin2003} Gondoin, P., Orr, A., Lumb, D., \& Siddiqui, H. 2003, A\&A, 397, 883
	\bibitem[Grafton-Waters et al.(2021)]{Grafton-Waters2021} Grafton-Waters, S., Branduardi-Raymont, G., Mehdipour, M., et al. 2021, A\&A, 649, A162
	\bibitem[Grafton-Waters et al.(2020)]{Grafton-Waters2020} Grafton-Waters, S., Branduardi-Raymont, G., Mehdipour, M., et al. 2020, A\&A, 633, A62
	\bibitem[Green et al.(2012)]{Green2012} Green, J. C., Froning, C. S., Osterman, S., et al. 2012, ApJ, 744, 60
	\bibitem[Harrison et al.(2013)]{Harrison2013} Harrison, F. A., Craig, W. W., Christensen, F. E., et al. 2013, ApJ, 770, 103
	\bibitem[Ingram \& van der Klis(2013)]{Ingram2013} Ingram, A. \& van der Klis, M. 2013, MNRAS, 434, 1476
	\bibitem[Jansen et al.(2001)]{Jansen2001} Jansen, F., Lumb, D., Altieri, B., et al. 2001, A\&A, 365, L1
	\bibitem[Kaastra (2017)]{Kaastra2017} Kaastra, J. S. 2017, A\&A, 605, A51
	\bibitem[Kaastra \& Bleeker(2016)]{KB2016} Kaastra, J. S. \& Bleeker, J. A. M. 2016, A\&A, 587, A151
	\bibitem[Kaastra et al.(2014)]{Kaastra2014} Kaastra, J. S., Kriss, G. A., Cappi, M., et al. 2014, Science, 345, 64
	\bibitem[Kaastra et al.(2018)]{Kaastra2018} Kaastra, J. S., Mehdipour, M., Behar, E., et al. 2018, A\&A, 619, A112
	\bibitem[Kaastra et al.(1996)]{Kaastra1996} Kaastra, J. S., Mewe, R., \& Nieuwenhuijzen, H. 1996, in UV and X-ray Spectroscopy of Astrophysical and Laboratory Plasmas, 411–414
	\bibitem[Kaastra et al.(2018)]{SPEX305} Kaastra, J. S., Raassen, A. J. J., de Plaa, J., \& Gu, L. 2018, SPEX X-ray spectral fitting package
	\bibitem[Kaastra et al.(2002)]{Kaastra2002} Kaastra, J. S., Steenbrugge, K. C., Raassen, A. J. J., et al. 2002, A\&A, 386, 427
	\bibitem[Keck et al.(2015)]{Keck2015} Keck, M. L., Brenneman, L. W., Ballantyne, D. R., et al. 2015, ApJ, 806, 149
	\bibitem[Kinkhabwala et al.(2002)]{Kinkhabwala2002} Kinkhabwala, A., Sako, M., Behar, E., et al. 2002, ApJ, 575, 732
	\bibitem[Kriss et al.(2019)]{Kriss2019} Kriss, G. A., Mehdipour, M., Kaastra, J. S., et al. 2019, A\&A, 621, A12
	\bibitem[Lamer et al.(2003)]{Lamer2003} Lamer, G., Uttley, P., \& McHardy, I. M. 2003, MNRAS, 342, L41
	\bibitem[Lobban et al.(2020)]{Lobban2020} Lobban, A. P., Turner, T. J., Reeves, J. N., Braito, V., \& Miller, L. 2020, MNRAS, 494, 5056
	\bibitem[Longinotti et al.(2019)]{Longinotti2019} Longinotti, A. L., Kriss, G., Krongold, Y., et al. 2019, ApJ, 875, 150
	\bibitem[Longinotti et al.(2013)]{Longinotti2013} Longinotti, A. L., Krongold, Y., Kriss, G. A., et al. 2013, ApJ, 766, 104
	\bibitem[Lyubarskii (1997)]{Lyubarskii1997} Lyubarskii, Y. E. 1997, MNRAS, 292, 679
	\bibitem[Magdziarz \& Zdziarski(1995)]{Magdziarz1995} Magdziarz, P. \& Zdziarski, A. A. 1995, MNRAS, 273, 837
	\bibitem[Mao et al.(2022)]{Mao_Prep} Mao, J., Kaastra, J. S., Mehdipour, M., et al. 2022, A\&A, 665, A72
	\bibitem[Markowitz et al.(2009)]{Markowitz2009} Markowitz, A., Reeves, J. N., George, I. M., et al. 2009, ApJ, 691, 922
	\bibitem[Markowitz et al.(2014)]{Markowitz2014} Markowitz, A. G., Krumpe, M., \& Nikutta, R. 2014, MNRAS, 439, 1403
	\bibitem[Mehdipour et al.(2011)]{Mehdipour2011} Mehdipour, M., Branduardi-Raymont, G., Kaastra, J. S., et al. 2011, A\&A, 534, A39
	\bibitem[Mehdipour et al.(2018)]{Mehdipour2018} Mehdipour, M., Kaastra, J. S., Costantini, E., et al. 2018, A\&A, 615, A72
	\bibitem[Mehdipour et al.(2016a)]{Mehdipour2016} Mehdipour, M., Kaastra, J. S., \& Kallman, T. 2016a, A\&A, 596, A65
	\bibitem[Mehdipour et al.(2017)]{Mehdipour2017} Mehdipour, M., Kaastra, J. S., Kriss, G. A., et al. 2017, A\&A, 607, A28
	\bibitem[Mehdipour et al.(2016b)]{Mehdipour55482016} Mehdipour, M., Kaastra, J. S., Kriss, G. A., et al. 2016b, A\&A, 588, A139
	\bibitem[Mehdipour et al.(2022)]{Mehidpour2022} Mehdipour, M., Kriss, G. A., Costantini, E., et al. 2022, ApJ, 934, L24
	\bibitem[Mehdipour et al.(2021)]{Mehdipour_Prep} Mehdipour, M., Kriss, G. A., Kaastra, J. S., et al. 2021, A\&A, 652, A150
	\bibitem[Murphy et al.(1996)]{Murphy1996} Murphy, E. M., Lockman, F. J., Laor, A., \& Elvis, M. 1996, ApJS, 105, 369
	\bibitem[Nardini et al.(2015)]{Nardini2015} Nardini, E., Reeves, J. N., Gofford, J., et al. 2015, Science, 347, 860
	\bibitem[Parker et al.(2019)]{Parker2019} Parker, M. L., Longinotti, A. L., Schartel, N., et al. 2019, MNRAS, 490, 683
	\bibitem[Petrucci et al.(2013)]{Petrucci2013} Petrucci, P. O., Paltani, S., Malzac, J., et al. 2013, A\&A, 549, A73
	\bibitem[Ponti et al.(2004)]{Ponti2004} Ponti, G., Cappi, M., Dadina, M., \& Malaguti, G. 2004, A\&A, 417, 451
	\bibitem[Porquet et al.(2004)]{Porquet2004} Porquet, D., Kaastra, J. S., Page, K. L., et al. 2004, A\&A, 413, 913
	\bibitem[Risaliti et al.(2005)]{Risaliti2005} Risaliti, G., Bianchi, S., Matt, G., et al. 2005, ApJ, 630, L129
	\bibitem[Rivers et al.(2011)]{Rivers2011} Rivers, E., Markowitz, A., \& Rothschild, R. 2011, ApJS, 193, 3
	\bibitem[Stru\"{u}der et al.(2001)]{Struder2001} Str\"{u}der, L., Briel, U., Dennerl, K., et al. 2001, A\&A, 365, L18
	\bibitem[Titarchuk (1994)]{Titarchuk1994} Titarchuk, L. 1994, ApJ, 434, 570
	\bibitem[Tombesi et al.(2013)]{Tombesi2013} Tombesi, F., Cappi, M., Reeves, J. N., et al. 2013, MNRAS, 430, 1102
	\bibitem[Turner et al.(2018)]{Turner2018} Turner, T. J., Reeves, J. N., Braito, V., et al. 2018, MNRAS, 481, 2470
	\bibitem[Uttley \& McHardy(2005)]{Uttley2005} Uttley, P. \& McHardy, I. M. 2005, MNRAS, 363, 586
	\bibitem[Vaughan et al.(2003)]{Vaughan2003} Vaughan, S., Edelson, R.,Warwick, R. S., \& Uttley, P. 2003, MNRAS, 345, 1271
	\bibitem[Walton et al.(2021)]{Walton2021} Walton, D. J., Balokovi\'{c}, M., Fabian, A. C., et al. 2021, MNRAS, 506, 1557
	\bibitem[Wang et al.(2022)]{Wang_Prep} Wang, Y., Kaastra, J., Mehdipour, M., et al. 2022, A\&A, 657, A77
	
	\end{thebibliography}

	\newpage	
	
	\appendix
	\section{Alternative models}
	\label{Sec:Alt_Mod}
	
	\begin{figure*}
		\centering
		\begin{subfigure}{0.5\linewidth}
			\includegraphics[width=1\linewidth]{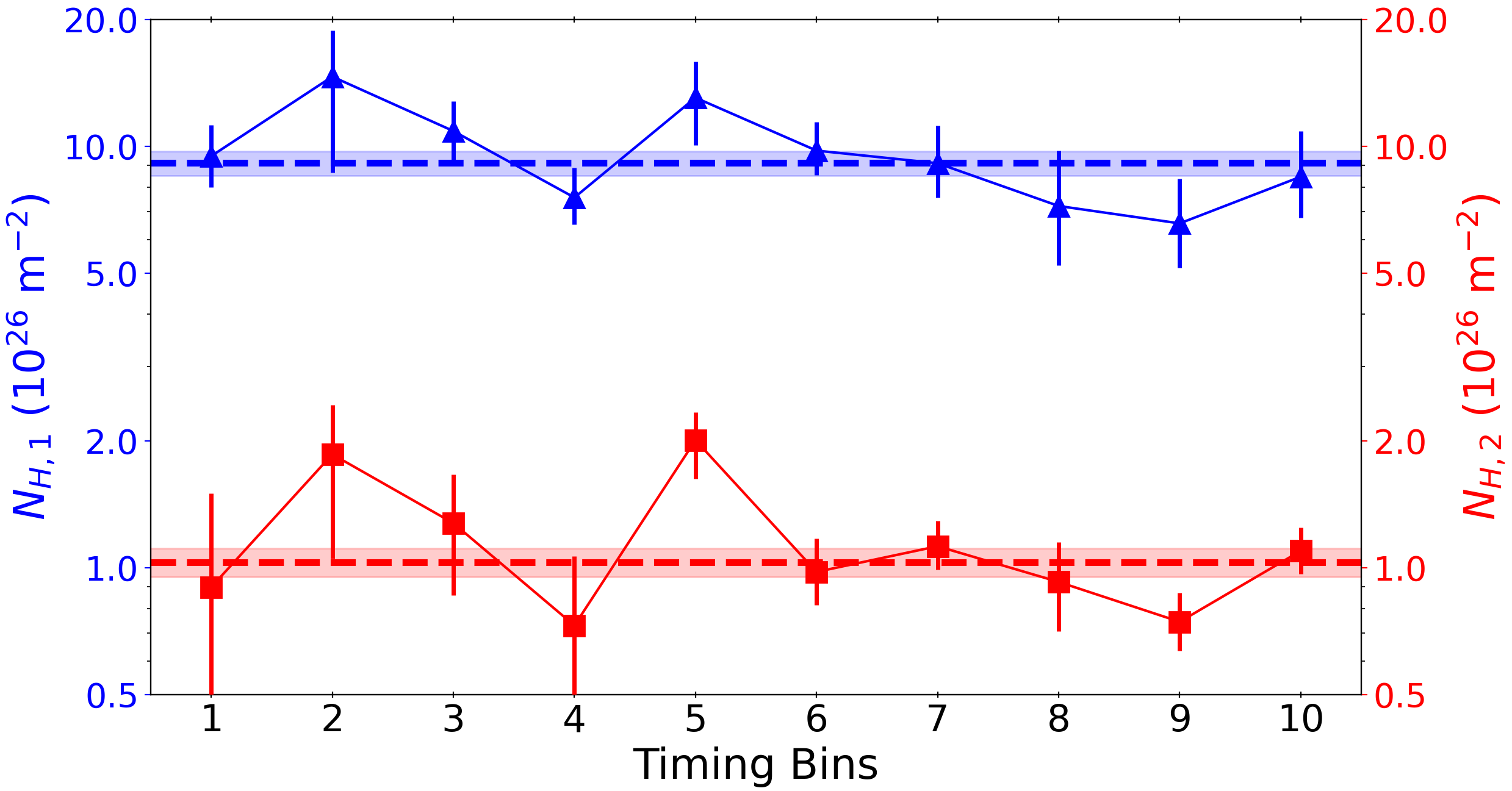}
		\end{subfigure}%
		\begin{subfigure}{0.5\linewidth}
			\includegraphics[width=1\linewidth]{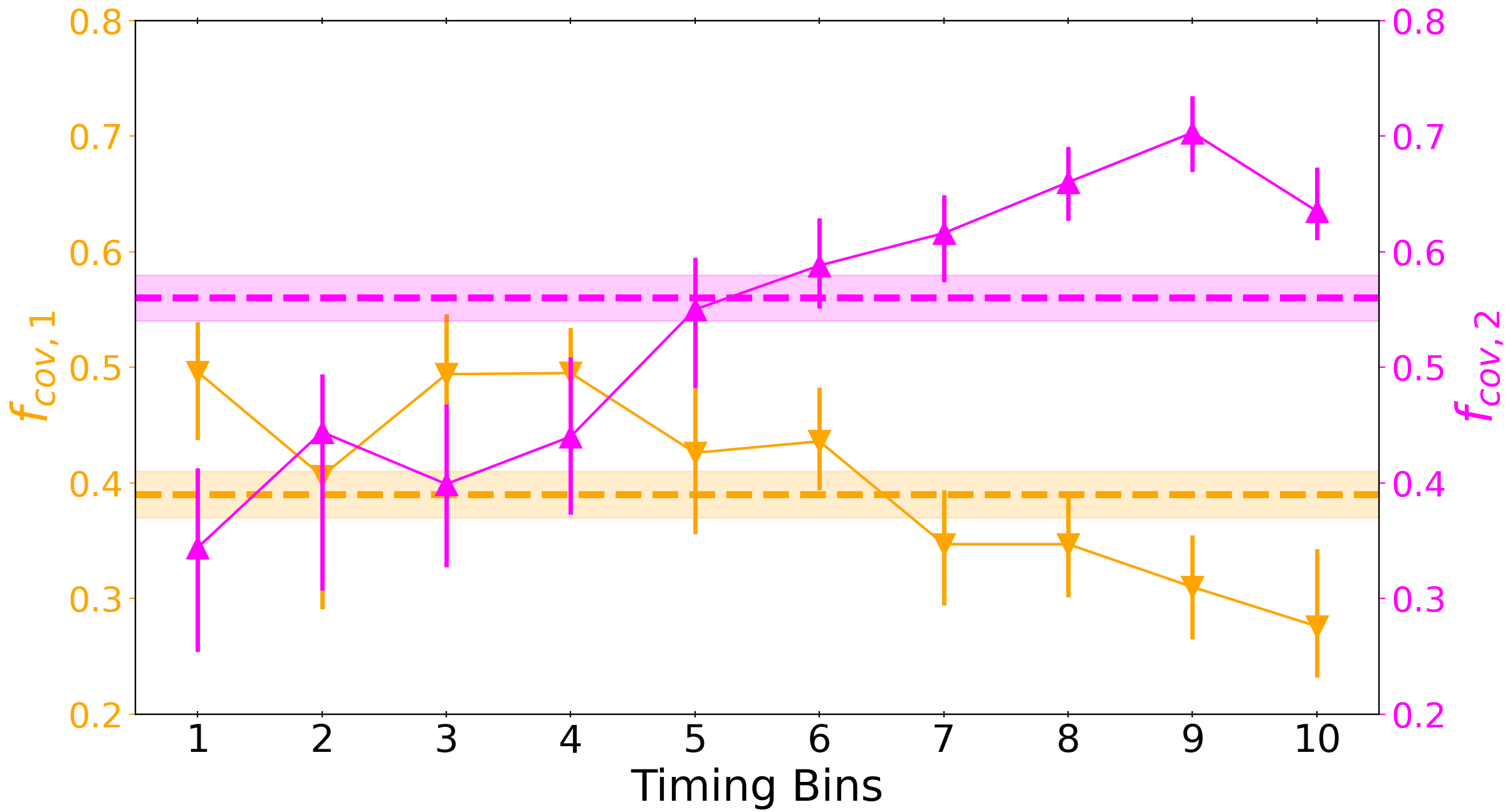}
		\end{subfigure}
		\caption{Comparing the parameters of the two obscurer model; details are discussed in Appendix \ref{Sec:Alt_Mod}. \textit{Left:} Column densities of the two obscurer components, blue for obscurer component 1 ($N_{H,1}$) and red for component 2 ($N_{H,2}$). \textit{Right:} Covering fractions of the two obscurer components, orange for obscurer 1 ($f_{cov,1}$) and magenta for obscurer 2 ($f_{cov,2}$). The dashed lines in each panel show the time-averaged best fit parameter values, with their uncertainties displayed as the shaded areas.}
		\label{Fig:Two_Obscuer_Pars}
	\end{figure*}

	In \citetalias{Mao_Prep}, we examined three different models applied to the November 2019 observation, and four models to the December 2019 observation, of NGC 3227 (see Tables 2 and 3, and Sect. 3, in \citetalias{Mao_Prep} for details). In the modelling carried out in this paper (Sects. \ref{Sec:Spectral_Fitting} and \ref{Sec:Results_Discussion}), we assumed that the neutral reflector was located close enough to the SMBH such that changes in the reflected continuum in 2019 were caused by changes in the intrinsic power law \citep{Gandhi2015}. However, in addition to the modelling in Sect. \ref{Sec:Spectral_Fitting}, we also fitted the M2 model from Table 2 in \citetalias{Mao_Prep}, coupling the \texttt{REFL} component normalisation and photon index in 2019 to the 2016 power law values in both observations. This model produced a better, albeit not significantly different, C-statistic compared to M1, but the parameter values were consistent between the models; a similar result was found in \citetalias{Mao_Prep}. As a result of this fitting, we found that these changes on the \texttt{REFL} component do not affect the properties of the obscurer. Therefore, no actions to update the current model were taken, and the results from Sects. \ref{Sec:Spectral_Fitting} and \ref{Sec:Results_Discussion} explain the properties of the obscurer and variations in the observed continuum for Obs 1 and Obs 2.
	
	The third model applied to the 2019 observations in \citetalias{Mao_Prep} (M3 in Table 2 and M4 in Table 3) was the two obscurer model, fitted here with two \texttt{PION} components. Similarly to the modelling in Sect. \ref{Sec:Spectral_Fitting} of this paper, the column densities, ionisation parameters and covering fractions of the obscurer components were fitted, along with the continuum parameters. After fitting the time-averaged spectra with the two obscurer model, we found large differences with the parameters in the current model from Fig. \ref{Fig:TB_Compare} and Table \ref{Table:Full_Spec_Results}.
	
	As a result of this, therefore, we fitted the two obscurer time-averaged best fit models to each TB spectrum in Obs 1 and Obs 2. For Obs 2, as expected, there is little change in the parameters between TBs and they are consistent with the time-averaged values, similar to Fig. \ref{Fig:TB_Compare}. For Obs 1, on the other hand, there are changes in parameter values between TBs, such as $N_{\rm pow}$ and $N_{\rm comt}$, which increase during the observation, similar to the model in Sect. \ref{Sec:Results_Discussion} and Fig. \ref{Fig:TB_Compare}. However, the difference between the one and two obscurer models is clearer when comparing the obscurer parameters. For both obscurers in the two obscurer model, the column densities stay fairly consistent between TBs, and compared to the time-average values, as displayed in the left side of Fig. \ref{Fig:Two_Obscuer_Pars}. 
	However, it is the covering fractions that appear to change in Obs 1: for obscurer 1, $f_{cov, 1}$ decreases over the duration of the observation, while $f_{cov, 2}$ increases. This comparison is displayed in the right panel of Fig. \ref{Fig:Two_Obscuer_Pars}. These results suggest that the covering fraction could be related to the continuum changes, rather than the column density; the opposite of what we initially discussed in Sect. \ref{Sec:Var_Cause}. On the other hand, when the TB errors are taken into account, the covering fraction values overlap, suggesting that the covering fraction correlations are not significant throughout Obs 1. Similarly with the model in Sects. \ref{Sec:Spectral_Fitting} and \ref{Sec:Results_Discussion}, the ionisation parameters of both obscurer components do not show a clear trend with over the Obs 1 and were difficult to constrain. As a result, we do no show this plot. 
	
	Further investigations were then implemented, similar to Fig. \ref{Fig:Test_NH_Cont_Fixed} in Sect. \ref{Sec:Var_Cause} - we fixed either the obscurer or continuum parameters to their time-averaged values and fitted the other. From this test, we find that both the column densities and covering fractions in both obscurers decrease, however the first obscurer component changes are more significant. These parameters show an anti-correlation with the continuum parameters $N_{\rm pow}$ and $N_{\rm comt}$, similar to the result found in Figs. \ref{Fig:TB_Compare} and \ref{Fig:Test_NH_Cont_Fixed}; albeit this time the covering fractions also vary. This test implies that both obscurer parameters can vary in accordance with the continuum, and are therefore valid to explain the change in the obscurer properties. Therefore, from this argument, we conclude that the observed variability in Obs 1 is likely caused by the intrinsic properties of the continuum, but changes caused by the obscurer cannot be ruled out as we see $N_{\rm H}$ and $f_{\rm cov}$ vary independently of the continuum. This is the same conclusion we found for the one obscurer component model in Sect. \ref{Sec:Conclusions}.
	
	In summary, the model fitted to the EPIC-PN spectra in Obs 1 and Obs 2 (Sects. \ref{Sec:Spectral_Fitting} and \ref{Sec:Results_Discussion}) explains the properties of the continuum and obscurer of NGC 3227 in 2019. There is no significant consequence on the obscurer parameters when coupling either the 2016 or 2019 power law to the reflection component, and similar trends are seen between the one and two obscurer models. Therefore, we argue that the model in this paper is the correct one, compared to the alternative models discussed in this Appendix.
	
	\section{\texttt{KNAK} corrections}
	\label{Appen:KNAK_Corr}
	
\begin{figure*}[!tbp]
		\centering
		\begin{subfigure}{0.5\linewidth}
			\includegraphics[width=1\linewidth]{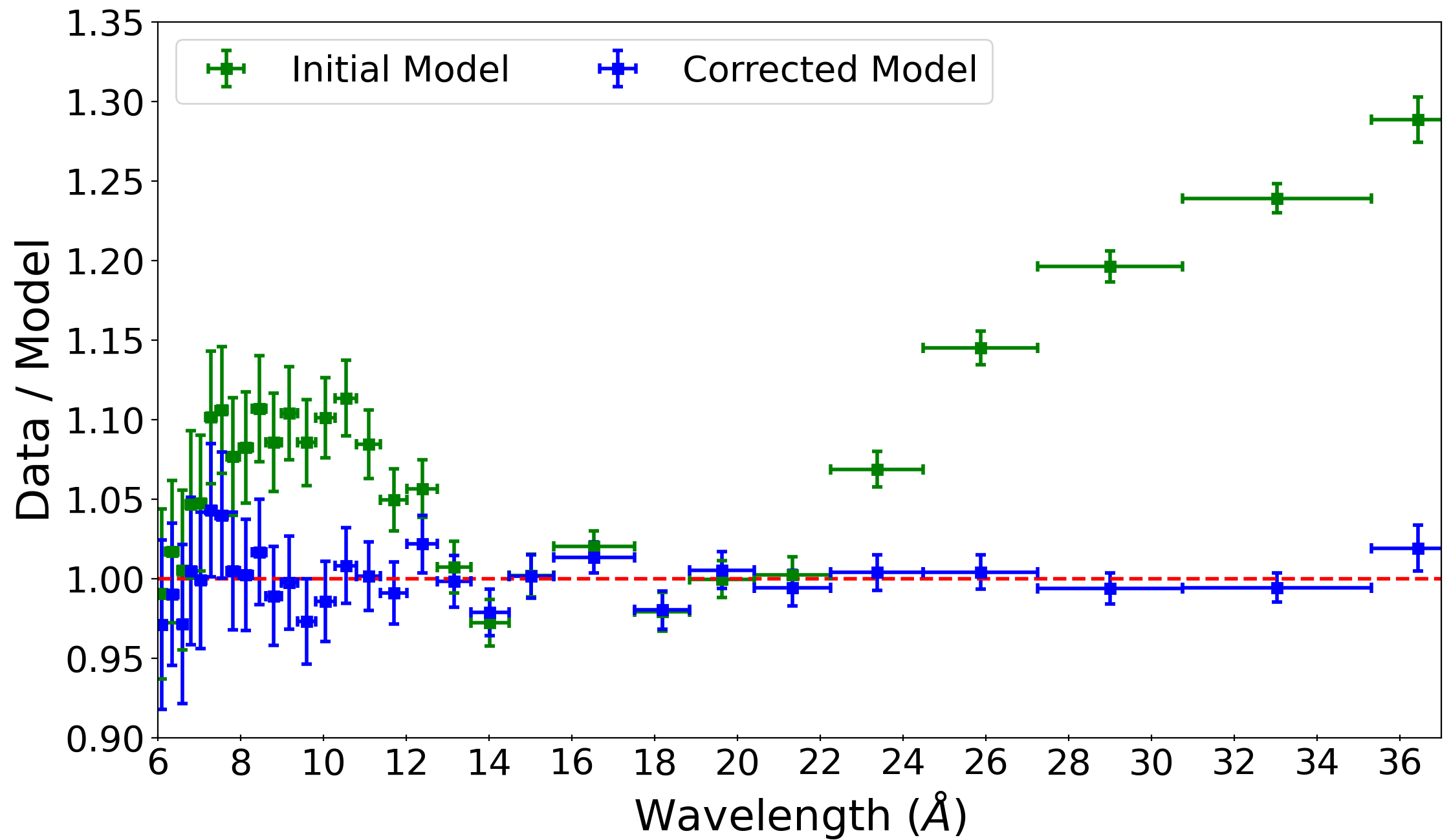}
		\end{subfigure}%
		\begin{subfigure}{0.5\linewidth}
			\includegraphics[width=1\linewidth]{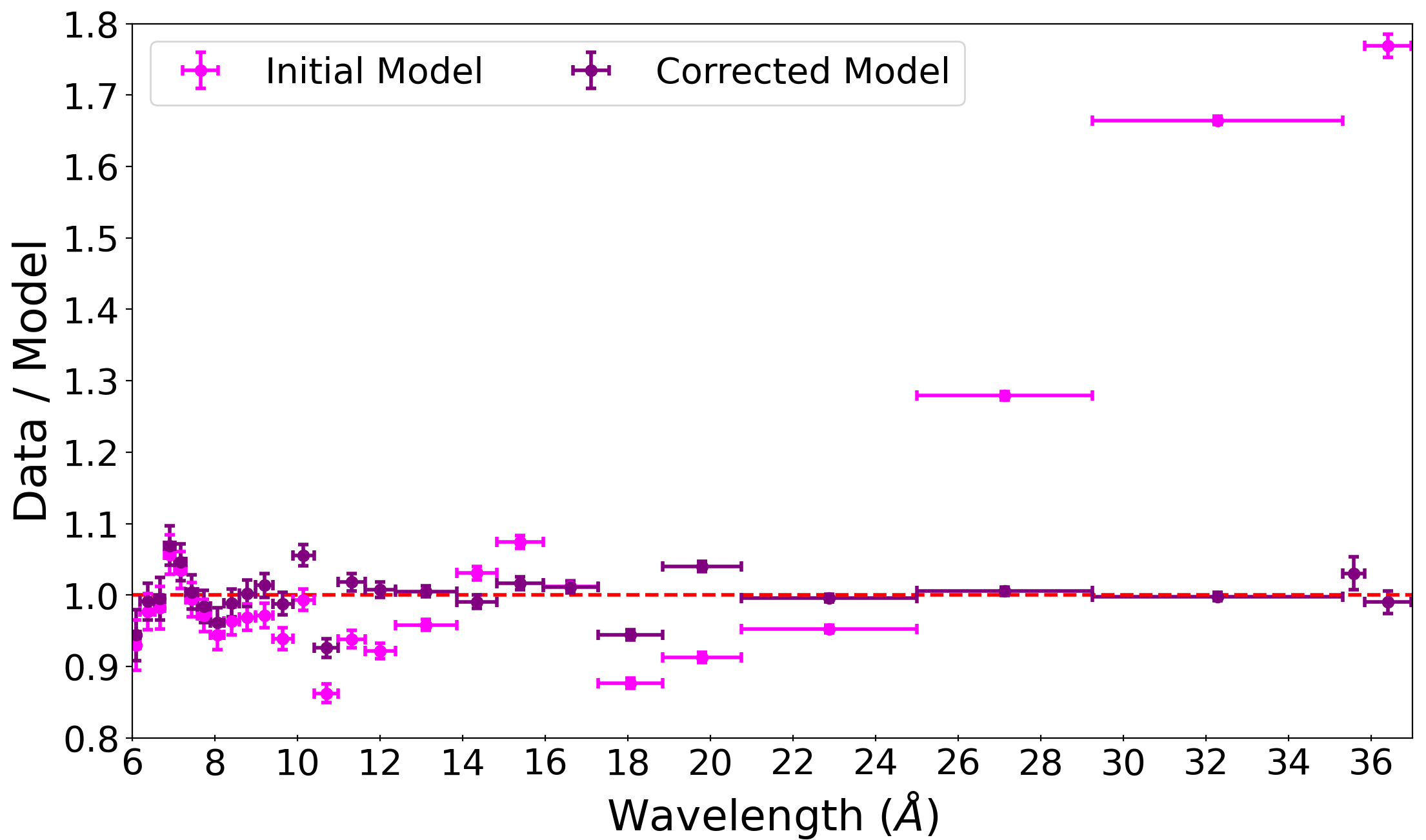}
		\end{subfigure}
		\begin{subfigure}{0.5\linewidth}
			\includegraphics[width=1\linewidth]{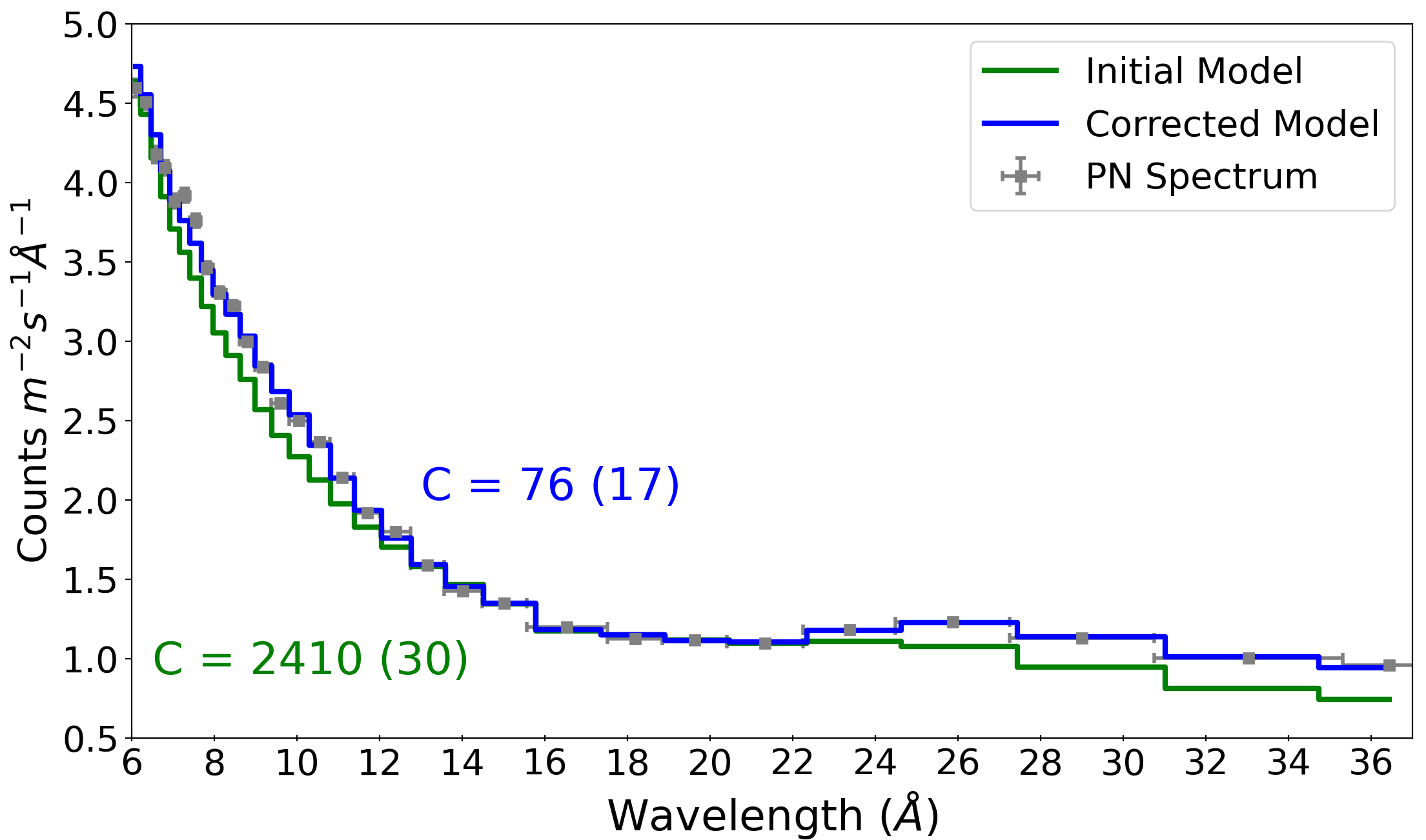}
		\end{subfigure}%
		\begin{subfigure}{0.5\linewidth}
			\includegraphics[width=1\linewidth]{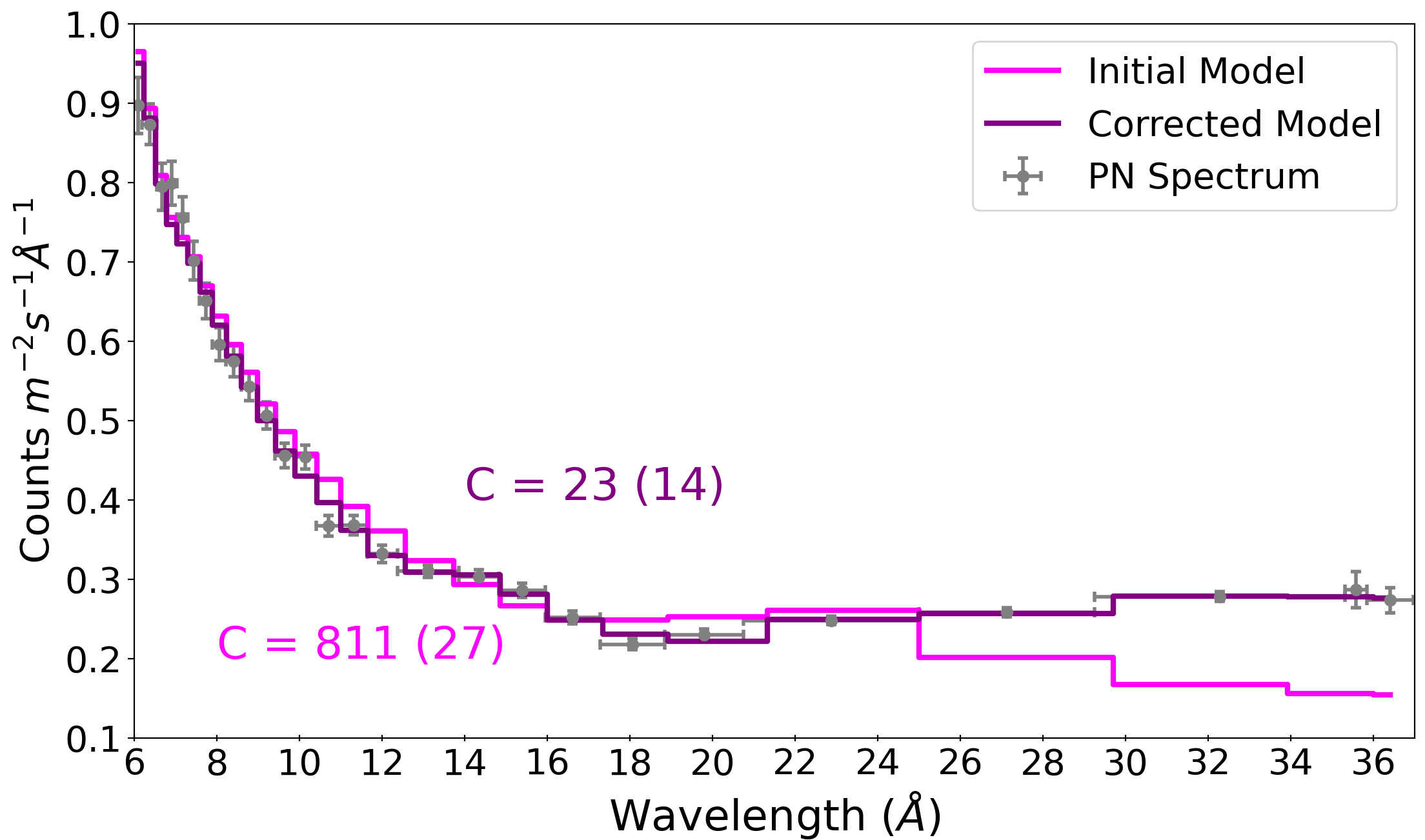}
		\end{subfigure}
		\caption{Steps taken to correct for the poor initial model fit caused by the RGS and EPIC-PN cross calibration issue (see Appendix \ref{Appen:KNAK_Corr} for details). \textit{Top:} The ratio between the data and model was used in the \texttt{KNAK} model ($\lambda_N$ and $T_N$) for the correction. \textit{Bottom:} The EPIC-PN spectrum between 6 - 36 \AA\ with the two models on top. The \texttt{KNAK} corrected model was fitted to the data here. In Obs 1 (left) green is the initial model from \citetalias{Mao_Prep} and blue is the corrected model; in Obs 2 (right) pink and purple are the initial and corrected models, respectively. The \texttt{KNAK} component only works for a wavelength grid, and therefore the units of this figure are in Angstroms rather than keV.}
		\label{Fig:Ratio_Compare}
	\end{figure*}

	This section explains how we corrected for the cross-calibration issue between the RGS and EPIC-PN instruments, presented in Sect. \ref{Sec:Fit_Spec}. We first obtained the ratio between the data ($D_{\rm PN}$) and the initial model ($M_{\rm PN}$) from \citetalias{Mao_Prep} for the EPIC-PN spectra ($R_{\rm PN} = \frac{D_{\rm PN}}{M_{\rm PN}}$; from Fig. \ref{Fig:Model_Compare}), between 6 and 37 \AA\ (as this corresponds to the RGS band, shown in Fig. \ref{Fig:Ratio_Compare}). Here we assumed that the model from \citetalias{Mao_Prep} fitted the RGS data very well, such that the ratio between the data ($D_{\rm RGS}$) and model ($M_{\rm RGS}$) was approximately one ($R_{\rm RGS} = \frac{D_{\rm RGS}}{M_{\rm RGS}} \sim 1$). Next, we applied a \texttt{KNAK} component to the model \citep[similar to][]{Porquet2004, Detmers2010}, which is a piecewise broken power law. \texttt{KNAK} splits the model into continuous segments, multiplying the model by a factor $T_N$, set to $R_{\rm PN}$ taken from Fig. \ref{Fig:Ratio_Compare} (top panel), for a given wavelength ($\lambda_N$), where \textit{N} are the segment number (the maximum is nine per \texttt{KNAK} component). \texttt{KNAK} only works for a wavelength grid, hence Fig. \ref{Fig:Ratio_Compare} is in units of Angstroms rather than in keV (see the \texttt{SPEX} manual for more details\footnote{\url{https://www.sron.nl/astrophysics-spex/manual}}). The \texttt{KNAK} model describes the instrumental effects and not the physics of NGC 3227, therefore when applying the \texttt{KNAK} component we kept the initial model parameters fixed. In order to be consistent outside the selected range, the lowest wavelength ($\lambda_1$) was fixed to 0.1 \AA\ and $T_1$ was coupled to $T_2$. For the opposite end, $\lambda_9$ was fixed at 100 \AA\ and $T_9$ was coupled to $T_8$. For the remaining seven segments, we chose $R_{\rm PN}$ and its respective wavelength by eye from Fig. \ref{Fig:Ratio_Compare}. Once the \texttt{KNAK} parameters ($T_N$ and $\lambda_N$) were fitted, we kept them fixed throughout the rest of the EPIC-PN modelling (Sect. \ref{Sec:Fit_Spec}). We repeated the above for Obs 2, as shown in the right side of Fig. \ref{Fig:Ratio_Compare}.
	
	The top panels of Fig. \ref{Fig:Ratio_Compare} shows the ratio between the data and the model for Obs 1 (left) and Obs 2 (right). The ratios were taken from Fig. \ref{Fig:Model_Compare}, where the initial model was taken directly from \citetalias{Mao_Prep}, and the corrected model was achieved after the \texttt{KNAK} corrections. The bottom panels in Fig. \ref{Fig:Ratio_Compare} display the EPIC-PN spectrum from each observation in the RGS wavelength range. Again, the initial and corrected models are shown in the bottom panels of Fig. \ref{Fig:Ratio_Compare}, where the C-statistic and d.o.f (in parentheses) are shown. After making the corrections in each observation, the \texttt{KNAK} parameters were fixed for the remainder of the spectral fitting (Sect. \ref{Sec:Fit_Spec}).
	
	In both observations (Fig. \ref{Fig:Ratio_Compare}), there are significant differences between the data and the model at longer wavelengths. However, for Obs 1 (left side), there are strong residuals between 7 and 12 \AA. The difference between the data and model in the initial fits for each observation, although taken less than a month a part, could be due to how well the initial model from \citetalias{Mao_Prep} was constrained, or whether there are some unresolved features present in the EPIC-PN data that were fitted with RGS. Furthermore, as Obs 1 has a larger flux compared to Obs 2, any significant differences between the RGS and EPIC-PN spectra will be more obvious, and hence why there are more residuals in Obs 1 compared to Obs 2. This is why we carried out the \texttt{KNAK} corrections.
	
	\section{Additional plots}
	\label{Appen:Add_Plots}
\begin{figure*}[!tbp]
		\centering
		\begin{subfigure}{0.5\linewidth}
			\includegraphics[width=1\linewidth]{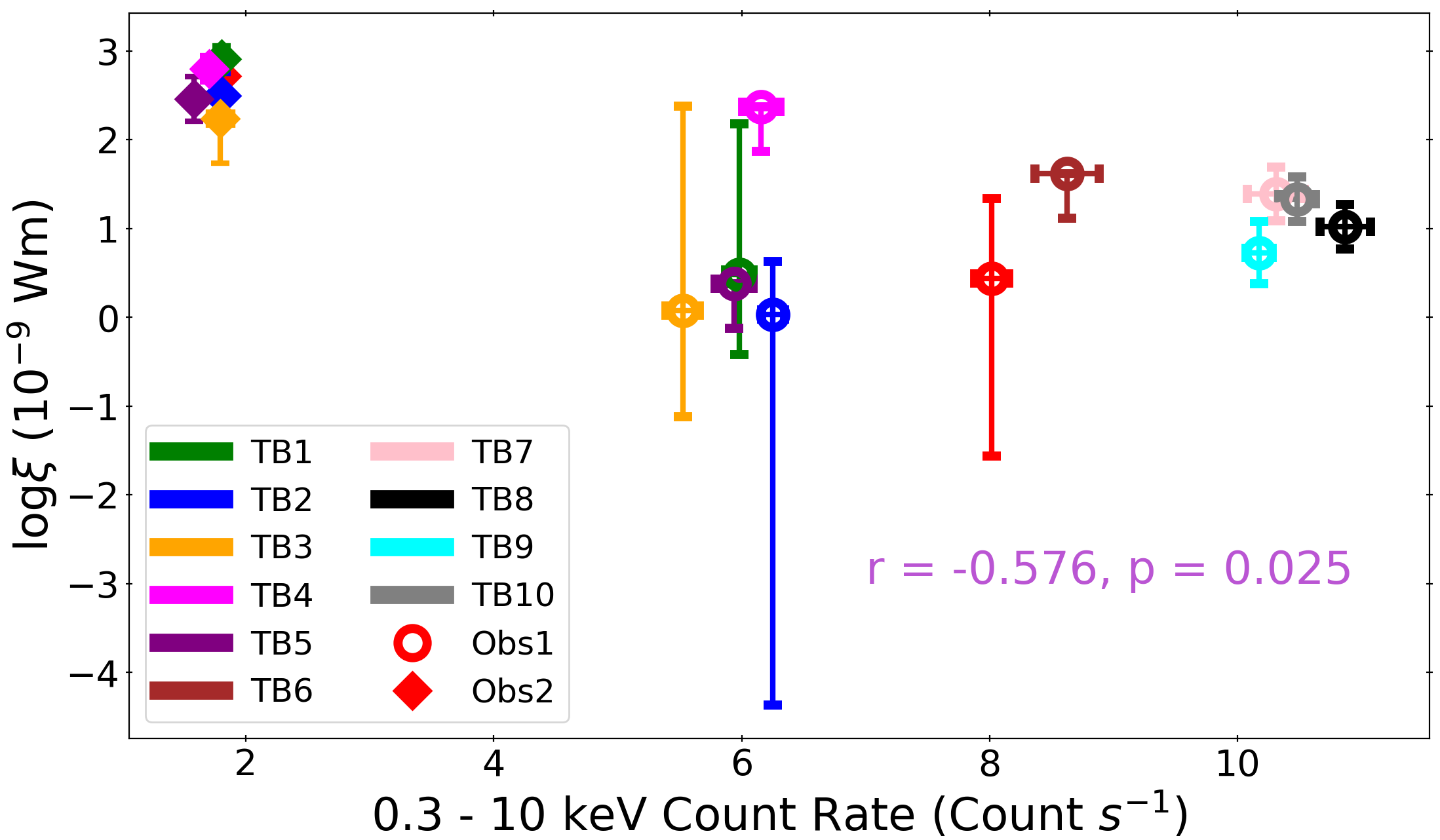}
		\end{subfigure}%
		\begin{subfigure}{0.5\linewidth}
			\includegraphics[width=1\linewidth]{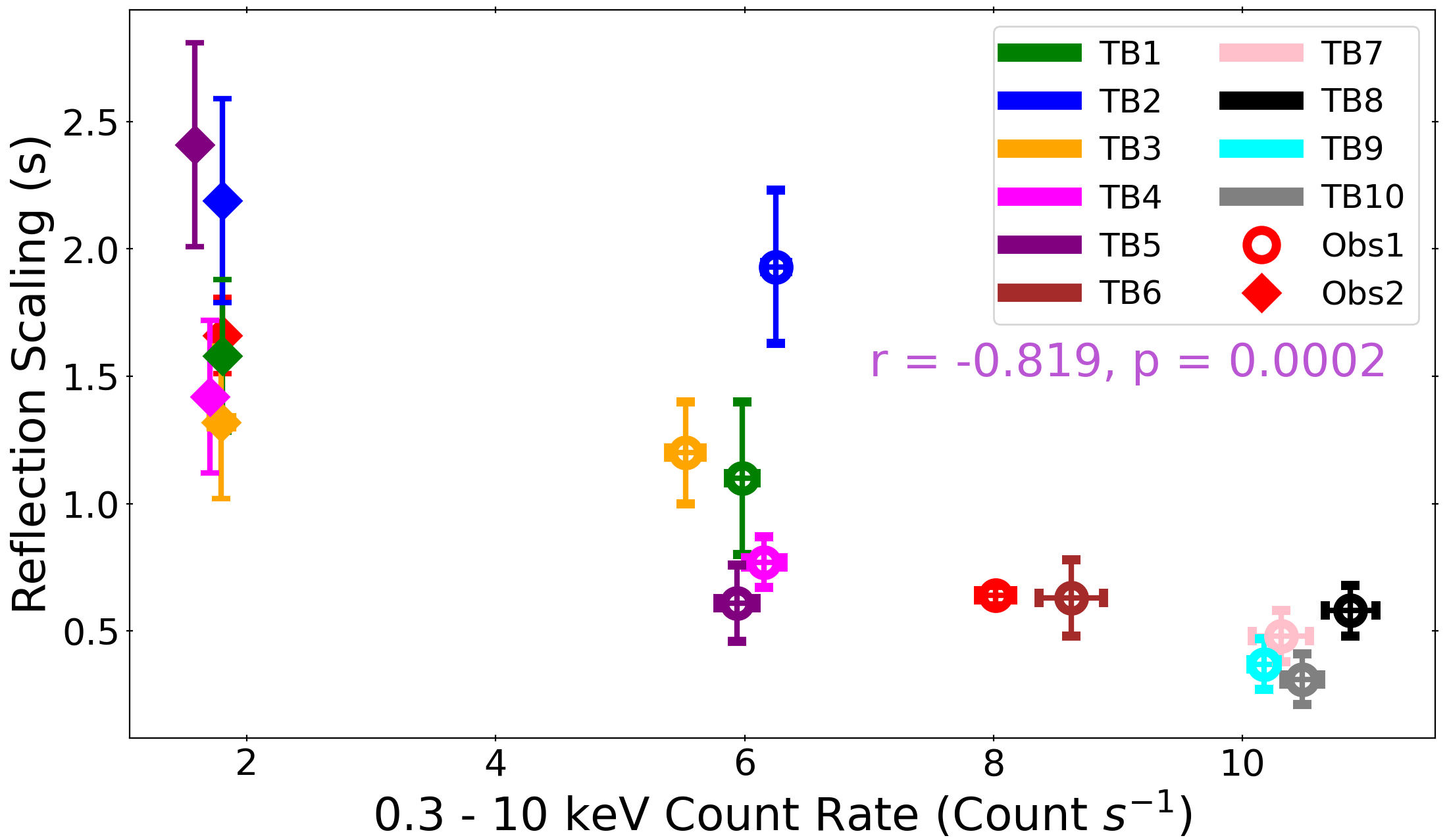}
		\end{subfigure}
		\begin{subfigure}{0.5\linewidth}
			\includegraphics[width=1\linewidth]{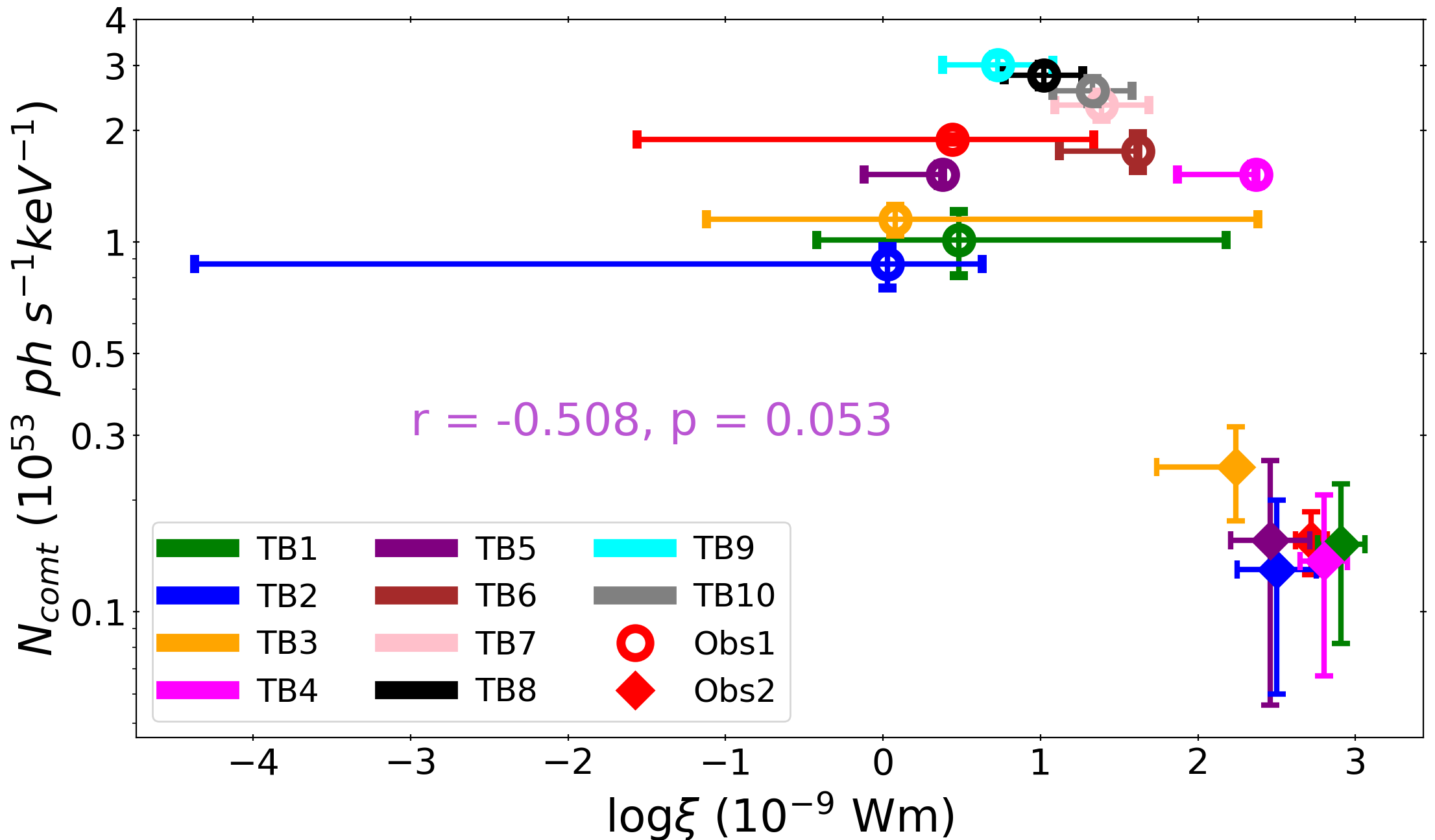}
		\end{subfigure}%
		\begin{subfigure}{0.5\linewidth}
			\includegraphics[width=1\linewidth]{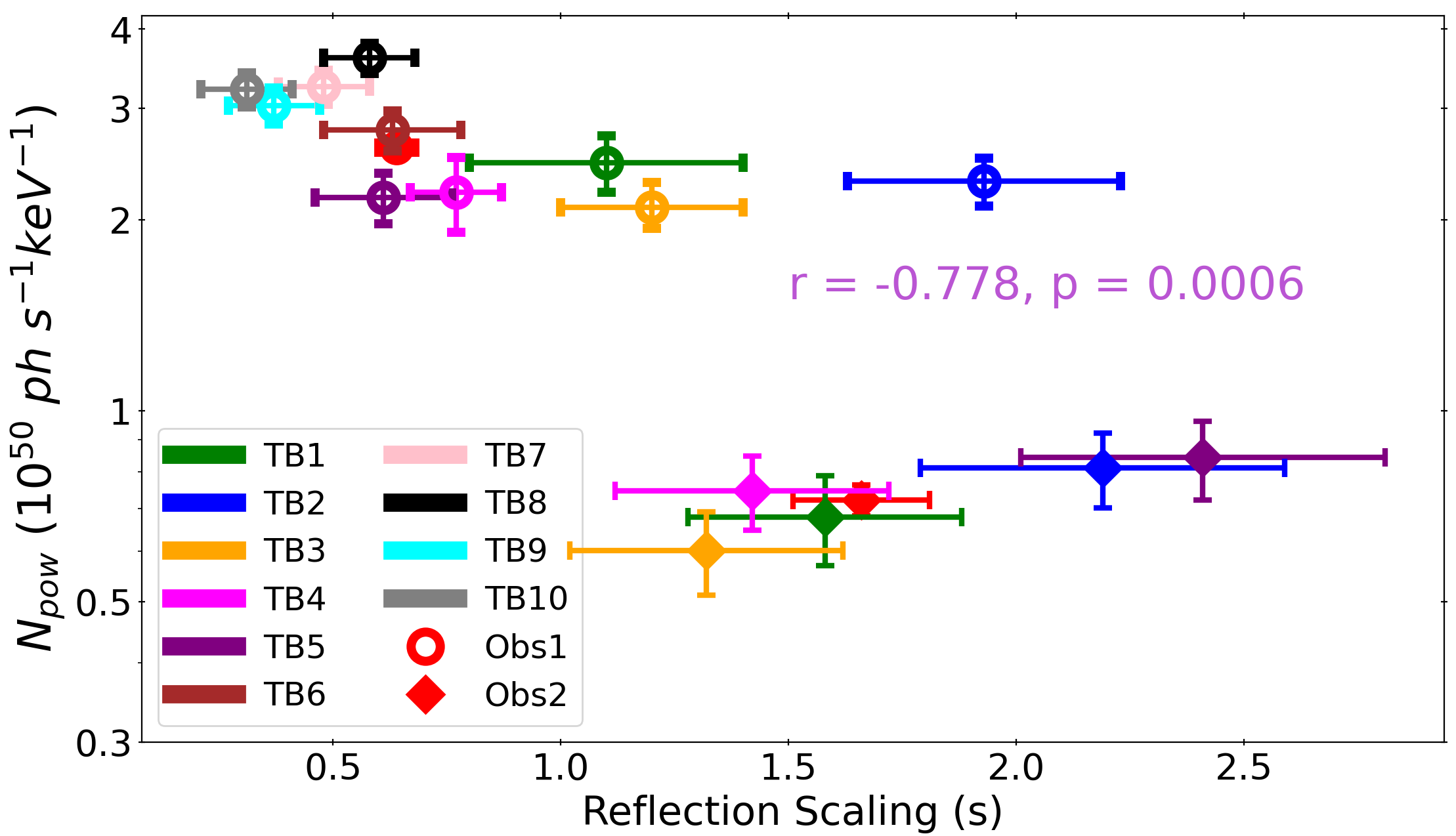}
		\end{subfigure}
		\begin{subfigure}{0.5\linewidth}
			\includegraphics[width=1\linewidth]{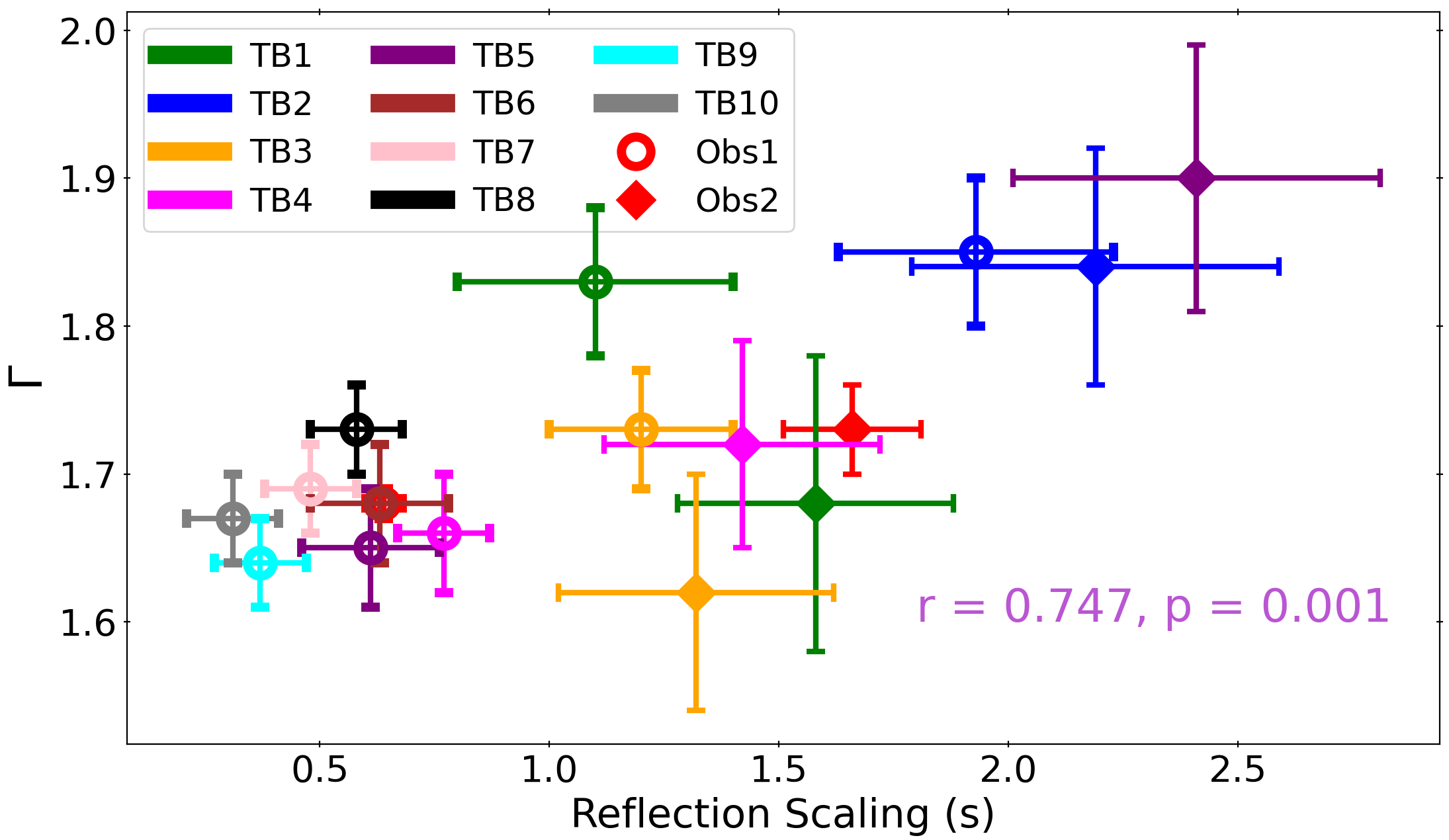}
		\end{subfigure}%
		\begin{subfigure}{0.5\linewidth}
			\includegraphics[width=1\linewidth]{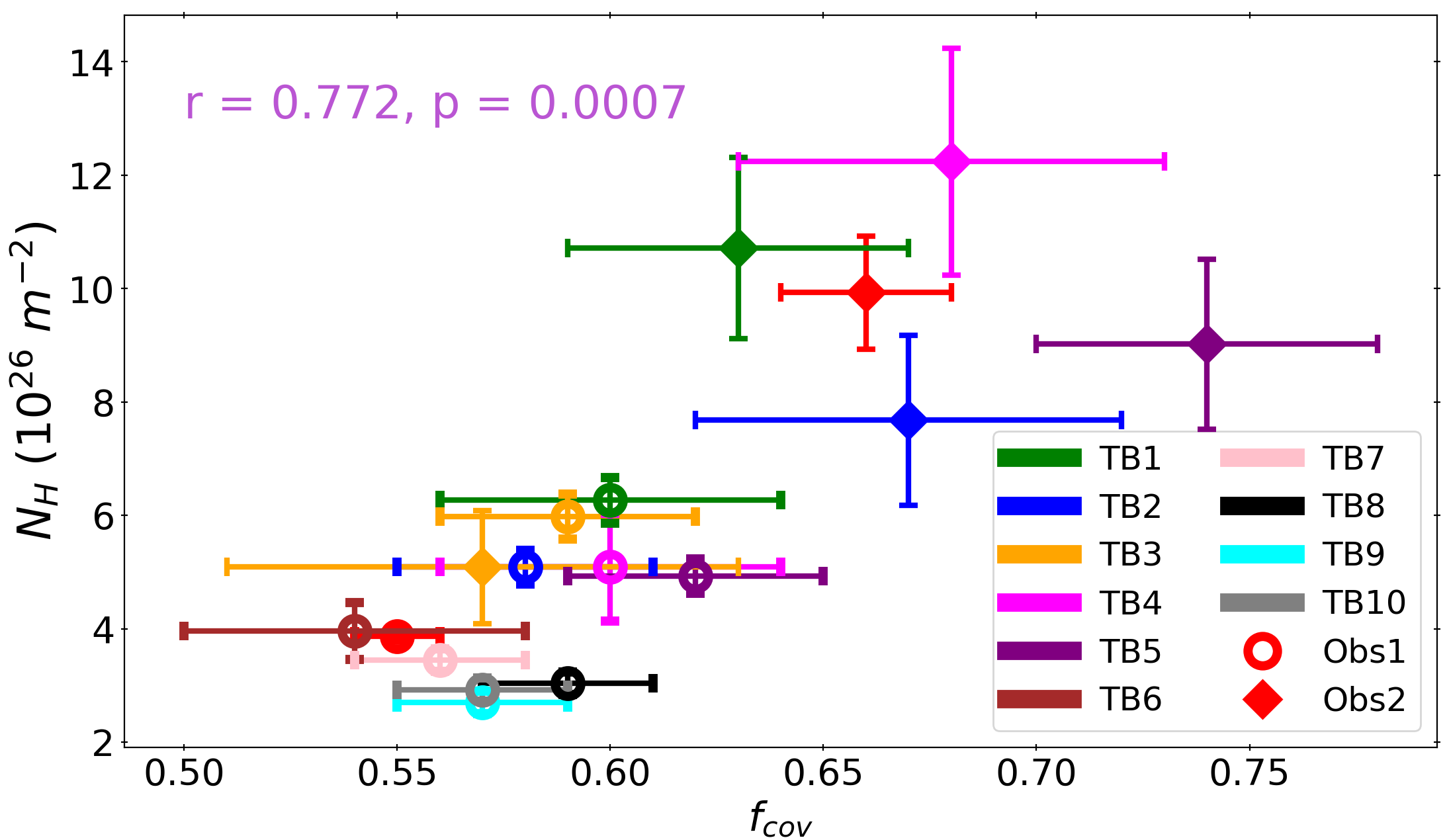}
		\end{subfigure}
		\begin{subfigure}{0.5\linewidth}
			\includegraphics[width=1\linewidth]{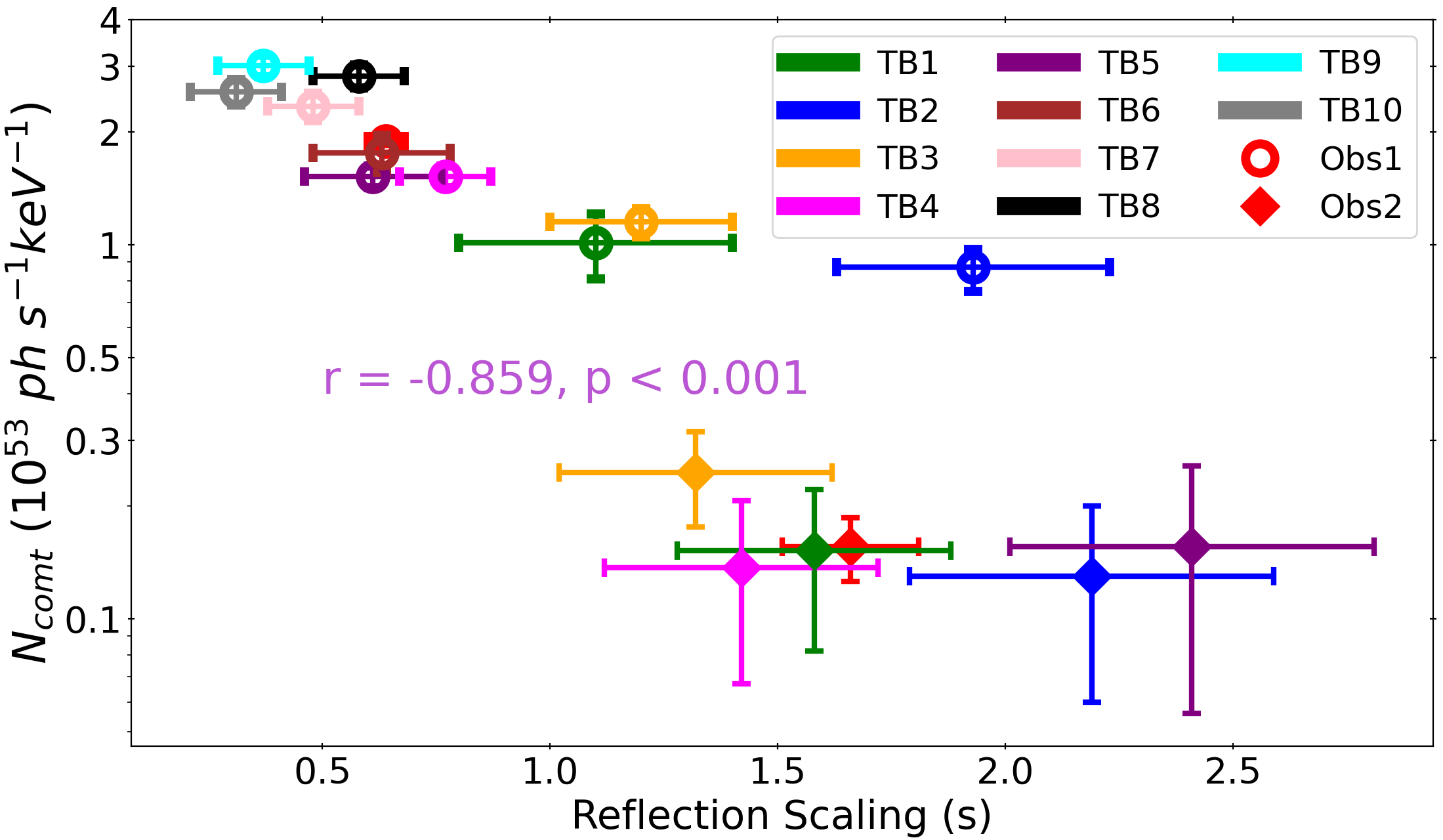}
		\end{subfigure}%
		\begin{subfigure}{0.5\linewidth}
			\includegraphics[width=1\linewidth]{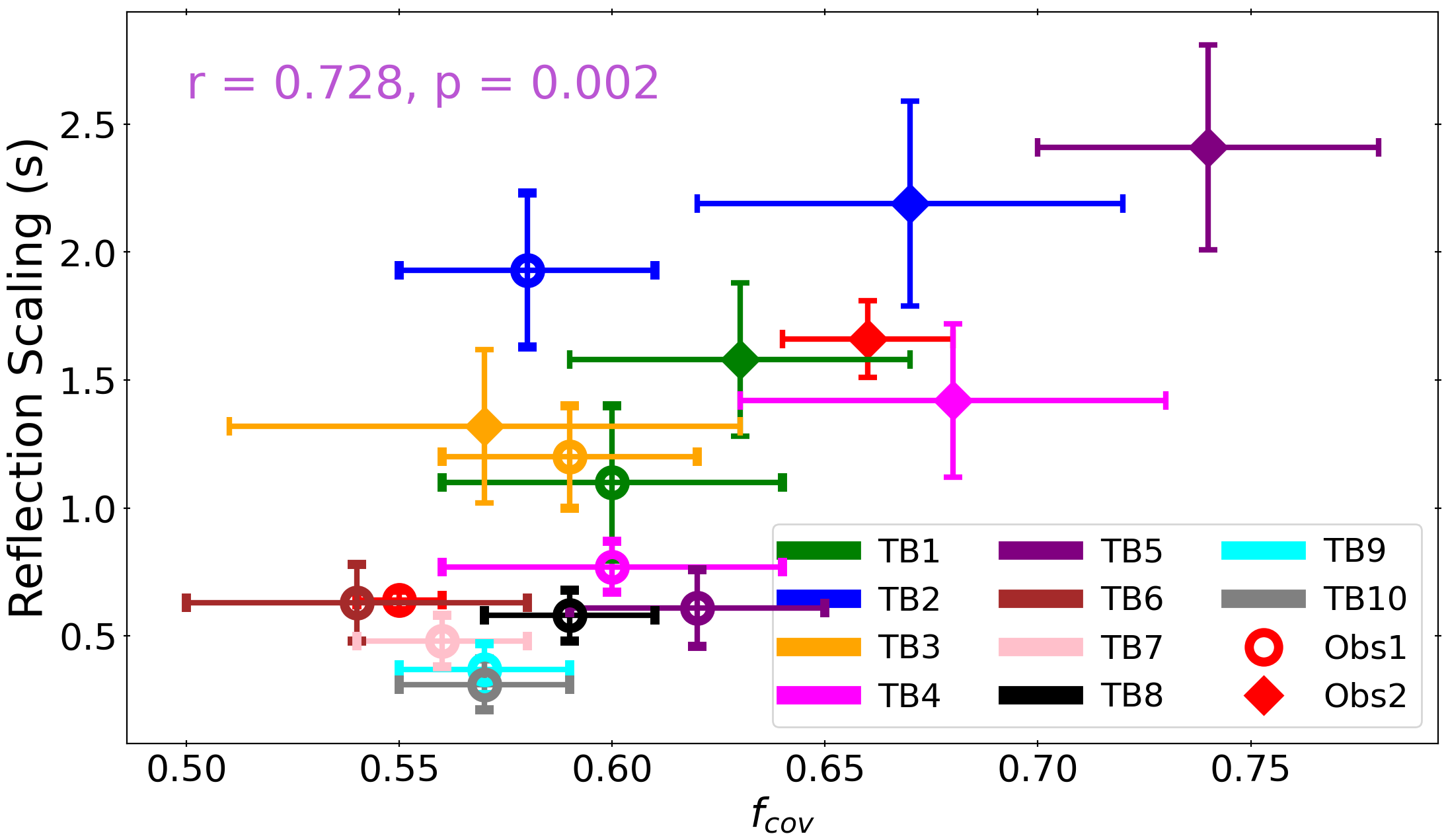}
		\end{subfigure}
		\caption{Comparing the relations between parameters in the model, in addition to the parameters discussed in Sect. \ref{Sec:Par_Comp} and Fig. \ref{Fig:Parameter_Compare}. Ionisation parameter (top left) and reflection scaling parameter (top right) against the 0.3 - 10 keV count rates. The remaining panels display the correlations between the different parameters in the model. In each panel, the Pearson rank (\textit{r}) and \textit{p}-value is shown to display the significance of each correlation.}
		\label{Fig:Appen_Parameter_Compare}
	\end{figure*}

	Figure \ref{Fig:Appen_Parameter_Compare} displays additional correlations between the fitted parameters in the models for Obs 1 and Obs 2, further to the parameter comparisons in Fig. \ref{Fig:Parameter_Compare} (Sect. \ref{Sec:Par_Comp}).
	
	In addition to Fig. \ref{Fig:Parameter_Compare}, we found some, albeit less conclusive, trends between parameters which we present in Fig. \ref{Fig:Appen_Parameter_Compare} in Appendix \ref{Appen:Add_Plots}. These include $N_{pow}$, $N_{\rm comt}$, $\Gamma$, and $f_{\rm cov}$ against $s$, $N_{\rm H}$ against $f_{\rm cov}$, and $s$ compared to the 0.3 - 10 keV count rates. Furthermore, Fig. \ref{Fig:Appen_Parameter_Compare} displays $\xi$ against the 0.3 - 10 keV count rates, and $N_{\rm comt}$ against $\xi$. The former comparison shows no strong correlation, as discussed in Sect. \ref{Sec:TB_Comp}, and the latter comparison does not show a reliable trend, with some outliers. In each panel of Fig. \ref{Appen:Add_Plots}, the Pearson rank and \textit{p}-values are shown to display how well correlated the parameters are, and the significance of each trend. In the majority of the panels in Fig. \ref{Appen:Add_Plots}, the correlations are not that significant as $p = 0.001 - 0.02$, however some panels in Fig. \ref{Fig:Appen_Parameter_Compare} have values that are lower. In some panels, the absolute values of $r$ values are greater than 0.7 (both in the positive and negative directions) implying that these parameters are somewhat well correlated or anti-correlated with each other, however less significantly compared to the parameters in Fig. \ref{Fig:Parameter_Compare}. The panels comparing $\log \xi$ with either the count rates or $N_{\rm comt}$ values have $r = -0.5$ to $-0.6$, suggesting that the ionisation state does not correlate or change with the continuum; similar to the bottom panel in Fig. \ref{Fig:TB_Compare}.
	
	\section{Expansion velocity derivation}
	\label{Appen:Expansion}
	
	Following on from Sect. \ref{Sec:Var_Cause}, one way in which the column density of the obscurer could change is if the cloud expanded, causing $N_{\rm H}$ to decrease. Assuming that the obscurer cloud is spherical, it will have a mass given by
	\begin{equation}
	M = 4 \pi r^2 m_p \mu N_H f_{cov},
	\end{equation}
	where $m_p$ is the proton mass, $\mu$ is the mean atomic mass per proton ($\sim 1.4$ for solar values), $r$ is the radius of the cloud, and $N_{\rm H}$ and $f_{\rm cov}$ are the column density and covering fraction of the obscurer from the modelling in Sect. \ref{Sec:Spectral_Fitting}. If the obscuring cloud does expand during Obs 1, and we assume the mass of the obscurer stays constant, we can equate the masses from TB1 and TB10 as follows
	\begin{equation}
	M = 4 \pi r^2 m_p \mu N_{H, 1} f_{cov, 1} = 4 \pi R^2 m_p \mu N_{H, 10} f_{cov, 10},
	\end{equation}
	where $r$ and $R$ are the radii of the obscurer in TB1 and TB10, respectively.
	
	A quick rearrangement gives $R^2 = A r^2$ where $A = \frac{N_{H, 1} f_{cov, 1} }{N_{H, 10} f_{cov, 10}}$. Substituting in the column density and covering fraction values for TB1 and TB10 from Table \ref{Table:Full_Spec_Results}, the result yields $R = \sqrt{2.25}\ r$.
	
	Here, we set $r = \frac{N_{H,1}}{n_H}$ for TB1, where $n_H$ is the hydrogen number density. From \citetalias{Mao_Prep} (Table 4), $n_H = 0.1 - 2.7 \times 10^{15}$ m\textsuperscript{-3} for a spherical cloud. Taking the average of $n_{\rm H}$ gives an obscurer radius in TB1 of $r = 2.24 \times 10^{11}$ m, and therefore yields a radius after expansion in TB10 of $R = 3.36 \times 10^{11}$ m. So the expansion distance travelled during Obs 1 would therefore be $D = R - r = 1.12 \times 10^{11}$ m.
	
	Finally, the expansion velocity can be estimated if we take the time to be $t = 103$ ks, and as such $v_{\rm exp} = D/t \sim 1120$ km s\textsuperscript{-1}. Therefore, the obscurer would have to expand with a velocity of around 1120 km s\textsuperscript{-1}, which is in the middle of the estimated crossing velocity ($v_{\rm cross}$) of the obscurer (between 680 and 1470 km s\textsuperscript{-1}), to explain the observed changes in the column density between TB1 and TB10. This expansion scenario to explain the observed decrease in $N_{\rm H}$ for Obs 1 is rather unlikely as the expansion speed is similar to the travelling velocity.
\end{document}